\documentstyle[11pt]{article}
\newcommand{\anti}[1]{\overline{#1}}
\newcommand{\winkel}{\;\hbox{$\rlap{)}\!\!\!<$}\,}
\newcommand{\ohalf}{{1\over2}}
\newcommand{\bra}[1]{\langle \, #1 \, \vert \>}
\newcommand{\ket}[1]{\> \vert \, #1 \, \rangle}
\newcommand{\kket}[1]{#1 \,\rangle}
\newcommand{\unity}{{\rm 1 \>\! \llap{I}}}

\newcommand{\half}{{1\over 2}}
\newcommand{\thalf}{{3\over 2}}

\begin{document}
\title{CORRELATED $\pi\pi$ AND $K\anti{K}$ EXCHANGE
IN THE BARYON-BARYON INTERACTION}

\author{A. Reuber, K. Holinde,
H.-C. Kim\thanks{Present address: Institut f\"ur Theoretische Physik
                     II, Ruhr-Universit\"at Bochum, 44780 Bochum,
                     Germany}, and J. Speth \\ Institut f\"ur
                     Kernphysik (Theorie),\\ Forschungszentrum
                     J\"ulich GmbH,\\ D-52425 J\"ulich, Germany}

\date{\today}

\maketitle

\begin{abstract}
The exchange of two correlated pions or kaons provides the main part
of the intermediate-range attraction between two baryons.
Here, a dynamical model for correlated two-pion and
two-kaon exchange in the baryon-baryon interaction is
presented, both  in the scalar-isoscalar ($\sigma$) and the
vector-isovector ($\rho$) channel.
The contribution of correlated $\pi\pi$ and $K\anti{K}$ exchange is
derived  from the amplitudes for the transition
of a baryon-antibaryon  state ($B\anti{B'}$) to a $\pi\pi$ or
$K\anti{K}$ state in the pseudophysical region by applying
dispersion theory and unitarity.
For the $B\anti{B'}\to \pi\pi, K\anti{K}$ amplitudes a microscopic
model is constructed, which is  based on the hadron-exchange picture.
 The Born terms include contributions from baryon-exchange as well as
$\rho$-pole diagrams. The correlations between the two
pseudoscalar mesons are taken into account by means of
$\pi\pi$-$K\anti{K}$ amplitudes derived likewise from a meson-exchange
model, which is in line with the empirical $\pi\pi$ data.
The parameters of the $B\anti{B'}\to \pi\pi, K\anti{K}$
model, which are related to each other by the assumption of
$SU(3)$ symmetry,  are determined by the
adjustment to the quasiempirical $N\anti{N}\to \pi\pi$
amplitudes in the pseudophysical region.
It is found that correlated $K\anti{K}$ exchange plays an
important role in the $\sigma$-channel for baryon-baryon
states with non-vanishing strangeness.
The strength of correlated $\pi\pi$ plus $K\anti{K}$ exchange in the
$\sigma$-channel decreases with the strangeness of the baryon-baryon
system becoming more negative.
Due to the admixture of baryon-exchange processes to the
$SU(3)$-symmetric $\rho$-pole contributions
 the results for correlated $\pi\pi$-exchange in the vector-isovector
channel deviate from what is expected in the
naive $SU(3)$ picture for genuine $\rho$-exchange.
In present models of the hyperon-nucleon interaction contributions
of correlated $\pi\pi$ and $K\anti{K}$ exchange are parametrized for
simplicity
by single $\sigma$ and $\rho$ exchange.
Shortcomings of this effective description, e.g.\ the missing
long-range contributions, are pointed out by comparison with
the dispersiontheoretic results.
\end{abstract}

\section{Introduction}

The study of the role of strangeness degrees of freedom in low energy
nuclear physics is of high current interest since it should lead to a
deeper understanding of the relevant strong interaction mechanisms in
the non-perturbative regime of QCD. For example, the system of a
strange baryon (hyperon $Y$) and a nucleon ($N$) is in principle an
ideal testing ground to investigate the importance of $SU(3)_{\rm
flavor}$ symmetry for the hadronic interactions. This symmetry is
obviously broken already by the different masses of hadrons sitting in
the same multiplet. However the important question arises whether (on
the level of hadrons) it is broken not only kinematically but also
dynamically, e.g.\ in the values of the coupling constants at the
hadronic vertices. The answer cannot be given at the moment since the
present empirical information about the $YN$ interaction is too scarce
and thus prevents any definite conclusions.  Hopefully the situation
will be improved by experiments of elastic $\Sigma^\pm p$, $\Lambda
p$, and even $\Xi p$, currently performed at KEK~\cite{Ieiri,Imai}.

Existing meson exchange models of the $YN$ interaction assume for the
hadronic coupling constants at least $SU(3)$ symmetry, in case of
models A and B of the J\"ulich group~\cite{Holz} even $SU(6)$ of the
static quark model.  This symmetry requirement provides relations
between coupling constants of a meson multiplet to the baryon current,
which strongly reduce the number of free model
parameters. Specifically, coupling constants at the strange vertices
are then connected to nucleon-nucleon-meson coupling constants, which
in turn are fixed between close boundaries by the wealth of empirical
$NN$ scattering information. All $YN$ interaction models can reproduce
the existing empirical $YN$ scattering data. Therefore at present the
assumption of SU(3) symmetry for the coupling constants is not in
conflict with experiment.

However, the treatment of the scalar-isoscalar meson sector, which
provides the intermediate range baryon-baryon interaction, is
conceptionally not very convincing so far. The
one-boson-exchange models of the Nijmegen group start from the
existence of a broad scalar-isoscalar $\pi\pi$ resonance
($\epsilon$-meson, $m_\epsilon=760\,MeV$, $\Gamma_\epsilon=640\,MeV$),
which is hidden in the experiment (e.g.\ $\pi N\to\pi\pi N$) under the
strong $\rho^0$-signal and can therefore not be identified
reliably. For practical reasons the exchange of this broad
$\epsilon$-meson is then approximated by the exchange of a sum of two
mesons with sharp mass $m_1$ and $m_2$; the smaller mass is around
$500\,MeV$ and thus corresponds to the phenomenological $\sigma$-meson
in conventional OBE-models. The $\epsilon$-meson is then treated as
$SU(3)$ singlet (model D~\cite{NijII}) or as member of a full nonet of
scalar mesons (model F~\cite{NijIII} and NSC~\cite{NijIV}). In the
$SU(3)$ framework the $\epsilon$-coupling strength is equal for all
baryons in model D, while it depends on four open $SU(3)$ parameters
in models F and NSC,
which are adjusted to $NN$ and $YN$ scattering data. In the latest
Nijmegen model NSC~\cite{NijIV} the scalar meson nonet includes apart
from the $\epsilon$ the isoscalar $f_0(975)$, the isovector $a_0(980)$
and the strange mesons $\kappa$, which the authors
identify~\cite{deSwMaui} with scalar $q^2\bar q^2$-states predicted by
the MIT bag model~\cite{MIT}. This interpretation is however doubtful,
at least for the $f_0(975)$ and $a_0(980)$. According to a recent
theoretical analysis of the $\pi\pi$, $\pi\eta$, and $K\anti{K}$
system~\cite{Janssen} in the meson exchange framework the $f_0(975)$
is a $K\anti{K}$ molecule bound in the $\pi\pi$ continuum, while
$a_0(980)$ is dynamically generated by the $K\anti{K}$ threshold. Thus
both mesons do not appear to be genuine quark model resonances, with
the consequence that $SU(3)$ relations should not be applied to these
mesons. (The non-strange members of the scalar nonet are
expected to be at higher energies).

In the Bonn potential~\cite{MHE} the intermediate range attraction is
provided by uncorrelated (Fig.~\ref{fig:1_2}a,b) and correlated
(Fig.~\ref{fig:1_2}c) $\pi\pi$ exchange processes with $NN$, $N\Delta$
and $\Delta\Delta$ intermediate states. It is known from the study of
the $\pi\pi$ interaction that the $\pi\pi$ correlations are important
mainly in the scalar-isoscalar and vector-isovector channel. The Bonn
potential includes such correlations, however only in a rough way,
namely in terms of sharp mass $\sigma'$ and $\rho$ exchange. One
disadvantage of such a simplified treatment is that this
parametrization cannot be transported into the hyperon sector in a
well defined way. Therefore in the $YN$ interaction models of the
J\"ulich group~\cite{Holz}, which start from the Bonn $NN$ potential,
the coupling constants of the fictitious $\sigma'$-meson at the
strange vertices ($\Lambda\Lambda\sigma'$, $\Sigma\Sigma\sigma'$) are
essentially free parameters.  In view of the little empirical
information about the $YN$ interaction this feature is not
satisfactory. This is especially true for an extension of the $YN$
models to baryon-baryon channels with strangeness $S=-2$. So far there
is no empirical information about these channels (apart from some data
on $\Xi$- and $\Lambda\Lambda$-hypernuclei). Still there is a large
interest in these channels initiated by the prediction of the
H-dibaryon by Jaffe~\cite{Jaffe}. The H-dibaryon is a deeply bound
6-quark state with the same quark content as the $\Lambda\Lambda$
system ($uuddss$) and with $^1S_0$ quantum numbers. For the
experimental search it is important to know whether conventional
deuteron-like $\Lambda\Lambda$ states exist. An analysis of possible
$S=-2$ bound states in the meson exchange framework could provide
valuable information in this regard, but requires a coupled channels
treatment of $\Lambda\Lambda$, $\Sigma\Sigma$, and $N\Xi$ channels. An
extension of the J\"ulich $YN$ models to those channels is only of
minor predictive power since the strength of the important
$\Xi\Xi\sigma'$ vertex is completely undetermined and cannot be fixed
by empirical data.

These problems can be overcome by an explicit evaluation of correlated
$\pi\pi$ exchange processes in the various baryon-baryon channels. A
corresponding calculation has been done for the $NN$ case
(Fig.~\ref{fig:1_2}c)~\cite{Kim}.  Starting point was a fieldtheoretic
model for both the $N\anti{N}\to\pi\pi$ Born amplitudes and the
$\pi\pi$-$K\anti{K}$ interaction~\cite{Lohse}. With the help of
unitarity and dispersion relations the amplitude for the correlated
$\pi\pi$ exchange in the $NN$ interaction has been determined, showing
characteristic discrepancies to $\sigma'$ and $\rho$ exchange in the
(full) Bonn potential.

For the correct description of the $\pi\pi$ interaction in the
scalar-isoscalar channel the coupling to the $K\anti{K}$ channel is
essential, which is obvious from the interpretation of the $f_0(975)$
as a $K\anti{K}$ bound state.  Apart from the $\pi\pi$-$K\anti{K}$
interaction model the $K\anti{K}$ channel is not considered in
Ref.~\cite{Kim}, i.e. the coupling of the kaon to the nucleon is not
taken into account.
In fact, this approximation is justified in the $NN$
system~\cite{Durso}; it is however not expected
to work in channels involving hyperons.

The aim of the present paper is a microscopic derivation of correlated
$\pi\pi$ as well as $K\anti{K}$ exchange processes in the various
baryon-baryon channels with $S=0, -1, -2$ (Fig.~\ref{fig:1_1}). The
$K\anti{K}$ channel is treated on an equal footing with the $\pi\pi$
channel in order to determine reliably the influence of $K\anti{K}$
correlations. Our results replace the phenomenological $\sigma'$ and
$\rho$ exchange in the Bonn $NN$ and J\"ulich $YN$ models by
correlated processes and in this way eliminate undetermined model
parameters (e.g.\ $\sigma'$ coupling constants). Corresponding
interaction models thus have more predictive power and should make a
sensible treatment of $S=-2$ baryon-baryon channels possible.

The formal treatment is similar to that of
Refs.~\cite{Kim,ALV,LinSerot} dealing with correlated $\pi\pi$
exchange in the $NN$ interaction. Due to the inclusion of the
$K\anti{K}$ channel and different baryon masses (e.g. in the
$N\Lambda$ channel) generalizations are however required at some
places. Starting point is a field-theoretic model for the
baryon-antibaryon ($B\anti{B'}$) $\to\pi\pi,\,K\anti{K}$ Born
amplitudes in the $J^P=0^+,1^-$ channels. Besides various baryon
exchange terms the model includes in complete consistency to the
$\pi\pi$-$K\anti{K}$ interaction model~\cite{Lohse,Pearce_piN} also a
$\rho$-pole term (cf. Fig.~\ref{fig:1_3}). These Born amplitudes are
analytically continued into the pseudophysical region below the
$B\anti{B'}$-threshold. The solution of a covariant scattering
equation with full inclusion of $\pi\pi-K\anti{K}$ correlations yields
the $B\anti{B'}\to\pi\pi,\,K\anti{K}$ amplitudes in the pseudophysical
region.  In the $N\anti{N}\to\pi\pi$ channel these amplitudes are then
adjusted to quasiempirical information~\cite{Hoehler2,Dumbrajs}, which
has been obtained by analytic continuation of $\pi N$ and $\pi\pi$
data. With the assumption of $SU(6)$ symmetry for the coupling
constants a parameter-free description of the other particle channels
can then be achieved.

Via unitarity relations the products of
$B\anti{B'}\to\pi\pi,\,K\anti{K}$ amplitudes fix the singularity
structure of the baryon-baryon amplitudes for $\pi\pi$ and $K\anti{K}$
exchange. Assuming analyticity for the amplitudes dispersion relations
can be formulated for the baryon-baryon amplitudes, which connect
physical amplitudes in the $s$-channel with singularities and
discontinuities of these amplitudes in the pseudophysical region of
the $t$-channel processes. With a suitable subtraction of uncorrelated
contributions, which are calculated directly in the $s$-channel and
therefore guaranteed to have the correct energy behavior, we finally
obtain the amplitudes for correlated $\pi\pi$ and $K\anti{K}$ exchange
in the baryon-baryon system.

In the next chapter we describe the underlying formalism which is
used to derive correlated $\pi\pi$ and $K\anti{K}$ exchange potentials
for the baryon--baryon amplitudes. Furthermore we present our
microscopic model for the required $B\anti{B'}\to\pi\pi,\,K\anti{K}$
amplitudes. Sect.3 contains our results and also a comparison with
those obtained from other models. The paper ends with some concluding
remarks.

\section{Formalism}


\subsection{Kinematics and amplitudes}

\label{sec:kap3_1}

The kinematics of a two-body scattering process $A+B \to C+D$ (cf.\
Fig.~\ref{fig:3_1_1})   is uniquely determined by the 4-momenta
$p_A,p_B,p_C,p_D$ of the particles.  Taking into account the
on-mass-shell relations ($p_X^2=M_X^2, X=A,\ldots,D$) and the
conservation of the total 4-momentum ($p_A+ p_B = p_C + p_D$) only two
independent Lorentz-scalars can be built out of these momenta.  For
these Lorentz-scalars one usually introduces the three Mandelstam
variables
\begin{equation}
\begin{array} {rcccl}
s&=&(p_A+p_B)^2&=&(p_C+p_D)^2 \quad,\\ t&=&(p_C-p_A)^2&=&(p_B-p_D)^2
\quad,\\
u&=&(p_D-p_A)^2&=&(p_B-p_C)^2 \quad,
\end{array}
\end{equation}
which are related by
\begin{equation}
s+t+u=M_A^2 +M_B^2 +M_C^2 +M_D^2 \equiv \Sigma \quad.
\label{eq:3_1_0}
\end{equation}

By crossing the scattering process $A+B\to C+D$ is closely related to
two other processes, as indicated by Fig.~\ref{fig:3_1_1}:
\begin{equation}
\begin{array} {lcl@{\qquad}c@{\qquad}lcl@{\qquad\qquad}l}
A,p_a& +& B,p_B &\to& C,p_C& +& D,p_D & \mbox{`$s$-channel'} \;, \\
A,p_A& +& \anti{C},-p_C &\to& \anti{B},-p_B& +& D,p_D &
\mbox{`$t$-channel'}\;, \\ A,p_A &+& \anti{D},-p_D &\to& C,p_C& +&
\anti{B},-p_B & \mbox{`$u$-channel'}\;.
\end{array}
\end{equation}
Here, the channels are named according to the Mandelstam variable
which denotes the squared total energy in the center-of-mass (c.m.)
system.

For the $s$-channel process the particle 4-momenta in the c.m.\ system
read:
\begin{equation}
p_A=\left( \begin{array}{c} E_A \\ \vec p_s \end{array} \right),\quad
p_B=\left( \begin{array}{c} E_B \\ -\vec p_s \end{array}
\right),\qquad p_C=\left( \begin{array}{c} E_C \\ \vec q_s \end{array}
\right),\quad p_D=\left( \begin{array}{c} E_D \\ -\vec q_s \end{array}
\right)
\end{equation}
with $E_X=\sqrt{M_X^2 + \vec p^{\;2}_X}$.  The modulus of the relative
momentum $\vec p_s$ ($\vec q_s$) of the initial (final) state can be
expressed in terms of $s$:
\begin{eqnarray}
{\vec p_s}\,^2 &=& { \left[ s - (M_A+M_B)^2 \right] \; \left[ s -
(M_A-M_B)^2 \right]
\over 4s}\quad,
\nonumber \\
{\vec q_s}\,^2 &=& { \left[ s - (M_C+M_D)^2 \right] \; \left[ s -
(M_C-M_D)^2 \right]
\over 4s}\quad.
\label{eq:3_1_1}
\end{eqnarray}
The scattering angle $\vartheta_s = \winkel(\vec p_s,\vec q_s)$ is
related to the Mandelstam variables by
\begin{equation}
\cos\vartheta_s= {  s (t-u) + (M_A^2 - M_B^2) (M_C^2 - M_D^2)
\over 4s |\vec p_s| |\vec q_s|}
\label{eq:3_1_2}
\end{equation}

For the $t$-channel process $A+\anti{C} \to \anti{B}+D$  the
c.m. 4-momenta of the particles are
\begin{equation}
p_A=\left( \begin{array}{c} E_A \\ \vec p_t \end{array} \right),\quad
-p_C=\left( \begin{array}{c} E_C \\ -\vec p_t \end{array}
\right),\qquad p_D=\left( \begin{array}{c} E_D \\ \vec q_t \end{array}
\right),\quad -p_B=\left( \begin{array}{c} E_B \\ -\vec q_t
\end{array} \right).
\end{equation}
The analogue of Eqs.~\ref{eq:3_1_1} for the modulus of the relative
momenta $\vec p_t$, $\vec q_t$ and the scattering angle $\vartheta_t =
\winkel(\vec p_t,\vec q_t)$ now reads:
\begin{eqnarray}
{\vec p_t}\,^2 &=& { \left[ t - (M_A+M_C)^2 \right] \; \left[ t -
(M_A-M_C)^2 \right]
\over 4t}\quad,
\nonumber \\
{\vec q_t}\,^2 &=& { \left[ t - (M_B+M_D)^2 \right] \; \left[ t -
(M_B-M_D)^2 \right]
\over 4t}\quad,
\label{eq:3_1_2a}
 \\
\cos\vartheta_t &=& {  t (u-s) + (M_A^2 - M_C^2) (M_D^2 - M_B^2)
\over 4t |\vec p_t| |\vec q_t|}\quad.
\label{eq:3_1_3}
\end{eqnarray}

Instead of the particle 4-momenta,
usually the total 4-momentum and the following three linear combinations
 are used  to characterize the
 kinematics of a two-body scattering process:
\begin{equation}
\Delta\equiv p_C-p_A=p_B-p_D, \quad P\equiv\ohalf(p_A+p_C),
\quad Q\equiv\ohalf(p_B+p_D) \quad.
\label{eq:3_1_3a}
\end{equation}
The scalar products of these three momenta can again be expressed in
terms of the Mandelstam variables
\renewcommand{\arraystretch}{1.4}
\begin{equation}
\begin{array}{rcl@{\qquad\qquad}rcl}
\Delta^2&=&t\quad,&
P\cdot\Delta&=&\ohalf(M_C^2 - M_A^2)\quad, \\ P^2&=&\ohalf(M_A^2 +
M_C^2)-t/4 \quad,& Q\cdot\Delta&=&\ohalf(M_B^2 - M_D^2) \quad, \\
Q^2&=&\ohalf(M_B^2 + M_D^2)-t/4 \quad,& P\cdot Q&=&(s-u)/4 \quad.
\end{array}
\label{eq:3_1_3d}
\end{equation}
\renewcommand{\arraystretch}{1.}

In covariant field-theory the scattering amplitude $T$ for a general
process with $n_i$ ($n_f$) particles in the initial (final) state
$\ket{i}$ ($\ket{f}$) is related to the (unitary) $S$
matrix by
\begin{equation}
\bra{f}S\ket{i} = \bra{f}\kket{i} - i {(2\pi)^4\over \sqrt{N_f N_i}}
	 \delta^{(4)}(P_f-P_i) \bra{f} T\ket{i}\quad,
\label{eq:3_1_10}
\end{equation}
where $P_i$ ($P_f$) denotes the total 4-momentum in the initial
(final) state.  The factors $N_x$ ($x=i,f$) are given by
\begin{equation}
N_x = (2\pi)^{3n_x} \prod_{j=1}^{n_x} {2E_j\over (2 M_j)^{b_j}}\quad,
\end{equation}
where the j-th particle of channel $x$ has mass $M_j$, momentum $\vec
p_j$, energy $E_j=(M_j^2 + \vec p_j\,^2)^{1/2}$ and spin $s_j$.  (Spin
and isospin quantum numbers are suppressed for the moment.)  The
exponent $b_j$ is given by
\begin{equation}
b_j=\left\{
\begin{array}{cl}
0 & \mbox{if\ } 2s_j \mbox{\ even}, \\ 1 & \mbox{if\ } 2s_j \mbox{\
odd}.
\end{array}
\right.
\end{equation}

For a two-body scattering process $A+B\to C+D$ Eq.~\ref{eq:3_1_10}
reads
\addtolength{\jot}{.5\baselineskip}
\begin{eqnarray}
\lefteqn{
\bra{C p_C, D p_D}S\ket{Ap_A,Bp_B}\; =\; \bra{C p_C, D p_D}
\kket{Ap_A,Bp_B}}
\nonumber \\
&& - \; {i\over (2\pi)^2}
\mbox{$ \sqrt{(2M_A)^{b_A} (2M_B)^{b_B}(2M_C)^{b_C}(2M_D)^{b_D}\over 16
 E_A E_B E_C E_D}$}
	 \delta^{(4)}(p_C+p_D-p_A-p_B)
\nonumber\\
&&
\qquad\qquad \qquad\qquad \bra{C p_C, Dp_D}T\ket{Ap_A,Bp_B}.
\label{eq:3_1_11}
\end{eqnarray}
\addtolength{\jot}{-.5\baselineskip}

Particles with spin (and helicity $\lambda_X$) are described in the
helicity basis according to the conventions of Jacob and
Wick~\cite{JW}. By separating off the helicity spinors $u_{X}(\vec
p_X,\lambda_X)$ of the particles from the scattering amplitudes one
obtains the transition matrix ${\cal M}$. If, for instance, all four
particles of the process $A+B\to C+D$ are spin-1/2 baryons, the
transition matrix ${\cal M}$ is a $16\times16$ matrix in spinor space
and is defined by
\begin{eqnarray}
\lefteqn{
\bra{C \vec p_C \lambda_C, D \vec p_D \lambda_D}
T \ket{A \vec p_A \lambda_A, B \vec p_B \lambda_B } = } \nonumber \\
&&
\anti{u}_{C}(\vec p_C,\lambda_C) \anti{u}_{D}(\vec p_D,\lambda_D)
 {\cal M}_{AB \to CD}(P,Q) u_{A}(\vec p_A,\lambda_A) u_{B}(\vec
p_B,\lambda_B)\quad.
\label{eq:3_1_11a}
\end{eqnarray}

Now the transition matrix ${\cal M}$ can be constructed as a linear
combination of the so-called {\em kinematic covariants} ${\cal O}_i$,
which are like ${\cal M}$ operators in spinor space.
\begin{equation}
{\cal M}(P,Q) = \sum_i c_i(s,t) {\cal O}_i(P,Q)\quad.
\label{eq:3_1_12}
\end{equation}
The ${\cal O}_i$ are built up from the Dirac $\gamma$-matrices and the
momenta $P$ and $Q$ (cf.\ Eq.~\ref{eq:3_1_3a}) in such a way that
their matrix elements are Lorentz-invariant quantities.  The {\em
invariant amplitudes} $c_i(s,t)$ are Lorentz-scalars.

The number of independent kinematic covariants for a given scattering
process corresponds to the number of independent helicity
amplitudes and is determined by the dimension of the spinor space and
invariance principles for the underlying interaction.  For the
scattering of four spin-1/2 baryons ($A+B\to C+D$) there are in
general eight independent kinematic covariants. For the elastic
scattering ($A+B\to A+B$) their number is reduced to six due to time
reversal invariance.For the `superelastic' scattering of four
identical particles ($A+A\to A+A$) the number of independent kinematic
covariants is further reduced to five due to the symmetry under
particle exchange.

The set of kinematic covariants is not unique.  However, for the
forthcoming it is essential that the ${\cal O}_i$ are chosen in such a
way that the invariant amplitudes $c_i(s,t)$ do not contain any
kinematic singularities, but only `physical' singularities demanded by
unitarity.  In the case of four spin-1/2 particles this condition is
fulfilled by the set of eight covariants given in Ref.~\cite{ScadJon}.
This set is based on the so-called Fermi-covariants $S,P,V,A,T$:
\begin{equation}
\begin{array}{lcccl@{\quad\qquad}lcccl}
 S &\equiv& {\cal O}_S &=& \unity_4 \otimes \unity_4 \quad,& T
 &\equiv&{\cal O}_T &=& \ohalf \sigma_{\mu\nu} \otimes
 \sigma^{\mu\nu}\quad,
\\
 P &\equiv&{\cal O}_P &=& \gamma_5 \otimes \gamma_5\quad, & && {\cal
 O}_6 &=& \unity_4 \otimes \gamma_\mu P^\mu \,-\, \gamma_\mu Q^\mu
 \otimes \unity_4\quad,
\\
 V &\equiv&{\cal O}_V &=& \gamma_\mu \otimes \gamma^\mu \quad,& &&
 {\cal O}_7 &=& \gamma_5 \otimes \gamma_5 \gamma_\mu P^\mu\quad,
\\
 A &\equiv&{\cal O}_A &=& \gamma_5 \gamma_\mu \otimes
 \gamma_5\gamma^\mu\quad, & && {\cal O}_8 &=& \gamma_5 \gamma_\mu
 Q^\mu \otimes \gamma_5\quad.
\label{eq:3_1_13}
\end{array}
\end{equation}
\smallskip
These covariants are of the form ${\cal O}_i(P,Q) = {\cal
O}^{(1)}_i(Q)\otimes {\cal O}^{(2)}_i(P)$ and the matrix elements have
to be evaluated according to
\begin{eqnarray}
\lefteqn{
\anti{u}_{C}(\vec p_C,\lambda_C) \anti{u}_{D}(\vec p_D,\lambda_D)
 {\cal O}_i(P,Q) u_{A}(\vec p_A,\lambda_A) u_{B}(\vec p_B,\lambda_B)=
} \nonumber \\ &&\!\!\!
\left[ \anti{u}_{C}(\vec p_C,\lambda_C)
{\cal O}^{(1)}_i(Q)u_{A}(\vec p_A,\lambda_A)\right] \;
\left[ \anti{u}_{D}(\vec p_D,\lambda_D) {\cal O}^{(2)}_i(P)u_{B}
(\vec p_B,\lambda_B)\right].
\end{eqnarray}

\subsection{Dispersion relations for baryon-baryon amplitudes of
$\pi\pi$ and $K\anti{K}$ exchange}
\label{sec:kap3_4}

For a general two-body scattering process the physical regions of the
Mandelstam variables for the $s$-, $t$- and $u$-channel reaction are
non-overlapping.  Therefore the transition matrices in the three
channels can be interpreted as independent branches of one operator
${\cal M}$ defined in the various kinematic regions. In the
$t$-channel ($A+\anti{C}\to\anti{B}+D$), for instance, the scattering
amplitude is related to the invariant amplitudes by
\begin{eqnarray}
\lefteqn{
\bra{\anti{B} -\vec p_B \lambda_B, D \vec p_D \lambda_D}
T \ket{ A \vec p_A \lambda_A, \anti{C} -\vec p_C \lambda_C} = }
\nonumber \\ &&
\hspace{-.6cm}\sum_{i=1}^8 c_i(s,t)
\left[\anti{v}_{C}(-\vec p_C,\lambda_C) {\cal O}^{(1)}_i(Q)u_{A}
(\vec p_A,\lambda_A)\right]
\left[\anti{u}_{D}(\vec p_D,\lambda_D){\cal O}^{(2)}_i(P)v_{B}
(-\vec p_B,\lambda_B)\right].
\nonumber \\
\end{eqnarray}
By introducing the concept of analyticity the invariant
amplitudes in the three channels become closely related: The invariant
amplitudes $c_i(s,t)$ are supposed to be analytic functions (except
for the physical singularities) in the whole complex $st$-plane.
Therefore, if all these physical singularities are known the
$c_i(s,t)$ can be deduced at any point in the complex Mandelstam plane
by the formulation of dispersion integrals.

The singularity structure of the invariant amplitudes is completely
determined by the unitarity of the $S$-matrix.  In terms of the
scattering amplitude unitarity of the $S$-matrix is expressed as
\begin{equation}
i\left[
\bra{f}T\ket{i} - \bra{f}T^\dagger\ket{i}
\right]
=
\sum_n {(2\pi)^4\over N_n} \delta^{(4)}(P_i-P_n) \bra{f}T^\dagger\ket{n}
\bra{n}T\ket{i}\quad,
\label{eq:3_3_1b}
\end{equation}
where $P_n$ and $P_f=P_i$ denote the total 4-momenta.  The summation in
Eq.~\ref{eq:3_3_1b} is to be understood over all {\em physical} states
$n$, i.e.\ states which are energetically accessible for the system
with energy $P_i^0$.

The singularities of the baryon-baryon ($A+B\to C+D$) amplitudes
generated by (correlated) $\pi\pi$ and $K\anti{K}$ exchange are most
easily derived using the unitarity relation for the $t$-channel
reaction $A+\anti{C}\to D+\anti{B}$.  For this, one has to restrict
the summation in Eq.~\ref{eq:3_3_1b} to physical $\pi\pi$ and
$K\anti{K}$ states; i.e., contributions of single-meson poles or of
heavier two-meson and multi-meson (e.g.\ $3\pi$, $4\pi$) channels are
disregarded.  In the c.m.\ system ($P_i=P_f=(\sqrt{t},\vec 0)$) the
summation over the states $n$ then reads
\begin{eqnarray}
\lefteqn{
\sum_n  {(2\pi)^4\over N_n} \delta^{(4)}(P_i-P_n) \ket{n}\bra{n}
=}
\nonumber \\
&& ={1\over (2\pi)^2}
\sum_{\mu\bar\mu=\pi\pi,K\bar K} N_{\mu\anti{\mu}}
\int d^3k
\delta(\sqrt{t}-2\omega_\mu(k)) {1\over 4\omega_\mu(k)^2}
\ket{\mu\anti{\mu} , \vec k} \bra{\mu\anti{\mu} , \vec k}
\nonumber \\
&& ={1\over 32\pi^2}
\sum_{\mu\bar\mu=\pi\pi,K\bar K} N_{\mu\anti{\mu}}
\sqrt{t-4m_\mu^2 \over t} \theta(t-4m_\mu^2)
\int d^2\hat k_{\mu\bar\mu}
\ket{\mu\anti{\mu} , \vec k_{\mu\bar\mu}}
\bra{\mu\anti{\mu} , \vec k_{\mu\bar\mu}}
\end{eqnarray}
with $\omega_\mu(k)=\sqrt{m_\mu^2+k^2}$ and the on-shell momentum
$k_{\mu\bar\mu}= \sqrt{t/4 - m_\mu^2}$.  The symmetry factor
$N_{\mu\anti{\mu}}$ is introduced in order to obtain the correct phase
space in case of identical particles:
\begin{equation}
N_{\mu\bar\mu} =
\left\{ \begin{array}{r@{\quad\mbox{if}\quad}l}
1/2 & \mu\anti{\mu}=\pi\pi \\ 1 &\mu\anti{\mu}= K\anti{K}
\end{array}
\right.
\label{eq:3_3_1a}
\end{equation}

Now, the kinematic covariants ${\cal O}_i$ in Eq.~\ref{eq:3_1_13} have
been chosen in such a way~\cite{ScadJon} that the left-hand side of
Eq.~\ref{eq:3_3_1b} yields the imaginary part of the invariant
amplitudes, i.e.\
\begin{eqnarray}
\lefteqn{
i\bra{D \anti{B}, \vec q\, \lambda_D \lambda_B} T-T^\dagger \ket{ A
\anti{C}, \vec p\, \lambda_A \lambda_C} = }
\nonumber \\
&&\hspace{-.5cm} -2 \sum_i {\sl Im}\left[ c_i (s,t)\right]
\anti{v}_{C}(-\vec p,\lambda_C) \; \anti{u}_{D}(\vec q,\lambda_D)
{\cal O}_i(P,Q) u_{A}(\vec p,\lambda_A) v_{B}(-\vec q,\lambda_B)\, .
\end{eqnarray}
Hence, the unitarity relation for the helicity amplitudes of the
process $A+\anti{C}\to D+\anti{B}$ becomes in the c.m.\ system:
\begin{eqnarray}
\lefteqn{
\sum_i {\sl Im}\left[ c_i (s,t)\right]
\anti{v}_{C}(-\vec p,\lambda_C) \; \anti{u}_{D}(\vec q,\lambda_D)
{\cal O}_i(P,Q) u_{A}(\vec p,\lambda_A) v_{B}(-\vec q,\lambda_B) = }
\nonumber \\ && =-{1\over 64\pi^2}
\sum_{\mu\bar\mu=\pi\pi,K\bar K} N_{\mu\anti{\mu}}
\sqrt{t-4m_\mu^2 \over t} \theta(t-4m_\mu^2)
\nonumber \\
&&
\quad\int d^2\hat k_{\mu\bar\mu}
\bra{\mu\anti{\mu} , \vec k_{\mu\bar\mu}}
T\ket{D \anti{B}, \vec q\, \lambda_D \lambda_B}^*
\bra{\mu\anti{\mu} , \vec k_{\mu\bar\mu}} T
\ket{ A \anti{C}, \vec p \,\lambda_A \lambda_C}\quad .
\label{eq:3_3_3}
\end{eqnarray}

{}From the beginning, the unitarity
relations~\ref{eq:3_3_1b},~\ref{eq:3_3_3}
are just defined above the kinematic threshold of the process
$A+\anti{C}\to D+\anti{B}$, that is for $t \geq t_0\equiv
\max\{M_A+M_C,M_B+M_D\}$.  However, they can be continued analytically
into the pseudophysical region ($4m_\pi^2\leq t \leq  t_0$) as
discussed in
Ref.~\cite{Kamp}.  Below the kinematic threshold the baryon-antibaryon
momenta become imaginary. According to Ref.~\cite{Kamp} the unitarity
relation for the $S$-matrix, which can be written symbolically as
$\left[S(p)\right]^* S(p)=1$, has to be continued to complex momenta
by $\left[S(p^*)\right]^* S(p)=1$.  As explained in
Ref.~\cite{BrownJack}, for imaginary momenta this is equivalent to
evaluating the expression $\left[S(p)\right]^*S(p)=1$ with real dummy
variables for the momenta and replacing these dummy variables at the
very end (after having performed all complex conjugations) by the
imaginary momenta.

The right-hand side of Eq.~\ref{eq:3_3_3} obviously vanishes below the
$\pi\pi$ threshold at $t=4m_\pi^2$.  Since for the processes
considered here the left-hand cuts, which are due to unitarity
constraints for the $u$-channel process, do not extend up to
$t=4m_\pi^2$ (for fixed $s$ lying inside the physical $s$-channel
region), the invariant amplitudes $c_i(s,t)$ are
real-analytic functions of $t$, i.e.\ $c_i(s,t^*)=c_i(s,t)^*$.
According to Eq.~\ref{eq:3_3_3} and
\begin{equation}
2i {\sl Im}\left[ c_i (s,t+i\epsilon)\right]=c_i(s,t+i\epsilon)-
c_i(s,t-i\epsilon) \quad,
\end{equation}
$c_i(s,t)$ has a branch cut along the real $t$-axis extending from
$t=4m_\pi^2$ to $t=+\infty$.  Corresponding statements hold for the
$K\anti{K}$ branch cut.

For c.m.\ energies below $1\,GeV$ the $\pi\pi$ interaction is dominated
by the $JI=00,11$ partial waves~\cite{Lohse}.  At low transfered
momenta, relevant in low-energy baryon-baryon scattering, correlations
between two exchanged pions (kaons) are therefore only considerable
when the exchanged $\pi\pi$ ($K\anti{K}$) system is in a state with
relative angular momentum $J=0$ and isospin $I=0$ (`$\sigma$-channel')
or $J=1$ and $I=1$ (`$\rho$-channel')~\cite{Kim}.
 In our approach only the correlated part of
two-pion and two-kaon exchange is evaluated by dispersiontheoretic
means. The uncorrelated part has to be calculated directly in the
$s$-channel in order to include all $t$-channel partial waves and to
guarantee the correct energy dependence of
this contribution.  In the $\Lambda N$ channel, for instance, the
iterative two-pion exchange with an $N\Sigma$ intermediate state
becomes complex above the $N\Sigma $ threshold. This behavior cannot
be reproduced with a single-variable dispersion relation in the
$t$-channel which will be applied to the correlated contribution (see
below).

Hence, the following dispersiontheoretic considerations can be limited
to the $\sigma$ and $\rho$ channel of $\pi\pi$ ($K\anti{K}$) exchange.
For this, the $B\anti{B'}\to \mu\anti{\mu}$ amplitudes on the
right-hand side of the unitarity relation~\ref{eq:3_3_3} are
decomposed into partial waves~\cite{GW}. Choosing
the coordinate system so that $\vec p$ points along the $\hat
z$-axis and $\vec q$ lies in the $\hat x \hat z$-plane the partial
wave decomposition gives
 \begin{eqnarray}
\lefteqn{
\int d^2\hat k_{\mu\bar\mu}
\bra{\mu\anti{\mu} , \vec k_{\mu\bar\mu}}  T
\ket{D \anti{B}, \vec q\, \lambda_D \lambda_B}^*
\bra{\mu\anti{\mu} , \vec k_{\mu\bar\mu}} T
\ket{ A \anti{C}, \vec p \,\lambda_A \lambda_C} =
}
\nonumber \\
&&\!\!\!\!\!
\sum_J {2J+1 \over 4\pi}
d^J_{\lambda_A-\lambda_C,\lambda_D-\lambda_B}(\cos\vartheta_t)
\bra{\mu\anti{\mu}} T^J (t) \ket{D \anti{B}, \lambda_D \lambda_B}^*
\bra{\mu\anti{\mu}} T^J (t) \ket{A \anti{C}, \lambda_A \lambda_C},
\nonumber \\
\label{eq:3_4_3}
\end{eqnarray}
where the on-shell momenta $p=p(t)$ and $q=q(t)$ (see
Eq.~\ref{eq:3_1_2a}) are suppressed as arguments of the partial wave
decomposed $T$ matrix elements.  Note that the right-hand side depends
on the Mandelstam variable $s$ only via the angle
$\vartheta_t=\vartheta_t(s,t)$ (see Eq.~\ref{eq:3_1_3})  between
$\vec p$
and $\vec q$.  Now, by restricting the sum over $J$ in
Eq.~\ref{eq:3_4_3} to $J=0$ ($J=1$) the contribution $c_i^{(J)}(s,t)=
c^\sigma_i (s,t)$ ($c^\rho_i (s,t)$) of the $\pi\pi$ and $K\anti{K}$
intermediate states to the discontinuity of the invariant amplitudes
$c_i(s,t)$ in the $\sigma$ ($\rho$) channel can be isolated.  From the
unitarity relation~\ref{eq:3_3_3} we obtain
\begin{eqnarray}
\lefteqn{
\sum_i {\sl Im}\left[ c^{(J)}_i (s,t)\right]
\anti{v}_{C}(-\vec p,\lambda_C) \; \anti{u}_{D}(\vec q,\lambda_D)
{\cal O}_i(P,Q) u_{A}(\vec p,\lambda_A) v_{B}(-\vec q,\lambda_B) = }
\nonumber \\ && =\sum_{\mu\bar\mu=\pi\pi,K\bar K}
d^J_{\lambda_A-\lambda_C,\lambda_D-\lambda_B}(\cos\vartheta_t)
H^J_{\mu\anti{\mu}} (t) \nonumber \\ && \hspace{2.7cm}
\bra{\mu\anti{\mu}} T^J (t) \ket{D \anti{B}, \lambda_D \lambda_B}^*
\bra{\mu\anti{\mu}} T^J (t) \ket{A \anti{C}, \lambda_A \lambda_C}
\quad,
\label{eq:3_4_4}
\end{eqnarray}
with the abbreviation
\begin{equation}
H^J_{\mu\anti{\mu}} (t) \equiv -{2J+1 \over 256\pi^3}
N_{\mu\anti{\mu}}
\sqrt{t-4m_\mu^2 \over t} \theta(t-4m_\mu^2)\quad.
\end{equation}
Eq.~\ref{eq:3_4_4} is a system of linear equations for the
discontinuities ${\sl Im}\left[ c^{(J)}_i (s,t)\right]$.  Its solution
provides the discontinuities as linear combinations of the following
products of $B\anti{B'}\to\mu\anti{\mu}$ helicity amplitudes:
\begin{equation}
F_{\lambda_D\lambda_B,\lambda_A\lambda_C}^{J}\equiv
\sum_{\mu\bar\mu=\pi\pi,K\bar K} H^J_{\mu\anti{\mu}} (t)
\bra{\mu\anti{\mu}} T^J (t) \ket{D \anti{B}, \lambda_D \lambda_B}^* \;
\bra{\mu\anti{\mu}} T^J (t) \ket{A \anti{C}, \lambda_A \lambda_C} \quad.
 \end{equation}
{}From the symmetry properties of the
$B\anti{B'}\to\mu\anti{\mu}$ helicity amplitudes,
which are due to the parity invariance of the
underlying strong interaction (see e.g.\ Ref.~\cite{JW}),
\begin{equation}
\bra{\mu\anti{\mu}}T^J(k,q)\ket{B\anti{B'}, \lambda\lambda'}
= \bra{\mu\anti{\mu}}T^J(k,q)\ket{B\anti{B'},-\lambda -\lambda'}\quad,
\end{equation}
 the following relations can be deduced:
\begin{equation}
F_{\lambda_D\lambda_B,\lambda_A\lambda_C}^{J}=
F_{(-\lambda_D)(-\lambda_B),\lambda_A\lambda_C}^{J}=
F_{\lambda_D\lambda_B,(-\lambda_A)(-\lambda_C)}^{J}=
F_{(-\lambda_D)(-\lambda_B),(-\lambda_A)(-\lambda_C)}^{J}\quad,
\end{equation}
Therefore only four linear independent
$F_{\lambda_D\lambda_B,\lambda_A\lambda_C}^{J}$ exist for $J>0$.  For
$J=0$ there is only one independent $B\anti{B'}\to\mu\anti{\mu}$
helicity amplitude,
$\bra{\mu\anti{\mu}}V^J(k,q)\ket{B\anti{B'}, ++}$,
 and consequently only
one independent $F_{\lambda_D\lambda_B,\lambda_A\lambda_C}^{J}$.
Using the explicit representation of helicity spinors in
Appendix~\ref{sec:AnhA} the evaluation of the matrix elements of the
kinematic covariants ${\cal O}_i$ in Eq.~\ref{eq:3_4_4} is
straightforward.

In case of unequal baryon masses $M_B\neq M_{B'}$ the analytic
structure of the $B\anti{B'}\to\mu\anti{\mu}$ helicity amplitudes is
much more involved compared to the case when the baryon masses are
equal. This complicates an analytic continuation of the amplitudes to
the pseudophysical region. The $\Lambda\anti{\Sigma}$ channel is the
only channel where this problem arises. In order to facilitate an easy
handling of the expressions we treat this channel approximately by
setting the mass of the $\Lambda$ and of the $\Sigma$ equal to the
average mass $(M_\Lambda+M_\Sigma)/2$. This approximation is justified
for two reasons: First the mass difference between $\Lambda$ and
$\Sigma$ hyperon is small ($77 MeV$) on the baryonic scale. Second,
since we evaluate the uncorrelated part directly in the $s$-channel
the important energy dependence of this contribution is not affected
by our approximation. The correlated contribution on the other hand is
known to have a rather smooth energy dependence which is supposed not
to change drastically by our approximation.

In the following we restrict ourselves therefore to the case where
\begin{equation}
M_A=M_C\equiv M \quad, \qquad M_B=M_D\equiv M' \quad.
\end{equation}

Solving the system of linear equations~\ref{eq:3_4_4}  yields that in
the $\sigma$ channel ($J=0$) all discontinuities ${\sl Im}\left[
c^\sigma_i (s,t)\right]$ vanish except for the scalar component
\begin{eqnarray}
&&{\sl Im}\left[ c^\sigma_S (t)\right] = {M' M \over q p}
F_{++,++}^{J=0}\quad,
\nonumber
\\
&&{\sl Im}\left[ c^\sigma_P (s,t)\right]= {\sl Im}\left[ c^\sigma_V
(s,t)\right]= {\sl Im}\left[ c^\sigma_A (s,t)\right]= {\sl Im}\left[
c^\sigma_T (s,t)\right]= {\sl Im}\left[ c^\sigma_6 (s,t)\right]=0 \quad,
\nonumber \\
\label{eq:3_4_5}
\end{eqnarray}
In contrast, in the $\rho$ channel only the axial-vector component
vanishes:
\renewcommand{\arraystretch}{1.5}
\begin{equation}
\begin{array}{lcl}
{\sl Im}\left[ c^\rho_S (s,t)\right]&=&\!\!\!\!\!\!\!  -
\begin{array}[t]{lcl}
  {1\over2} g(t) \sqrt{t} q p \cos\vartheta_t (s,t) & [& -2\sqrt{t}(M'
+ M) (2 F_{++,++}^{J=1} + F_{+-,+-}^{J=1} )
\\ &&
+\sqrt{2}(t + 4M' M) (F_{++,+-}^{J=1} + F_{+-,++}^{J=1})],
\end{array}
\\
{\sl Im}\left[ c^\rho_P (s,t)\right]&=&
\begin{array}[t]{lcl}
 g(t) q p \cos\vartheta_t (s,t) &[& (M' + M)(t - 4M' M)
F_{+-,+-}^{J=1} \\ && -2\sqrt{2} \sqrt{t} (p^2 F_{++,+-}^{J=1} + q^2
F_{+-,++}^{J=1}) ] \quad,
\end{array}
\\
{\sl Im}\left[ c^\rho_V (t)\right]&=&
\begin{array}[t]{lcl}
 g(t) \sqrt{t} &[& -\sqrt{t} (M' + M)(t/4- M'^2 - M^2 + M' M)
F_{+-,+-}^{J=1}
\\ &&
+2\sqrt{2} M' M (p^2 F_{++,+-}^{J=1} + q^2 F_{+-,++}^{J=1}) ] \quad,
\end{array}
\\
{\sl Im}\left[ c^\rho_A (s,t)\right]&=&0 \quad,
\\
{\sl Im}\left[ c^\rho_T (t)\right]&=&
\begin{array}[t]{lcl}
  {1\over4} g(t) t &[& (M' + M)(t - 4M' M) F_{+-,+-}^{J=1} \\ &&
-2\sqrt{2} \sqrt{t}(p^2 F_{++,+-}^{J=1} + q^2 F_{+-,++}^{J=1}) ]
\quad,
\end{array}
\\
{\sl Im}\left[ c^\rho_6 (t)\right]&=&
\begin{array}[t]{lcl}
 g(t) \sqrt{t} &[& (M'^2 - M^2)\sqrt{t} F_{+-,+-}^{J=1} \\
&&-2\sqrt{2} (p^2 M' F_{++,+-}^{J=1} - q^2 M F_{+-,++}^{J=1}) ]\quad,
\end{array}
\end{array}
\nonumber \\
\label{eq:3_4_6}
\end{equation}
\renewcommand{\arraystretch}{1.}
with
\begin{equation}
g(t)\equiv {M' M\over 2 q^2 p^2 (M' + M) t}\quad.
\end{equation}
The discontinuity of the pseudoscalar and the tensor component are
obviously related by
\begin{equation}
{\sl Im}\left[ c^\rho_P (s,t)\right]= {4q p \cos\vartheta_t \over t}
{\sl Im}\left[ c^\rho_T (t)\right]
\stackrel {(\ref{eq:3_1_3})}{=} {u-s\over t}{\sl Im}
\left[ c^\rho_T (t)\right]
\quad.
\label{eq:3_lindep}
\end{equation}
Consequently only four of the five nonvanishing discontinuities are
linear independent in alignment with the number of independent
$F_{\lambda_D\lambda_B,\lambda_A\lambda_C}^{J=1}$.

Except for poles (corresponding to single-particle exchange) and cuts
the invariant amplitudes $c_i^{(J)}(s\; \mbox{fixed},t)$ are
real-analytic functions of $t$. Therefore, fixed-$s$ dispersion
relations can
be formulated for the  $c_i^{(J)}(s,t)$.
Since we want to restrict the dispersiontheoretic evaluation to the
contribution of (correlated) $\pi\pi$ and $K\anti{K}$ exchange to the
baryon-baryon amplitudes we take into account only those singularities
which are generated by  $\pi\pi$ and $K\anti{K}$
intermediate states, namely the discontinuities of
Eqs.~\ref{eq:3_4_5}--\ref{eq:3_4_6}  due to the
$\pi\pi$ and $K\anti{K}$ unitarity cut (`right-hand cut').
The left-hand cuts, which are due to unitarity constraints
for the $u$-channel reaction, can be neglected in the baryon-baryon
channels  considered here, since they start at large, negative
$t$-values (from which they extend to $-\infty$) and are therefore far
away from the physical region relevant for low-energy $s$-channel
processes. For identical baryons (e.g.\ $NN\to NN$) this is only true
if the dispersion relations are applied only to the direct
baryon-baryon amplitude and the antisymmetrization of the amplitudes
is not taken into
account from the very beginning but just for the final $s$-channel
amplitudes. Otherwise, crossing of the exchange diagram would result
in a $u$-channel cut starting at $u=4m_\pi^2$ which could not be
neglected in the dispersion integrals~\cite{LinSerot}.

In this work
the $B\anti{B'}\to\mu\anti{\mu}$ amplitudes, which enter
in Eq.~\ref{eq:3_4_4}, are derived from a microscopic model which is
based
on the hadron-exchange picture (see Sect.~\ref{sec:kap4}).
Of course,  this model has a limited range of validity: for energies
far beyond $t'_{max}\approx100m_\pi^2$ it cannot provide reliable
results.
The dispersion integral for the invariant amplitudes extending in
principle along the whole $\pi\pi$ right-hand cut has therefore to be
limited to an upper bound ($t'_{max}$).
In addition left-hand cuts and unphysical cuts introduced for instance
by the
form factor prescription of the microscopic model for the
$B\anti{B'}\to\mu\anti{\mu}$ amplitudes are neglected.
Because of these approximations of the exact expressions, which are
necessary in order to obtain a solution of the physical problem,
the formulation of either a dispersion relation or a
subtracted dispersion relation might lead to different results for the
amplitudes although both should be mathematically equivalent.
However, the ambiguity which dispersion relation to choose can be
avoided by demanding that the analytic structure of the resulting
$c_i^{(J)}(s,t)$ should agree as far as possible with the expressions
for sharp $\sigma$ and $\rho$ exchange in the baryon-baryon
interaction.

The transition amplitude ${\cal M}^\sigma$ for the exchange of a
scalar $\sigma$ meson with mass $m_\sigma$ between two $J^P=1/2^+$
baryons $A$ and
$B$ follows from the interaction Lagrangians ($X=A,B$)
\begin{equation}
{\cal L}_{XX\sigma}(x)= g_{XX\sigma} \anti{\psi}_{X}(x) \psi_{X}(x)
 \phi_\sigma(x)
\end{equation}
(See Appendix~\ref{sec:AnhA} for the hadronic field operators.)
The result is
\begin{equation}
{\cal M}^\sigma(t)=g_{AA\sigma} g_{BB\sigma} { F^2_\sigma(t)\over t -
m_\sigma^2} {\cal O}_S \quad,
\label{eq:3_5_sig}
\end{equation}
where a form factor $F_\sigma(t)$
 has to be applied at each vertex since the exchanged $\sigma$ meson is
far away from its mass-shell.
This form factor is parametrized in the conventional monopole form
\begin{equation}
F_\sigma(t)={\Lambda_\sigma^2 - m_\sigma^2 \over \Lambda_\sigma^2 -
t}\quad.
\label{eq:3_5_2}
\end{equation}
with a  cutoff mass $\Lambda_\sigma$ assumed to be uniform for both
vertices.

The construction of the amplitude for $\rho$ exchange in the transition
 $A+B \to C+D$ (with $M_A=M_C\equiv M$ and $M_B=M_D\equiv M'$) starts
from the interaction Lagrangians
\begin{eqnarray}
{\cal L}_{XY\rho}(x)=&& g_{XY\rho} \anti{\psi}_{X}(x) \gamma_\mu
 \psi_Y(x) \phi^\mu_\rho(x)
\nonumber \\
&+& {f_{XY\rho} \over 4 M_N} \anti{\psi}_{X}(x) \sigma_{\mu\nu}
 \psi_Y(x) (\partial^\mu \phi^\nu_\rho(x) - \partial^\nu
 \phi^\mu_\rho(x))
\nonumber \\
&& (+ \mbox{\rm h.c., if } X\neq Y)
\label{eq:BBvcoup2}
\end{eqnarray}
with $(XY)=(AC),(BD)$.
According to the conventional Feynman rules, using Eq.~\ref{eq:3_1_3d}
and the generalized
Gordon decomposition~\cite{HolzDipl} for the spinors of two Dirac
particles $X$ and $Y$,
\begin{equation}
\anti{u}_{X}(\vec p\,',\lambda')
\left[(M_X+M_{Y}) \gamma^\mu - (p'+p)^\mu
-i \sigma^{\mu\nu} (p'-p)_\nu \right]
u_Y(\vec p,\lambda)=0 \quad,
\end{equation}
 the transition amplitude ${\cal
M}^\rho$ comes out to be
\begin{eqnarray}
{\cal M}^\rho(P,Q)= {-1 \over t - m_\rho^2}
& \Biggl[ &
{f_{AC\rho}f_{BD\rho}\over 4 M_N^2} (s-u)\;{\cal O}_S
\nonumber \\ &&
+ G_{AC\rho} G_{BD\rho} \;{\cal O}_V
\nonumber \\ &&
+ {G_{AC\rho} f_{BD\rho}-G_{BD\rho} f_{AC\rho} \over 2M_N}\; {\cal O}_6
\nonumber \\ &&
+ {G_{AC\rho} f_{BD\rho}+G_{BD\rho} f_{AC\rho} \over 2M_N}\; {\cal P}_2
\Biggr] \quad,
\label{eq:AnhG1}
\end{eqnarray}
with
\begin{equation}
G_{XY\rho} \equiv
g_{XY\rho} + {M_X+M_{Y}\over 2 M_N}
 f_{XY\rho}
\end{equation}
and the 4-momenta $P$ and $Q$ defined in Eq.~\ref{eq:3_1_3a}.
${\cal P}_2$ is one of the so-called perturbative covariants
introduced in Ref.~~\cite{ALV} as
\begin{equation}
{\cal P}_2=   -\unity_4 \otimes \gamma_\mu P^\mu \,-\,
		\gamma_\mu Q^\mu \otimes \unity_4
\quad.
\end{equation}
${\cal P}_2$ can be expanded in terms of the ${\cal O}_i$ of
Eq.~\ref{eq:3_1_13} since they
form a complete set of kinematic covariants.
For the given baryon masses one finds
\begin{equation}
{\cal P}_2 = {1\over 2(M+M')}
\left[
(u-s)({\cal O}_S + {\cal O}_P)
-4MM'{\cal O}_V
+ t {\cal O}_T
+2(M-M') {\cal O}_6
\right]\quad.
\end{equation}
After replacing ${\cal P}_2$ in Eq.~\ref{eq:AnhG1}
the final result for ${\cal M}^\rho$ reads
\begin{eqnarray}
 {\cal M}^\rho(P,Q)= {-F^2_\rho(t)\over t - m_\rho^2}
&
\Biggl\{&
{g_{AC\rho}f_{BD\rho}+ f_{AC\rho}g_{BD\rho}\over 4M_N M_{tot} } (u-s)
\; {\cal O}_S
\nonumber \\&&
\left[
g_{AC\rho}g_{BD\rho}+ g_{AC\rho}f_{BD\rho} {M'^2 \over M_N M_{tot} } +
f_{AC\rho}g_{BD\rho} {M^2 \over M_N M_{tot} }
\right] {\cal O}_V
\nonumber \\&&
\left(
g_{AC\rho}f_{BD\rho}{M'\over M_N M_{tot}} - f_{AC\rho}g_{BD\rho} {M_A
\over M_N M_{tot} }
\right) {\cal O}_6
\nonumber \\ &&
{G_{AC\rho}f_{BD\rho}+ f_{AC\rho}G_{BD\rho}\over 4M_N M_{tot} }
\left[(u-s)\, {\cal O}_P \;+ t  \,{\cal O}_T \;\right]
\Biggr\}
\nonumber \\
\label{eq:3_5_rho}
\end{eqnarray}
with $M_{tot}=M+M'$
and the form factor $F_\rho(t)$ parametrized according to
Eq.~\ref{eq:3_5_2}.

By comparison of the discontinuities in the $\sigma$ and $\rho$
channel in Eqs.~\ref{eq:3_4_5},~\ref{eq:3_4_6} with the transition
amplitudes
for sharp $\sigma$ and $\rho$ exchange it follows that
$c_S^\sigma,c_V^\rho$ and $c_6^\rho$ obey an unsubtracted dispersion
relation,
\begin{equation}
c^{(J)}_i (s,t) = {1\over\pi} \int_{4m_\pi^2}^{t'_{max}} { {\sl
Im}\left[ c^{(J)}_i (s,t')\right] \over t'-t} dt' \quad.
\label{eq:3_4_1}
\end{equation}
The tensor component of sharp $\rho$ exchange is proportional to $t$
(cf.~\ref{eq:3_5_rho}).
In order to generate this factor $t$ also for the tensor component of
correlated $\pi\pi$ and $K\anti{K}$ exchange a subtracted dispersion
relation (subtraction point $t_0$ and subtraction constant
$c_T^\rho(s,t_0)=0$) is assumed for the invariant amplitude
$c_T^\rho(s,t)$:
\begin{equation}
c^\rho_T (s,t) = {t\over\pi} \int_{4m_\pi^2}^{t'_{max}} { {\sl
Im}\left[ c^\rho_T (s,t')\right]/t' \over t'-t} dt' \quad.
\label{eq:3_4_2a}
\end{equation}
Similarly, the  $u-s$ dependence of the (pseudo-)scalar component of
sharp $\rho$ exchange (cf.\ Eq.~\ref{eq:3_5_rho}) can be reproduced
by assuming a subtracted dispersion relation for
$ c^\rho_S (s,t)$ and $c^\rho_P (s,t)$ with
$t_0(s)=\Sigma-2s$ and $c^\rho_{S,P} (s,t_0(s)) = 0$:
\begin{equation}
c^\rho_{S,P} (s,t) = {u-s\over\pi} \int_{4m_\pi^2}^{t'_{max}} { {\sl
Im}\left[ c^\rho_{S,P} (s,t')\right]/(u'-s) \over t'-t} dt' \quad,
\label{eq:3_4_2b}
\end{equation}
where according to Eq.~\ref{eq:3_1_0} $s+t+u=s+t'+u'=\Sigma$.

\subsection{Baryon-baryon interaction arising from correlated $\pi\pi$
and $K\anti{K}$ exchange}

The invariant amplitudes constructed  in
the preceding Section using dispersion theory still contain the
uncorrelated contributions of $\pi\pi$
and $K\anti{K}$ exchange.
Investigating the problem of baryon-baryon scattering requires the
knowledge of the on-shell scattering amplitude $T$, which is usually
obtained as a solution of a scattering equation $T=V+VGT$ that
iterates the interaction kernel $V$.
In general, $V$ contains besides other terms one-pion and one-kaon
exchange contributions
as well as the contribution $V_{2\pi}$ from two-pion and two-kaon exchange. But
iterating $\pi$ and $K$ exchange in the second order term $VGV$ also
generates  two-pion and two-kaon exchange contributions to the
scattering amplitude. In order to avoid double counting these
`iterative' contributions therefore have to be left out from the
dispersiontheoretically calculated $V_{2\pi}$. As stated above we even
go beyond this and subtract all uncorrelated contributions from
$V_{2\pi}$. By this the dispersiontheoretic calculations can
be restricted to the $\sigma$ and $\rho$ channel (since only there
significant correlations occur in the kinematic region considered),
whereas the uncorrelated contributions are evaluated in the $s$-channel
and therefore contain all $t$-channel partial waves.

In order to eliminate the uncorrelated contributions from  $V_{2\pi}$
we determine the discontinuities
${\sl Im} [c^{(J)}_{i,Born}(s,t)]$
generated from the $B\anti{B'}\to\mu\anti{\mu}$ Born amplitudes $V^J$
(i.e., no $\pi\pi-K\anti{K}$ correlations included)
using as before the unitarity relation~\ref{eq:3_4_4} (with $T^J$
replaced by $V^J$)
and subtract them finally from the full discontinuities
${\sl Im}[c^{(J)}_{i}(s,t)]$.
(The contributions of the $\rho$-pole diagram to the
$B\anti{B'}\to\mu\anti{\mu}$ Born amplitudes must  not be subtracted
since the corresponding $s$-channel processes are not included
explicitly in $V$.)
Hence, for the invariant amplitudes of {\em correlated} $\pi\pi$ and
$K\anti{K}$ exchange, $\tilde c^{(J)}_{i}(s,t)$,
the (unsubtracted) dispersion relation~\ref{eq:3_4_1} has to be
modified to
\begin{equation}
\tilde c^{(J)}_{i} (s,t) = {1\over\pi} \int_{4m_\pi^2}^{t'_{max}}
{ \rho^{(J)}_i (s,t') \over t'-t} dt'\quad,\qquad
(J)=\sigma,\rho\quad,
\label{eq:3_5_1}
\end{equation}
where the spectral function $\rho^{(J)}_i $ is given by
\begin{equation}
\rho^{(J)}_i (s,t') \equiv
{\sl Im}\left[ c^{(J)}_i (s,t')\right] - {\sl Im}\left[
c^{(J)}_{i,Born} (s,t')\right]\quad.
\label{eq:3_5_1a}
\end{equation}
Corresponding expressions hold for the subtracted dispersion
relations~\ref{eq:3_4_2a} and \ref{eq:3_4_2b}.

Now the baryon-baryon helicity amplitudes arising from correlated
$\pi\pi$ and $K\anti{K}$ exchange can be evaluated according to
Eqs.~\ref{eq:3_1_11a},~\ref{eq:3_1_12}
\begin{eqnarray}
\lefteqn{
\bra{C D , \vec q\,\lambda_C \lambda_D}
V^{(J)}_{2\pi} \ket{A B, \vec p \,\lambda_A\lambda_B } =}
\nonumber \\
&&
\anti{u}_{C}(\vec q,\lambda_C)\; \anti{u}_{D}(-\vec q,\lambda_D)\;
 {\cal V}^{(J)}_{2\pi} (P,Q)\; u_{A}(\vec p,\lambda_A) \; u_{B}(-\vec
p,\lambda_B)
\nonumber
\end{eqnarray}
with
\begin{equation}
{\cal V}^{(J)}_{2\pi} (P,Q) \equiv \sum_i \tilde c_i^{(J)}(s,t)
{\cal O}_i
(P,Q)
\qquad \left( (J)=\sigma,\rho \right)\quad.
\label{eq:3_5_0}
\end{equation}
The partial wave decomposition of these matrix elements then proceeds
as usual (see e.g.\ Ref.~\cite{Holz}).

Of course, when iterating the baryon-baryon interaction kernel in a
scattering equation $V_{2\pi,corr}$ has to be known off-shell.
However, dispersion theory applies only to on-shell amplitudes and
does not provide any information on the off-shell behavior of the
amplitudes.
Therefore an arbitrary prescription for the off-shell extrapolation of
$V_{2\pi,corr}$ has to be defined~\cite{Kim}, which is
certainly a drawback of the dispersiontheoretical derivation of this
potential.
Nevertheless the characteristic features of correlated $\pi\pi$ and
$K\anti{K}$ exchange like the strength of $V_{2\pi,corr}$ in the
various baryon-baryon channels can already be discussed by means of the
unique
on-shell amplitudes. Therefore we postpone the discussion of how to
extrapolate $V_{2\pi,corr}$  off-shell to a subsequent work.

The dispersiontheoretic  amplitudes for correlated $\pi\pi$ and
$K\anti{K}$ exchange
(Eqs.~\ref{eq:3_4_5},~\ref{eq:3_4_6}) have been constructed in such a
way that their operator structure agrees as far as
possible with sharp $\sigma$ and $\rho$
exchange~\ref{eq:3_5_sig},~\ref{eq:3_5_rho}.
Therefore our results for the correlated exchange can be parametrized
in terms of $\sigma$ and $\rho$ exchange; i.e., the products of
coupling constants for $\sigma$ and $\rho$ exchange are replaced by
effective coupling strengths $G^{(J)}(s,t)$, which contain the full
$s$- and
$t$-dependence  of the dispersiontheoretic results.
In the $\sigma$ channel this gives for the elastic baryon-baryon
process $A+B\to A+B$
\begin{equation}
g_{AA\sigma}g_{BB\sigma} \quad\longrightarrow \quad G_{AB\to
AB}^\sigma (t)= {(t-m_\sigma^2)\over F^2_\sigma(t)} \tilde
c^\sigma_S(t) = {(t-m_\sigma^2)\over\pi F^2_\sigma(t)}
\int_{4m_\pi^2}^{t'_{max}} { \rho^\sigma_S(t') \over t'-t} dt'
\label{eq:3_effccsig}
\end{equation}
Note that sharp $\sigma$ exchange (Eq.~\ref{eq:3_5_sig}) would
correspond
to a spectral function
\begin{equation}
\rho^{\sigma}_S (s,t') = - g_{AA\sigma} g_{BB\sigma}
\delta( t' - m_\sigma^2)
\end{equation}
(except for form factors).
This suggests to interpret the spectral function as a function that
denotes the strength of an exchange process depending on the
invariant mass of the exchanged system (here: $\pi\pi$, $K\anti{K}$).

By comparing the coefficients of the kinematic covariants
$ {\cal O}_i$ in Eqs.~\ref{eq:3_5_rho} and \ref{eq:3_5_0} we obtain in
the $\rho$-channel:
\begin{eqnarray}
g_{AB\rho}g_{BD\rho} &\!\to \!& ^{VV}\!G_{AB\to CD}^\rho (t)=
{(t-m_\rho^2)\over F^2_\rho(t)} \left[ 4MM'{\tilde c^\rho_S(s,t)\over
u-s} - \tilde c^\rho_V(t) +(M'-M) \tilde c^\rho_6(t)\right],
\nonumber\\
g_{AB\rho}f_{BD\rho} &\!\to\!& ^{VT}\!G_{AB\to CD}^\rho (t)=
-{(t-m_\rho^2)\over F^2_\rho(t)} M_N
\left[ 4M {\tilde c^\rho_S(s,t)\over u-s}
+ \tilde c^\rho_6(t)\right],
\nonumber\\
f_{AB\rho}g_{BD\rho} &\!\to\!& ^{TV}\!G_{AB\to CD}^\rho (t)=
-{(t-m_\rho^2)\over F^2_\rho(t)} M_N
\left[ 4M'  {\tilde c^\rho_S(s,t)\over u-s}
- \tilde c^\rho_6(t)\right],
\nonumber\\
f_{AB\rho}f_{BD\rho} &\!\to\! & ^{TT}\!G_{AB\to CD}^\rho (t)=
{(t-m_\rho^2)\over F^2_\rho(t)} 4M_N^2 {\tilde c^\rho_S(s,t)- \tilde
c^\rho_P(s,t)\over u-s}.
\label{eq:3_effccrho}
\end{eqnarray}

Obviously the effective coupling strengths do not depend on $s$ but
only on $t$. This is only possible by choosing the subtracted
dispersion relation~\ref{eq:3_4_2b} for $\tilde c^\rho_{S,P}(s,t)$,
since the integrand of the dispersion integral becomes independent of
$s$
(${\sl Im}[c^\rho_{S,P}(s,t')]\propto \cos\vartheta_t(s,t')\propto
u'-s$).
Therefore $\tilde c^\rho_{S,P}(s,t)$ depends on $s$ only by the factor
$(u-s)$, which cancels out exactly when calculating the effective
coupling
strengths.
It should be emphasized that the parametrization of $V_{2\pi,corr}$
discussed here does (so far)  not contain any approximations.

\subsubsection{Isospin-Crossing}
\label{sec:kap3_6}

Up to now, in favor of a clear representation of the
dispersiontheoretical calculation,
 we have suppressed isospin degrees of freedom.
The isospin structure of the $B\anti{B'}\to\mu\anti{\mu}$ amplitudes
will be discussed in the next Sections when the microscopic model for
these amplitudes is presented.
However, when `crossing' is applied to the baryon-baryon ($s$-channel)
and baryon-antibaryon ($t$-channel) amplitudes it has to be kept in mind
that the total isospin in the various channels is constructed from
different combinations of particle isospins.
The total isospin $I_s$ of the $s$-channel process
$A+B\to C+D$ is composed out of  $[I_A\otimes I_B]_{I_s}$ or
$[I_C\otimes I_D]_{I_s}$
and that of the $t$-channel process
$A+\anti{C} \to D+\anti{B}$ out of
$[I_A\otimes I_{\bar C}]_{I_t}$ or
$[I_{D}\otimes I_{\bar B}]_{I_t}$.

Consequently, besides the analytic continuation of the invariant
amplitudes in $s$ and $t$ the recoupling of the particle isospins has
to be taken into account when crossing amplitudes.
Therefore, the isospin amplitudes
$T_s^{AB\to CD}(I_s)$ and $T_t^{A\anti{C} \to D\anti{B}}(I_t)$ of the
$s$- and $t$-channel processes being independent of the isospin
projections $m_s$ and $m_t$ are linearly related:
\begin{equation}
T_s^{AB\to CD}(I_s)= \sum_{I_t}
 X_{AB,CD}(I_s,I_t)
T_t^{A\anti{C} \to D\anti{B}}(I_t)\quad,
\end{equation}
where $X_{AB,CD}(I_s,I_t)$ is the so-called
isospin-crossing matrix. Note that our isospin-crossing matrix
differs from the $\tilde X_{AB,CD}(I_s,I_t)$ introduced in
Refs.~\cite{MartSpear,Neville}
since their $t$-channel process ($\anti{D}+B\to C+\anti{A}$)
differs from the one in Sect.~\ref{sec:kap3_1}:
\begin{equation}
X_{AB,CD}(I_s,I_t) = (-1)^{I_B+I_D+I_t-2I_s}\tilde X_{BA,DC}(I_s,I_t)
\quad.
\end{equation}
As shown in Ref.~\cite{MartSpear} the isospin-crossing matrix
$\tilde X_{AB,CD}(I_s,I_t)$ can be expressed in terms of
Clebsch-Gordan coefficients
\begin{eqnarray}
\lefteqn{
\tilde X_{AB,CD}(I_s,I_t)= }
 \nonumber \\ &&
\eta_A \eta_D \!\! \sum_{m_A,m_B, \atop m_C,m_D,m_t}\!
(-1)^{I_A+I_D+m_A+m_D}
\bra{I_AI_Bm_Am_B}\kket{I_sm_s}
\bra{I_CI_Dm_Cm_D}\kket{I_sm_s}
 \nonumber \\ && \hspace{5.5cm}
\bra{I_CI_Am_C(-m_A)}\kket{I_tm_t}
\bra{I_DI_B(-m_D)m_B}\kket{I_tm_t}
\nonumber \\
\label{eq:3_6_1}
\end{eqnarray}
with $m_s$ ($|m_s|\leq I_s$) being arbitrary.
For a particle $A$ with isospin $I_A$ and isospin projection $m_A$ the
particle state $\ket{A}$ and the isospin state $\ket{Im}_A$ might
differ in sign. The isospin state of the antiparticle
$\anti{A}$ is generated by applying the $G$-parity operator ${\cal G}$
to $\ket{Im}_{A}$~\cite{MartSpear}:
\begin{equation}
\ket{Im}_{\bar A}=\eta_A {\cal G} \ket{Im}_{A}\quad.
\label{eq:3_6_1b}
\end{equation}
With the phase convention used here
 for the $SU(3)$ field operators (e.g.\ when calculating the
isospin factors of the $B\anti{B'}\to\mu\anti{\mu}$ amplitudes in the
next Section) the phase $\eta_A$, which is independent of $m_A$, comes
out to be~\cite{Neville}
\begin{equation}
\eta_A= (-1)^{I_A-Y_A/2}\quad,
\label{eq:SU3phase}
\end{equation}
where $Y$ denotes the hypercharge of particle $A$. Note that this
phase convention differs from the one used in Ref.~\cite{MartSpear}.
For the baryon-baryon processes considered in this work the
isospin-crossing matrices are tabulated in Tab.~\ref{tab:3_1}.

By the partial wave decomposition the amplitude of correlated
 $\pi\pi$ and  $K\anti{K}$ exchange is separated into the
contributions of the $\sigma$- ($I_t=0$) and $\rho$-channel ($I_t=1$).
Suppressing the spin-momentum dependence the isospin amplitude can be
written as
\begin{equation}
T_s(I_s)= X(I_s,0) T_t^\sigma + X(I_s,1) T_t^\rho \quad.
\end{equation}
The column $X(I_s,0)$ ($X(I_s,1)$)  of the isospin-crossing matrix
agrees   except for a constant factor $F_\sigma$ ($F_\rho$)
 with the isospin factors for $t$-channel exchange of a $\sigma$
($\rho$) meson in the corresponding $s$-channel process.
Conventionally these constant factors $F_\sigma$ and $F_\rho$, which
are also tabulated in Tab.~\ref{tab:3_1}, are  extracted
from the isospin-crossing matrix and put into the spectral
functions~\ref{eq:3_5_1a}, so that the isospin factors for the
$s$-channel potential
$V_{2\pi}$ agree with the isospin factors of
$\sigma$ and $\rho$ exchange.

\subsection{A microscopic model for the
$B\anti{B'} \to \pi\pi,K\anti{K}$ transition amplitudes}
\label{sec:kap4}

In the preceding Sections
we have outlined the dispersiontheoretic calculation of correlated
$\pi\pi$ and $K\anti{K}$ exchange in the baryon-baryon interaction
starting from amplitudes for the transition of a
baryon-antibaryon ($B\anti{B'}$) state to two pions  ($\pi\pi$) or a
kaon and a antikaon  ($K\anti{K}$).
These amplitudes have to be known in the so-called pseudophysical
region, i.e.\
for energies   below the $B\anti{B'}$ threshold.
However, in case of  the process $N\anti{N}\to \pi\pi$, these
amplitudes can be derived from
empirical data by analytic continuation of $\pi N$ and $\pi\pi$
scattering amplitudes, which are  extracted from scattering data, into
the pseudophysical region~\cite{Hoehler2,Hoehler,Nielsen,Nielsen2}.
Corresponding analyses for the transitions $Y\anti{Y'}\to \pi\pi,
K\anti{K}$ are out of sight since the required empirical information
(e.g.\ $\pi \Lambda$ scattering data) does not exist.
An evaluation of correlated $\pi\pi$ and $K\anti{K}$ exchange  in the
$YN$ or $YY$ interaction therefore necessitates the construction of a
microscopic model
for the $B\anti{B'} \to \pi\pi,K\anti{K}$ amplitudes. This model can
be tested against the quasiempirical information for the $N\anti{N}\to
\pi\pi$ amplitudes and then has to be extrapolated to the other
channels of interest.
In addition, only the use of a microscopic model for
the $B\anti{B'} \to \pi\pi,K\anti{K}$ amplitudes allows a consistent
treatment of medium
modifications of the baryon-baryon interaction.

The microscopic model presented in the following is a generalization
of the hadron-exchange model  for the $N\anti{N}\to \pi\pi$
 transition amplitudes of Ref.~\cite{Pearce_piN}, where it was
applied to the analysis
of  correlated two-pion exchange in the $\pi N$ interaction.
A main feature of the model presented here is the completely
consistent treatment of its two components, namely the
 $B\anti{B'}\to\pi\pi, K\anti{K}$ Born
amplitudes and the $\pi\pi-K\anti{K}$ correlations.
Both components are derived in field theory from an ansatz for the
hadronic Lagrangians.

The amplitudes for the processes
 $B\anti{B'} \to \pi\pi,K\anti{K}$
are obtained from a scattering equation which can be written
in operator form as
\begin{equation}
\mbox{\bf T} = \mbox{\bf V} + \mbox{\bf T}\mbox{\bf G}\mbox{\bf V}\quad,
\label{eq:skizz_1}
\end{equation}
where the scattering amplitude $\mbox{\bf T}$, the Born amplitude
$\mbox{\bf V}$ and the
Greens operator $\mbox{\bf G}$ are operators in channel space of
$B\anti{B'}$, $\pi\pi$ and $K\anti{K}$.

For large $t'$, contributions
of the spectral functions of correlated
$\pi\pi$ and $K\anti{K}$ exchange to the dispersion
integrals~\ref{eq:3_4_1}--\ref{eq:3_4_2b}
values are suppressed  by the denominator $1/(t'-t)$ because in the
physical $s$-channel $t{\leq}0$.
Since the unitarity cuts of the $B\anti{B'}$
states start far above the branch points of the $\pi\pi$ and
$K\anti{K}$ cuts the contribution of the $B\anti{B'}$ Greens function
$G_{B\anti{B'}}$
to the $B\anti{B'}\to\pi\pi, K\anti{K}$ scattering amplitudes in
Eq.~\ref{eq:skizz_1} and to the spectral functions  can be neglected.
For the same reason the coupling to other mesonic channels like the
$\rho\rho$ channel can be renounced.
Of course, by these approximations the range of validity of the
microscopic model for the  $B\anti{B'}\to\mu\anti{\mu}$
($\mu\anti{\mu}=\pi\pi,K\anti{K}$)  amplitudes
is limited to low $t'$ values. Therefore instead of integrating along
the whole $\pi\pi$ unitarity cut up to
infinity the upper bound of the dispersion integrals, $t'_{max}$, is
set to a value somewhere below the $\rho\rho$ threshold at $t'\approx
120m_\pi^2$ so that convergence of the integrals is achieved.

Taking into account only $\pi\pi$ and $K\anti{K}$ intermediate states,
i.e.\ neglecting $G_{B\anti{B'}}$, we obtain for the two components of
Eq.~\ref{eq:skizz_1} which give the $B\anti{B'}\to \mu\anti{\mu}$
amplitudes:

\begin{equation}
\left( \begin{array}{c}
T_{B\anti{B'}\to\pi\pi} \\ T_{B\anti{B'}\to K\anti{K}}
\end{array} \right)
=
\left( \begin{array}{c}
V_{B\anti{B'}\to\pi\pi} \\ V_{B\anti{B'}\to K\anti{K}}
\end{array} \right)
+
\left( \begin{array}{cc}
T_{\pi\pi\to\pi\pi} & T_{K\anti{K}\to\pi\pi} \\ T_{\pi\pi\to
K\anti{K}} & T_{K\anti{K}\to K\anti{K}}
\end{array} \right)
\left( \begin{array}{c}
G_{\pi\pi}\; V_{B\anti{B'}\to\pi\pi} \\ G_{K\anti{K}}\; V_{B\anti{B'}\to
K\anti{K}}
\end{array} \right)\, .
\label{eq:skizz_2}
\end{equation}
In Fig.~\ref{fig:4_0a} this scattering equation for the
$B\anti{B'} \to\pi\pi,K\anti{K}$ amplitudes is represented
symbolically.

The correlations $T_{\pi\pi/K\anti{K}\to\pi\pi/K\anti{K}}$ are
generated with a realistic model~\cite{Pearce_piN} of the $\pi\pi-K
\anti{K}$ interaction.
This meson-exchange model is a modified version of the so-called
J\"ulich $\pi\pi$
model~\cite{Lohse}.
The Born amplitudes of this model for the elastic $\pi\pi$ and
$K\anti{K}$ channels as well as for the transition
 $\pi\pi\to K\anti{K}$ are shown in Fig.~\ref{fig:4_5}.
Besides the $t$-channel (and in case of $\pi\pi\to\pi\pi$ also
$u$-channel)
  exchanges of the vector mesons
$\rho,\omega,\phi,K^*$
the $s$-channel exchanges (pole graphs) of the $\rho$, the
scalar-isoscalar
$\epsilon$ and the isoscalar tensor meson $f_2$ are taken into
account.
The potentials derived from Fig.~\ref{fig:4_5} are iterated in a
coupled-channel calculation according to the prescription of
Blankenbecler and Sugar~\cite{BbS}.
The free parameters of the $\pi\pi-K\anti{K}$ model of
Ref.~\cite{Pearce_piN} were adjusted to
the empirical $\pi\pi$ phase shifts and inelasticities. A good
agreement with the empirical data was achieved(cf.\ Fig.~\ref{fig:4_6}).

The scattering equation~\ref{eq:skizz_2} is likewise solved
using the ansatz~\cite{BbS} of Blankenbecler and Sugar (BbS) to
reduce the
4-dimensional Bethe-Salpeter equation to a 3-dimensional
equation which simplifies the calculation while retaining unitarity.

The total conserved 4-momentum $P$ and the relative 4-momentum $k'$ of
the intermediate $\mu'\anti{\mu'}=\pi\pi, K\anti{K}$ state are
expressed in terms of the particle 4-momenta $k'(1)$ and $k'(2)$ by
\begin{equation}
P=k'(1) + k'(2) \quad, \qquad k' = (k'(1) - k'(2))/2
\quad.
\label{eq:bbsc}
\end{equation}
Corresponding relations hold for the relative 4-momenta $q$ of the
initial
$B\anti{B'}$ and $k$ of the final $\mu\anti{\mu}=\pi\pi, K\anti{K}$
state.
In the center-of-mass system we have
$P=(\sqrt{t'},\vec 0)$ with $\sqrt{t'}=E_B(q)+E_{B'}(q)$,
$\vec k' =\vec k'(1) = - \vec k'(2)$ and,  if the particles in the
initial and final state are on their mass-shell,
\begin{equation}
q_0=(E_B(q) - E_{B'}(q))/2
\quad,\qquad
k_0=k'_0=0
\label{eq:bbsm}
\end{equation}
with $ E_B(q):=\sqrt{ \vec q\,^2 + M_B^2}$.

Now, according to the prescription of Blankenbecler and Sugar, the
relativistic two-particle Greens function,  $G_{\mu'\anti{\mu}'}$, is
replaced by a 3-dimensional Greens function
$g_{\mu'\anti{\mu}'}\propto \delta(k'_0)$, which respects unitarity.
Due to the $\delta$ function $\delta(k'_0)$, which sets the two
intermediate particles (being of equal mass) equally off-mass-shell
($k'_0(1)=k'_0(2)$),
the integration over $k'_0$ can be carried out immediately and the
so-called Blankenbecler-Sugar equation for the helicity amplitudes is
obtained:
\begin{eqnarray}
\lefteqn{
\bra{\mu\bar\mu, \vec k} T(t) \ket{B\anti{B'},\vec q,\lambda_1
\lambda_2}
 =
\bra{\mu\bar\mu, \vec k} V(t) \ket{B\anti{B'},\vec q,\lambda_1
\lambda_2}}
\nonumber \\
&& +{1 \over (2\pi)^3 } \sum_{\mu' \bar{\mu}'=\pi\pi,K\bar{K}}
N_{\mu' \bar{\mu}'}
 \int d^3k' {T_{ \mu' \bar{\mu}' \to \mu\bar\mu }(\vec k,\vec k';t)
\bra{\mu'\bar\mu', \vec k'} V (t) \ket{B\anti{B'},\vec q,\lambda_1
\lambda_2}
 \over \omega_{k'} (t- 4\omega^2_{k'} + i \epsilon) }\quad,
\nonumber \\
\label{eq:bbsma}
\end{eqnarray}
where $N_{\mu' \bar{\mu}'}$ is defined as in Eq.~\ref{eq:3_3_1a}
and $T_{ \mu'\bar{\mu}'\to\mu\bar\mu}\equiv{\cal M}_{\mu'\bar{\mu}'\to
\mu\bar\mu}$.

Because of the rotational invariance of the underlying interactions
the BbS equation can be decomposed into partial waves~\cite{GW}:
\begin{eqnarray}
\lefteqn{
\bra{\mu\bar\mu}  T^J(k,q;t) \ket{B\anti{B'},\lambda_1 \lambda_2}
 =
\bra{\mu\bar\mu} V^J(k,q;t) \ket{B\anti{B'},\lambda_1 \lambda_2}}
\nonumber \\
&& + {1 \over (2\pi)^3 }
\sum_{\mu' \bar{\mu}'} N_{\mu' \bar{\mu}'}  \int {k'}^2 dk'
{T^J_{ \mu' \bar{\mu}' \to \mu\bar\mu }(k,k';t)
\bra{\mu'\bar\mu'} V^J(k',q,t) \ket{B\anti{B'},\lambda_1 \lambda_2}
 \over \omega_{k'} (t- 4\omega^2_{k'} + i \epsilon)}.
\nonumber \\
\label{eq:bbsn}
\end{eqnarray}
Here, $q,k,k'$ denote the modulus of the corresponding 3-momenta.

The extrapolation of the model for the $N\anti{N} \to \pi\pi$
amplitudes to the other particle channels is made under the assumption
that the hadronic interactions are, except for the particle masses,
$SU(3)_{flavor}$ symmetric.
That means that the coupling constants at the various hadronic vertices
are
related to each other by $SU(3)$ relations.
In this way it comes out that all free parameters of the model for the
$B\anti{B'} \to \mu\anti{\mu}$ amplitudes can be fixed by adjusting
the $N\anti{N} \to \pi\pi$ amplitudes to the quasiempirical
data~\cite{Hoehler2,Dumbrajs} (see Sect.~\ref{sec:5_1}).


\subsubsection{$B\anti{B'} \to \pi\pi,K\anti{K}$ Born amplitudes}
\label{sec:bornsp}

In our model, the  $B\anti{B'} \to \mu\anti{\mu}$ transition
potentials are build up from
a $\rho$-pole diagram  and all diagrams in which
a baryon out of the $J^P=1/2^+$ octet or the $J^P=3/2^+$ decuplet is
exchanged. Of course only those diagrams are considered which respect
the conservation of isospin and strangeness.
As an example Fig.~\ref{fig:4_0b} shows the Born amplitudes for the
transition $\Sigma\anti{\Sigma}\to\pi\pi,K\anti{K}$.
The particles occuring in this model are listed in Tab.~\ref{tab:4_0}
together with their
masses and their basic quantum numbers.

In order to start from a maximum $SU(3)$ symmetry our model differs
slightly from the one presented in Ref.~\cite{Pearce_piN} for the
$N\anti{N}\to\pi\pi$ amplitudes.
Here, we include the exchange of the $Y^* \equiv\Sigma(1385)$ in the
$N\anti{N} \to  K\anti{K}$ transition potential, which was neglected in
Ref.~\cite{Pearce_piN}. In addition, the form factors at the
hadronic vertices are chosen identical within a given $SU(3)$ multiplet.

Starting point for the derivation of the various Born amplitudes are
the following interaction Lagrangians, which are
characterized by the  $J^P$ quantum numbers of the hadrons
involved. For simplicity, we suppress here
the isospin or $SU(3)$ dependence of the Lagrangians.
\begin{equation}
\begin{array} {ll}
\half^+ \otimes \half^+ \otimes 0^-: &
{\cal L}_{B'Bp}(x)= {f_{B'Bp} \over m_{\pi^+}}
 \anti{\psi}_{B'}(x) \gamma_5\gamma^\mu \psi_B(x) \partial_\mu
 \phi_p(x)
\quad (+ \;\mbox{\rm h.c., if } B'\neq B)\,, \\ \\
\thalf^+ \otimes \half^+ \otimes 0^-:&
{\cal L}_{DBp}(x)= {f_{DBp} \over m_{\pi^+}} \anti{\psi}_B (x)
 (g_{\mu\nu} + x_\Delta \gamma_\mu \gamma_\nu) \psi^\mu_D(x)
\partial^\nu \phi_p(x)
\quad   + \mbox{\rm h.c.}\,,\\ \\
\half^+ \otimes \half^+ \otimes 1^-:&
{\cal L}_{B'B\rho}(x)=
\begin{array}[t]{l}
 g_{B'B\rho} \anti{\psi}_{B'}(x) \gamma_\mu
 \psi_B(x) \phi^\mu_\rho(x)\\
+ {f_{B'B\rho} \over 4 M_N} \anti{\psi}_{B'}(x) \sigma_{\mu\nu}
 \psi_B(x) (\partial^\mu \phi^\nu_\rho(x) - \partial^\nu
\phi^\mu_\rho(x))
 \\ \\
(+ \mbox{\rm h.c., if } B'\neq B) \,,
\end{array} \\ \\
0^- \otimes 0^- \otimes 1^-: &
{\cal L}_{p'p\rho}(x)= g_{p'p\rho} \phi_{p'} (x) \partial_\mu \phi_p
 \phi^\mu_\rho(x) \quad (+ \mbox{\rm h.c., if } p'\neq p) \,.
\\
\end{array} \nonumber \\
\label{eq:Lag}
\end{equation}

For the conventions used for the field operators and the Dirac
$\gamma$-matrices, see Appendix~\ref{sec:AnhA}.
The Lagrangian ${\cal L}_{DBp}$ includes an off-shell part, which is
proportional to $x_\Delta$. As the parameter $A$, which occurs in the
free Lagrangian~\cite{Read} and then in the non-pole part of the
propagator of a spin-3/2
particle (cf.\ Eq.~\ref{eq:AnhARS}), the  parameter $x_\Delta$
characterizing the
strength of the off-shell part of the $DBp$ coupling  is not
determined from first principles.
However, it is known~\cite{NEK} that fieldtheoretic amplitudes derived
with this  most general ansatz depend only on a certain combination of
$A$ and $x_\Delta$, namely
\begin{equation}
{1+4x_\Delta \over 2A+1} =: 1+4Z\quad.
\label{eq:bornspa}
\end{equation}
It follows that different  pairs of ($x_\Delta,A$) values, which give
the
same value of $Z$, describe the same interaction theory.
Therefore, without restricting the general validity of our results,
we can set $A=-1$ (i.e.\ omitting the non-pole part of the spin-3/2
propagator)
and select the interaction theory (characterized by $Z$) through
$x_\Delta$, which is finally adjusted to the quasiempirical data for
$N\anti{N}\to\pi\pi$.

In order to account for the extended structure of hadrons the vertex
functions resulting from the Lagrangians~\ref{eq:Lag} are modified by
phenomenological form factors.
For the baryon-exchange processes these form factors are parametrized
in the usual multipole form
\begin{equation}
F_{X}(p^2) = \left( {n_{X} \Lambda^2_{X} - M_X^2 \over n_{X}
\Lambda^2_{X} - p^2 }
\right)^{n_{X}}\quad,
\label{eq:BexFF}
\end{equation}
where $p$ denotes the 4-momentum and $M_X$ the mass of the exchanged
baryon $X$.
The  two parameters, the so-called cutoff mass $\Lambda_X$ and the
power $n_X$, are chosen uniquely for all $BB'p$ ($\Lambda_8$, $n_8$)
and for all $BDp$ ($\Lambda_{10}$, $n_{10}$)
vertices in order to keep the number of parameters of our model
as low as possible.
The dependence of the form factor~\ref{eq:BexFF} on the power $n_X$ is
quite weak~\cite{Pearce_piN}. We choose $n_8=1$ and $n_{10}=2$.
Finally, $\Lambda_8$ and $\Lambda_{10}$ are adjusted to the
quasiempirical data for the $N\anti{N}\to\pi\pi$ amplitudes in the
pseudophysical region.

For the $\rho$-pole diagram we parametrize the form factor at the
$\mu\mu\rho$ vertex in the same way as is done in the model of the
$\pi\pi$-$K\anti{K}$ interaction in Ref.~\cite{Pearce_piN}:
\begin{eqnarray}
F_{\mu\mu\rho}(\vec k\,^2) &=& \left( {n_{\mu\mu\rho}
\Lambda^2_{\mu\mu\rho} + (m^{(0)}_\rho)^2 \over n_{\mu\mu\rho}
\Lambda^2_{\mu\mu\rho} + 4\omega_\mu^2(\vec k\,^2) }
\right)^{n_{\mu\mu\rho}}
\label{eq:rhoFF}
\end{eqnarray}
where $\vec k$ is the relative 3-momentum of the two pseudoscalar
mesons and
 $k_{\mu\bar\mu}^2 (t) = t/4 - m_\mu^2$
is the squared on-shell momentum of the $\mu\anti{\mu}$ state.

For the dispersiontheoretic calculation of correlated $\pi\pi$ and
$K\anti{K}$ exchange the  $B\anti{B'}\to \mu\anti{\mu}$ Born amplitudes
are
evaluated only for  $B\anti{B'}$ states being on their mass-shell.
Therefore, there  is no need for a form factor at the $BB'\rho$ vertex
as to assure convergence of the scattering
equation~\ref{eq:bbsn}. Therefore, we disregard this form factor,
again in order to keep the parameters of the model as low as possible.

Now, taking into account that the $B\anti{B'}$ state is on mass-shell
and that due to the Blankenbecler-Sugar condition (Eq.~\ref{eq:bbsm})
the energy component
of the relative momentum of the $\mu\anti{\mu}$ state always vanishes
($k_0=0$) the 4-momenta at the external legs of the
$B\anti{B'}\to\mu\anti{\mu}$ Born diagrams read  in the center-of-mass
system:
\begin{equation}
q_B=\left(
\begin{array}{c}
E_B(q) \\ \vec q
\end{array}
\right)
,\quad\!\!\!\!
q_{\bar B'}=\left(
\begin{array}{c}
E_{B'}(q) \\ -\vec q
\end{array}
\right)
,\quad k(\mu)=\left(
\begin{array}{c}
\sqrt{t'} \\ \vec k
\end{array}
\right)
,\quad\!\!\!\!  k(\anti{\mu})=\left(
\begin{array}{c}
\sqrt{t'} \\ -\vec k
\end{array}
\label{eq:4born30}
\right).
\end{equation}
According to the usual Feynman rules~\cite{BjDr} (for the various
propagators, see also Appendix~\ref{sec:AnhA}) we obtain for the
spin-momentum parts of the $B\anti{B'}\to \mu\anti{\mu}$ Born
amplitudes
${\cal V}_{ B\anti{B'} \to \mu \anti{\mu}}(\vec k,\vec q;t)$:
\begin{itemize}
\item Exchange of a baryon $X$ with $j^P=\half^+$ and momentum $p=q-k$
\begin{equation}
{\cal V}^X_{ B\anti{B'} \to \mu \anti{\mu}}(\vec k,\vec q;t') =
\left(
-{f_{XB'\mu}\over m_{\pi^+}} \gamma_5 \gamma_\lambda k^\lambda(\anti
{\mu})
\right)
{-(\not\!p + M_X) \over p^2 - M_X^2 }
\left(
-{f_{XB\mu}\over m_{\pi^+}} \gamma_5 \gamma_\nu k^\nu(\mu)
\right)\quad,
\label{eq:4born32}
\end{equation}

\item Exchange of a baryon $X$ with $J^P=\thalf^+$ and momentum
$p=q-k$
\begin{eqnarray}
\lefteqn{
{\cal V}^X_{ B\anti{B'} \to \mu \anti{\mu}}(\vec k,\vec q;t') =
\left[
-{f_{XB'\mu}\over m_{\pi^+}} (g_{\lambda\nu} +
x_\Delta\gamma_\nu\gamma_\lambda) k^\nu(\anti{\mu})
\right]
S^{\lambda\rho}_X(p,A=-1)} \hspace{5.cm}
\nonumber \\
&&
\left[
-{f_{XB\mu}\over m_{\pi^+}} (g_{\rho\sigma} +
x_\Delta\gamma_\rho\gamma_\sigma) k^\sigma(\mu)
\right]\quad,
\label{eq:Del_exch}
\end{eqnarray}

\item $\rho$-pole graph with bare mass $m^{(0)}_\rho$
\begin{eqnarray}
\lefteqn{
{\cal V}^\rho_{ B\anti{B'} \to \mu \anti{\mu}}(\vec k,\vec q;t') =
\left(
i g^{(0)}_{BB'\rho} \gamma_\lambda + {f^{(0)}_{BB'\rho} \over 2M_N}
\sigma_{\lambda\nu} P^\nu
\right)
{ g^{\lambda\sigma} - P^\lambda P^\sigma/(m^{(0)}_\rho)^2
\over t' - (m^{(0)}_\rho)^2  }}\hspace{7cm}
\nonumber \\
&& \!\!\!\! \left[ -i g^{(0)}_{\mu\mu\rho} (k(\mu)-k(\anti{\mu}))_\sigma
\right]\,.
\label{eq:4born33}
\end{eqnarray}
\end{itemize}

These expressions can be further simplified by introducing the momenta
given in Eq.~\ref{eq:4born30} and contracting the $\gamma$-matrices.
Finally, the  corresponding $B\anti{B'} \to \mu \anti{\mu}$ helicity
amplitudes are obtained by applying
 $ {\cal V}_{B\anti{B'} \to \mu\bar\mu}(\vec k,\vec q;t)$
to the Dirac helicity spinors of the baryons (cf.\ Eqs.~\ref{eq:app2a}
and  \ref{eq:app3a} in the Appendix):
\begin{equation}
\bra{\mu\bar\mu, \vec k} V(t) \ket{B\anti{B'},\vec q ,\lambda_1
\lambda_2} =
\anti{v}_{B'}(-\vec q,\lambda_2) {\cal V}_{B\anti{B'} \to \mu\bar\mu}
(\vec k,\vec q;t)
u_{B}(\vec q,\lambda_1)\quad.
\label{eq:born_heli}
\end{equation}
The final results for the helicity amplitudes are  summarized in
Appendix~\ref{sec:AnhC}.

In our model the coupling constants at the various hadronic vertices
are related to
each other by $SU(3)$ arguments. The $SU(3)$ relations together with
the isospin factors of the various Born amplitudes are given in
Appendix~\ref{sec:borniso}.


As discussed in the previous chapters,
for the dispersiontheoretic calculation of correlated $\pi\pi$ and
$K\anti{K}$ exchange the $B\anti{B'} \to \pi\pi,K\anti{K}$ have to be
known in the pseudophysical region, i.e.\ for energies $t'$ below the
$B\anti{B'}$ threshold ($4m_\pi^2 \leq t' < (M_B + M_{B'})^2 $).
 Therefore, after having derived the analytic
expressions for these Born amplitudes in the physical region
($\sqrt{t'} >M_B+M_{B'}$) they have to be continued analytically as
functions of $t'$ (and $s$) into
the pseudophysical region.
For this all energy-dependent quantities occuring in the expressions for
the Born amplitudes (cf.\ Appendix~\ref{sec:AnhC})
have to be expressed as functions of $t'$ and $s$.

If we adopt  the approximation introduced in Sect.~\ref{sec:kap3_4},
namely  that the masses
of the baryon and of the antibaryon are equal ($M_B=M_{B'}$), the
square of the relative 4-momentum and the one-particle energies of the
$B\anti{B'}$
state  are given in the
center-of-mass system for physical values of $t$ by
\begin{equation}
\begin{array}{rcl}
q^2(t) &=& t/4 \, - M_B^2 \quad,\\ E_B=E_{B'} &=& \sqrt{t}/2\quad.
\end{array}
\end{equation}
The analytic continuation of these relations to the pseudophysical
region is obvious.
Note that if we would have allowed $M_B$ and $M_{B'}$ to be different
the
corresponding relations would look more involved:
\begin{equation}
\begin{array}{rcl}
q^2(t) &=& {
\left[t-(M_B+ M_{B'})^2\right]
\left[t-(M_B- M_{B'})^2\right]
\over  4t}\quad,
\\
E_{B} (t) &=& {(E_{B} + E_{B'})\over2} + {(E_{B}^2 - E_{B'}^2)\over
2(E_{B} + E_{B'})}
\\
&=& {t+M_B^2 - M_{B'}^2 \over 2\sqrt{t}} \quad,
\\ \\
E_{B'} (t) &=&{t+M_{B'}^2 - M_{B}^2 \over 2\sqrt{t}}\quad.
\end{array}
\label{eq:cont_3}
\end{equation}

Baryon-exchange diagrams in which the mass $M_X$ of the exchanged
baryon is
sufficiently
smaller than the mass $M_B=M_{B'}$ of the external baryons (e.g.
$N$-exchange in
$\Lambda\anti{\Lambda}\to K\anti{K}$ or $\Lambda$-exchange in
$\Sigma\anti{\Sigma}\to\pi\pi$ do not satisfy a Mandelstam
representation as was already pointed out in Ref.~\cite{Riska} for the
latter example.
 The nonvalidity of the Mandelstam representation becomes obvious when
extrapolating the corresponding Born amplitude in Eq.~\ref{eq:4born32}
to the
pseudophysical region.
For given $t< 4(M_B^2 - M_X^2)$, the propagator of baryon $X$,
\begin{eqnarray}
\left[ p^2 - M_X^2\right]^{-1}
 &=&  -\left[ t/4 -M_B^2 + M_X^2 + {\vec k}^2 - 2 \vec q \cdot \vec k
\right]^{-1}\quad,
\end{eqnarray}
 then acquires a singularity at $cos\theta=0 $
($\vec q \cdot \vec k$ imaginary) and the following off-shell momentum
of the $\mu\anti{\mu}$ state
\begin{equation}
{\vec k}^2 = M_B^2 - M_X^2 - t/4 >0
\quad.
\end{equation}
Since this problem would hinder an evaluation of correlated $\pi\pi$
and $K\anti{K}$ exchange and we do not see at present any proper
solution, we eliminate the problem by the following approximation:
In all baryon-exchange diagrams in which  the mass of the exchanged
baryon, $M_X$, is smaller than the mass of the external baryons,
$M_B$, $M_X$ is increased by hand to $M_B$.
Again, since the uncorrelated contributions of two-pion and two-kaon
exchange (e.g.\  iterative
two-pion exchange in the $\Sigma N$ channel with a $\Lambda N$
intermediate state )
are evaluated explicitly in  the $s$-channel, these
contributions are not affected by our approximation and thus have the
correct energy dependence.
Approximations are only made in the correlated part which have a much
weaker energy dependence than the uncorrelated contributions.

\section{Results and Discussion}
\label{sec:results}

\subsection{Determination of free parameters}
\label{sec:5_1}

During the construction of the microscopic model for the $B\anti{B'}
\to\mu\anti{\mu}$
amplitudes it proved to be essential to restrict the number of free
parameters as much as possible, which can then be fixed by adjusting the
model predictions to the quasiempirical $N\anti{N}\to\pi\pi$
amplitudes. Only in
this way one can hope to obtain a reasonable description of the other
baryon-antibaryon channels, for which no empirical data exist.

As shortly outlined in Sect.~\ref{sec:kap4}
the $\pi\pi-K\anti{K}$ interaction model has been
developed independently before, with all parameters adjusted to fit the
existing $\pi\pi$ scattering phase shifts. Therefore coupling constants
and
form factors at the $\pi\pi\rho^{(0)}$ and $KK\rho^{(0)}$ vertices
occurring in the
$\rho^{(0)}$ pole terms  are already determined. Assuming that the
bare $\rho$ meson couples universally to the isospin current we can
also fix
all vector couplings $g^{(0)}_{BB'\rho}$ to the baryonic vertices.
Corresponding tensor
couplings $f^{(0)}_{BB'\rho}$ will be related by $SU(3)$ symmetry,
with two parameters
remaining, namely the coupling constant $f^{(0)}_{NN\rho}$ and the
$F/(F+D)$ ratio $\alpha_v^m$.

The coupling of the pseudoscalar mesons $\pi$ and $K$ to the octet
baryons is
in the framework of $SU(3)$ symmetry likewise determined by two
parameters,
the $F/(F+D)$ ratio $\alpha_p$   and the coupling constant $g_{NN\pi}$.
(For the latter
we will take throughout  $g_{NN\pi}^2/4\pi= 14.3$). There is an
additional freedom
since $SU(3)$ symmetry can be either assumed for the pseudoscalar
coupling constants $g_{BB'p}$ or the pseudovector ones $f_{BB'p}$,
which are related
 by
\begin{equation}
 g_{B'Bp} = f_{B'Bp} \;{M_B + M_{B'}\over m_{\pi^+} } \quad.
\label{eq:ps_pv}
\end{equation}
These two possibilities are not equivalent since,
because of $SU(3)$ breaking of baryon masses, both sets of coupling
constants cannot be $SU(3)$ symmetric at the same time. In this work we
will assume $SU(3)$ symmetry for the pseudoscalar couplings, but we will
also check the influence of the alternative possibility.

Finally, under the assumption of $SU(3)$, the couplings $f_{BDp}$ of
$\pi$ and $K$
to the transition current between baryon octet and decuplet are
determined by only one parameter, $f_{N\Delta\pi}$. We will use
$f^2_{N\Delta\pi}/4\pi = 0.36$
in the following.

In addition we have form-factor parameters at the hadronic vertices.
In order to keep their number small we assume that the cutoff masses
$\Lambda_{BX\mu}$ are independent of the exchanged baryon $X$ within
one $SU(3)$
multiplet. Consequently we have two additional parameters: $\Lambda_8$
if $X$ is a
member of the baryon  octet  and $\Lambda_{10}$ if it is in a decuplet.
The power
$n_8=1$ and $n_{10}=2$ in the form-factor {\em ansatz} are sufficient
to ensure
convergence of the scattering equation. Finally, $x_\Delta$
(Eq.~\ref{eq:Lag})
characterizing the off-shell part of the $(3/2)^+\otimes (1/2)^+\otimes
0^-$ coupling is treated
as a free parameter in our $B\anti{B'}\to\mu\anti{\mu}$  model.

In order to reduce the number of parameters further we even assume
$SU(6)$ symmetry, which fixes $\alpha_p$ and $\alpha_v^m$ to be 0.4.
Thus we are left
with four free parameters
$f^{(0)}_{NN\rho}$, $x_\Delta$, $\Lambda_8$ and $\Lambda_{10}$, which
have been fixed
by adjusting our theoretical predictions for the $N\anti{N}\to\pi\pi$
amplitudes
to the quasiempirical results of H\"ohler and Pietarinen~\cite{Hoehler2,
Dumbrajs} given in
the form of Frazer-Fulco amplitudes $f_\pm^J(t)$, which, up to kinematic
factors, correspond to the partial wave decomposed helicity amplitudes
of Sect.~\ref{sec:kap3_4}, i.e.
 \begin{eqnarray}
f^J_+(t)&=&-{1\over 16 \pi^2} {qM_N\over (qk)^J }
\bra{\pi\pi}  T^J(t) \ket{N\anti{N},++}\times F^J \quad,
\nonumber \\
f^J_-(t)&=&-{1\over 8 \pi^2} {qM_N\over (qk)^J \sqrt{t} }
\bra{\pi\pi}  T^J(t) \ket{N\anti{N},+-}\times F^J\quad,
\end{eqnarray}
where  $k$ and $q$ are the on-shell momenta of pions and nucleons. The
factors $F^J = - 1/\sqrt{6}\, (-1/2)$ for $J=0\,(1)$ are due to the
transition from
isospin amplitudes used in this work to the Frazer-Fulco amplitudes,
which are defined in isospin space as coefficients of the independent
isospin operators $\delta_{\alpha\beta}$ and $\ohalf[\tau_\alpha,
\tau_\beta]$.
Since $\bra{\pi\pi}  T^{J=0}(t) \ket{N\anti{N},+-}$ vanishes identically
we have
only one amplitude, $f_+^0$, in the $\sigma$ channel whereas in the
$\rho$ channel
we have both $f_+^1$ and $f_-^1$.

Fig.~\ref{fig:5_2} shows the predictions of our microscopic model for
$f_+^0$, $f_+^1$
and $f_-^1$, in comparison to the quasiempirical results of
Ref.~\cite{Hoehler2} in the
pseudophysical region $t\geq4m_\pi^2$; Table~\ref{tab:5_param}
contains the chosen parameter
values. (Note that the present model differs somewhat from our former
model~\cite{Pearce_piN}, e.g.\ by the inclusion of $Y^*$ exchange;
therefore the values
differ slightly from those given in Ref.~\cite{Pearce_piN}). With only
four parameters
we obtain a very satisfactory reproduction of the quasiempirical data,
especially in the $\rho$ channel. Some discrepancies occur in the
$\sigma$ channel,
which however have only small influence on final results for
the correlated $\pi\pi$ exchange in the $s$-channel reactions ($NN$,
$\pi N$), as also
discussed below. Furthermore it should be kept in mind that in this
channel the quasiempirical information is plagued with considerable
uncertainties.
\subsection{The $B\anti{B'}\to\mu\anti{\mu}$  transition potentials}
\label{sec:5_2}

Having fixed all parameters in our microscopic model for the
$B\anti{B'}\to\mu\anti{\mu}$
amplitudes we will now first look at the transition potentials (Born
amplitudes) in the various baryon-antibaryon channels.
Figs.~\ref{fig:5_nnborn} and \ref{fig:5_ssborn}$B\anti{B'}\to\pi\pi,\,
K\anti{K}$ amplitudes
 show
the contributions of the various baryon-exchange processes to the
$N\anti{N}\to\pi\pi,K\anti{K}$
and $\Sigma\anti{\Sigma}\to\pi\pi,K\anti{K}$
 Born amplitudes above the $\mu\anti{\mu}$
thresholds, i.e. for $t\geq 4m_\mu^2$.
Contributions of the $\rho^{(0)}$  pole  terms are not shown since
they possess a singularity  at the bare rho mass $m_{\rho}^{(0)}$. This
pole will be
regularized only after iteration and coupling to the full $\pi\pi$
amplitude
and then leads to the resonance structure at the physical rho mass
$m_\rho$.

The notation of the (Born) amplitudes  follows that of the Frazer-Fulco
amplitudes, i.e.
\begin{eqnarray}
V^J_+(t)&=&\bra{\mu\anti{\mu}}  V^J(t) \ket{B\anti{B'},++}\quad,
\nonumber \\
V^J_-(t)&=&\bra{\mu\anti{\mu}}  V^J(t) \ket{B\anti{B'},+-}\quad.
\end{eqnarray}
The partial wave decomposed Born amplitudes are either even or odd
functions of the baryonic relative momenta
(cf.\ Appendix~\ref{sec:AnhC}).
Since the latter are
imaginary in the pseudophysical region the Born amplitudes being odd
functions (e.g.\ $V^0_+$) become likewise imaginary.

Apart from coupling constants and isospin factors the strengths of the
various contributions are strongly determined by mass ratios. Exchange
baryons with the same mass $M_X$ as the outer baryon-antibaryon pairs
(according to our approximation introduced in Sect.~\ref{sec:bornsp}
this case includes
also those exchange baryons whose physical mass is lower than that of
the
outer baryons) produce, in the $\pi\pi$ amplitudes $V_+^0$ and $V_+^1$,
a typical
structure at the $\pi\pi$ threshold. The strong rise of the amplitudes,
which
acquire a finite value at $t=4m_\pi^2$, is a direct signal of the
so-called
left-hand cut, which is generated by the singularity of the
corresponding
$u$-channel pole graph. This cut starts just below the $\pi\pi$
threshold and
extends from $t_0=4m_\pi^2(1-m_\pi^2/M_X)$ along the real axis to
$-\infty$.

Obviously, in the $\sigma$ channel, the various pieces interfere
constructively, whereas in the $\rho$ channel also destructive
interferences
occur. The contributions generated by a spin-3/2 exchange baryon have
opposite sign in $V^1_+$ and $V_-^1$ whereas both amplitudes are almost
equal
when a spin-1/2 baryon is exchanged.

The  $N\anti{N}\to \pi\pi$  Born amplitudes in Fig.~\ref{fig:5_nnborn}
built up by
nucleon and $\Delta$
exchange are noticeably larger than the $N\anti{N}\to K\anti{K}$ Born
amplitudes, which
are suppressed because of the high mass of the exchange baryons
$\Lambda$, $\Sigma$, $Y^*$.
Due to different mass ratios this is no longer true in the
hyperon-antihyperon
channels.
For instance, in the $\Sigma\anti{\Sigma}$ channel shown in
Fig.~\ref{fig:5_ssborn}  the coupling via $\Delta$ exchange to
$K\anti{K}$ even dominates,
due to the large coupling constant ($f^2_{\Sigma\Delta K}=
f^2_{N\Delta\pi}$)
and the small mass
difference between $\Sigma$ and $\Delta$ ($M_\Delta-M_\Sigma\approx 39\
MeV$).
Therefore already from these results it is to be expected
that correlated $K\anti{K}$ exchange processes play a minor role in the
$NN$ system
but are important for the    interactions involving hyperons.

\subsection{The $B\anti{B'}\to \mu\anti{\mu}$ transition amplitudes }
\label{sec:Kap53}

The $B\anti{B'}\to \mu\anti{\mu}$  helicity amplitudes, $T^J_\pm (t)$,
which are obtained from the
solution of the BbS scattering equation~\ref{eq:bbsn}, consist of Born
terms
and pieces containing meson-meson correlations.The latter part fixes
the phase of the $B\anti{B'}\to \mu\anti{\mu}$  amplitudes and
generates the discontinuities of
$T^J_\pm (t)$  along the unitarity cut. It contains $\pi\pi$ and
$K\anti{K}$ correlations
in the form of $\pi\pi$, $K\anti{K}$ amplitudes. Corresponding
predictions from
our field-theoretic model, for the scalar-isoscalar on-shell amplitude
$T^{J=0,I=0}(t)$  are shown in Fig.~\ref{fig:5_tmumu}. By comparison
of the $\pi\pi$ amplitude with the
$\pi\pi$ phase shifts (cf.\ Fig.~\ref{fig:4_6}) we can see immediately
that the vanishing
of the real part of the amplitude at $t=37 m_\pi^2$ and $t=71m_\pi^2$
corresponds
to the phase shifts acquiring the value of 90 resp.\  270 degrees,
respectively. The phase value of 180 degrees (and the corresponding
vanishing of the amplitude) occurs just below the $K\anti{K}$ threshold
($t=50.48 m_\pi^2$).
Remarkably, there is a steep rise of the imaginary part of the $\pi\pi$
amplitude at low energies, which has to vanish at the $\pi\pi$ threshold
for unitarity reasons. We mention that, in the $NN$ system, this part
provides the dominant contribution to correlated $\pi\pi$ exchange. It
should
be noted that both the real and imaginary part of the $K\anti{K}\to
K\anti{K}$
amplitude (which in Fig.~\ref{fig:5_tmumu}  are shown only in the
physical region above
the  $K\anti{K}$ threshold) acquire a non-vanishing value at the
$K\anti{K}$ threshold
due to the open $\pi\pi$ channel.

Figs.~\ref{fig:5_tnn} and \ref{fig:5_tss} show the resulting
$B\anti{B'}\to \pi\pi$
amplitudes for $B\anti{B'}=N\anti{N}, \Sigma\anti{\Sigma}$. Besides the
predictions of the full model (solid line) the figures also show the
effect which is obtained when neglecting the $B\anti{B'}\to K\anti{K}$
Born amplitudes
in the $\sigma$ channel (dashed line) and restricting oneself to the
$\rho$-pole amplitudes in the $\rho$ channel (dash-dotted line).
The $B\anti{B'}\to K\anti{K}$
amplitudes (dotted lines) are only shown for $J=0$ since for $J=1$ as
exemplified for the  $\Sigma\anti{\Sigma}$ channel they provide almost
negligible
contributions, mainly because of the weak $K\anti{K}$ interaction in
that
partial wave. Note that in the $\sigma$ channel, due to the imaginary
Born amplitudes, the role of the real and imaginary part of $T^0_+$ are
in a certain sense interchanged: ${\sl Im} [T^0_+]$ contains the Born
amplitudes
whereas the discontinuity along the unitarity cut is contained in
${\sl Re}[T^0_+]$.

Obviously, in all channels, the structure of the $\pi\pi$ amplitude of
Fig.~\ref{fig:5_tmumu}
can be recognized in those $B\anti{B'}\to\pi\pi$ amplitudes $T^0_+$,
which have been
evaluated without the $B\anti{B'}\to K\anti{K}$ Born amplitudes (dashed
curves). The
steep rise of ${\sl Im}[T^0_+]$ at the $\pi\pi$ threshold originates
from the analogous
behavior of the Born amplitudes (cf.\  Figs.~\ref{fig:5_nnborn} and
\ref{fig:5_ssborn}).
As follows from the
unitarity relation for the $B\anti{B'}\to\pi\pi$ amplitudes below the
$K\anti{K}$ threshold
($t_{thr}\approx 50.48 m_\pi^2$) the discontinuity of the $B\anti{B'}
\to\pi\pi$ amplitudes
(contained in ${\sl Re}[T^0_+]$) must vanish at $t\approx 50.43 m_\pi^2$
since there $T^{J=0,I=0}=0$.
Furthermore, those amplitudes $T^0_+$, which have been evaluated without
the  $K\anti{K}$ Born amplitudes, vanish at an additional, though for
each
channel different position in the energy region below the  $K\anti{K}$
threshold.
Interestingly this vanishing of the amplitudes is lost in the
hyperon-antihyperon
channels when the  $K\anti{K}$ Born amplitudes are included; in
case of the $N\anti{N}\to\pi\pi$ channel on the other hand the
resulting full
amplitude still vanishes, at a slightly different position
($t\approx 50.2m_\pi^2$).
Note that the quasiempirical $N\anti{N}\to\pi\pi$ amplitude $f^0_+$ in
Fig.~\ref{fig:5_2} has a
comparable structure, however the amplitude vanishes already at a
somewhat smaller value of $t\approx43m_\pi^2$. Fig.~\ref{fig:5_thr}
also clearly shows the cusp
structure of the amplitudes at the $K\anti{K}$ threshold, which is
however
much stronger in the hyperon-antihyperon channels.

The $B\anti{B'}\to K\anti{K}$ Born amplitudes do not play any role in
the $N\anti{N}\to\pi\pi$
channel, as noted already in Refs.~\cite{Durso,Pearce_piN}. In the
hyperon-antihyperon
channels on the other hand, we have already at small $t$-values
important
contributions to the $B\anti{B'}\to \pi\pi$ amplitudes $T^0_+$, which
lead to an
enhancement of the amplitudes below the $K\anti{K}$ threshold and to a
noticeably
different energy dependence.

The amplitudes $T^1_\pm$ in the $\rho$ channel are dominated by the
resonance
structure in the region of the physical $\rho$ mass at $t=30.4 m^2_\pi$.
We
stress that by multiplication with the $\pi\pi$ correlations in the
scattering equation also the baryon-exchange Born terms lead to
resonant contributions. However, in all channels, the discontinuity
of the amplitudes mainly arises (to at least $60\%$) from both $\rho$
pole
terms  and those contributions generated by them in the
scattering equation (dash-dotted curves in Figs.~\ref{fig:5_tnn} and
\ref{fig:5_tss}).

As shown in Fig.~\ref{fig:5_tss} for the $\Sigma\anti{\Sigma}\to K\anti
{K}$ transition,
the $B\anti{B'}\to K\anti{K}$
on-shell amplitudes (dotted curves) are completely unimportant in the
$\rho$ channel. On the other hand, in the $\sigma$ channel, the $ B\anti
{B'}\to K\anti{K}$
amplitudes can acquire large values especially near the $K\anti{K}$
threshold,
due to the corresponding behavior of the $K\anti{K}$  amplitude
$T^{J=0,I=0}$
(cf.\ Fig.~\ref{fig:5_tmumu}),
 which lead to non-negligible contributions to the spectral functions.

\subsection{The spectral functions}
\label{sec:5_4}

Based on the  $B\anti{B'}\to\mu\anti{\mu}$ amplitudes in the
pseudophysical region determined
in the last section we can now in principle evaluate the spectral
functions $\rho^{\sigma,\rho}_i $  (defined in Eq.~\ref{eq:3_5_1}) of
correlated $\pi\pi$ and
$K\anti{K}$
exchange for all baryon-baryon channels containing the octet baryons
$N,\Lambda,\Sigma$ and $\Xi$. In this work however we restrict
ourselves to those
channels in which experimental information is available at present or
in the near future. Besides the baryon-baryon channels with strangeness
$S=0$ and $S=-1$, for which numerous ($NN$) and scarce ($N \Lambda$,
$N\Sigma$)  scattering
data exist, we will also consider the $S=-2$ channels $\Lambda\Lambda$,
$\Sigma\Sigma$
and $N \Xi$,
which are relevant both for the description of $\Lambda\Lambda$-
resp.\ $\Xi$-hypernuclei
and for the study of the $H$-dibaryon~\cite{Jaffe} predicted in the
($S=-2$, $J=I=0$) channel.

Fig.~\ref{fig:5_5_1}a (Fig.~\ref{fig:5_5_1}b) shows the spectral
function
$\rho^{\sigma}_S(BB')$ in the $\sigma$ channel
predicted by the full microscopic model for the $S=0,-1$ $BB'$
channels $NN,N\Lambda,N\Sigma$
 ($S=-2$ channels $\Lambda\Lambda,\Sigma\Sigma,N\Xi$). Due to isospin
conservation the $\pi\pi$
resp.\ $K\anti{K}$ exchange contributes to the $N\Lambda$ and
$\Lambda\Lambda$ interaction only in the
$\sigma$ channel and to the $N\Lambda-N\Sigma$  transition only in the
$\rho$ channel. Up to
near the  $K\anti{K}$ threshold at $t=50.48 m^2_\pi$ $\rho^{\sigma}_S$
is negative in all
channels. Therefore, as expected, correlated $\pi\pi$ and $K\anti{K}$
exchange
provides attractive contributions in all channels. Especially at small
$t$-values, which determine the long-range part of the correlated
exchanges and therefore yield for low-energy processes in the
$s$-channel
the main contributions to the dispersion integral, the spectral
function  $\rho^{\sigma}_S(NN)$
 is by about a factor of 2 larger than the results for
$\rho^{\sigma}_S(N\Lambda)$
and $\rho^{\sigma}_S(N\Sigma)$, which have about the same size. Due to
the sizable
contributions of the $K\anti{K}$ Born amplitudes to the
hyperon-antihyperon
amplitudes found already in the last section,
$\rho^{\sigma}_S(N\Lambda)$  as well as
$\rho^{\sigma}_S(N\Sigma)$ show a noticeably different $t$-dependence
than $\rho^{\sigma}_S(NN)$.
Below the $K\anti{K}$ threshold the spectral functions are broadened
to larger
$t$-values, therefore the overall range of this correlated exchange is
reduced. This is even more true for the $S=-2$ channels $\Lambda\Lambda$
and $\Sigma\Sigma$
in Fig.~\ref{fig:5_5_1}b
since there the hyperon-antihyperon amplitudes enter quadratically.
The effect of the $B\anti{B'}\to K\anti{K}$ Born amplitudes is once
more shown in
Fig.~\ref{fig:5_5_1b}
for the $NN$ and $N\Sigma$ channel. Whereas they have small influence
on $\rho^{\sigma}_S(NN)$
they provide important contributions to $\rho^{\sigma}_S(N\Sigma)$
already near the
$\pi\pi$ threshold.

Fig.~\ref{fig:5_5_2} shows the corresponding spectral functions in the
$\rho$ channel
($\rho^\rho_S$, $\rho^\rho_V$, $\rho^\rho_T$ and $\rho^\rho_6$)
 for the various particle channels, which are non-vanishing. (Note
that $\rho^\rho_P$ depends linearly on the four functions shown, cf.\
Eq.~\ref{eq:3_lindep}, and
is therefore not shown). $\rho^\rho_6$ contributes only in $BB'$
channels
with different masses and therefore vanishes in the $NN$ and
$\Sigma\Sigma$ channel.
According to Eq.~\ref{eq:3_4_6} $\rho^\rho_S(s,t)$ depends on $s$
through the factor $\cos\vartheta_t(s,t)$.
In order to enable a comparison of the results for $\rho^\rho_S(s,t)$
in the
various particle channels we have throughout set $s$  on the threshold
of the $s$-channel process in question, i.e. $s=(M_B+M_{B'})^2$. Since
 the $\rho$ channel
is dominated by the resonant pieces in the $\pi\pi$ channel and the
$K\anti{K}$
channel does not play any role the $\rho^\rho_i(BB')$ possess almost
 the same
$t$-dependence in the various particle channels.

For the $ NN$ channel, the spectral functions for correlated $\pi\pi$
and $K\anti{K}$
exchange can be derived either from the $N\anti{N}\to\pi\pi$ amplitudes
of our
microscopic model or, alternatively, from the quasiempirical results
obtained in Refs.~\cite{Hoehler2,Dumbrajs}. As before we have then to
subtract the
uncorrelated pieces evaluated from the microscopic model for the
$N\anti{N}\to\pi\pi$ Born amplitudes. Therefore the results derived
from the
quasiempirical amplitudes depend on parameters (e.g.\ $g_{NN\pi}$) of
the
microscopic model.

In Fig.~\ref{fig:5_5_4} we show the $NN$  spectral functions in the
$\sigma$  as well as
in the $\rho$ channel obtained from our microscopic model (solid line)
and the quasiempirical amplitudes (dashed line). As expected already
from the comparison of the amplitudes in Sect.~\ref{sec:5_1}  the
results agree
quite well in the  $\rho$ channel. On the other hand some discrepancies
occur in the $\sigma$ channel. At smaller $t$-values, which determine
the
long-range part of the correlated exchanges, the theoretical model
yields somewhat more attraction than the quasiempirical results.
Larger discrepancies occur at higher $t$-values, which are however
of minor relevance for the correlated exchange in the $NN$ interaction.
Namely, above $t\approx30 m_\pi^2$   the correlated exchanges are of
shorter range
than $\omega$ exchange, which generates the strong repulsive inner part
of the $NN$  potential (as well as of the other baryon-baryon
potentials).
Therefore the short-ranged parts of the correlated $\pi\pi$
exchanges are
completely masked by the repulsive $\omega$ exchange and have a small
influence only on $NN$  observables. Furthermore one has to realize that
the quasiempirical results obtained by extrapolation of data from the
physical region of the $s$- and $u$-channel into the pseudophysical
 region
of the $t$-channel have considerable uncertainties.


\subsection{ The potential of correlated $\pi\pi$ and
$K\anti{K}$ exchange}

Using the spectral functions of the last chapter we can now evaluate
the dispersion integrals, Eqs.~\ref{eq:3_4_1}-\ref{eq:3_4_2b}, in order
to obtain the invariant
amplitudes in the $s$-channel. Eq.~\ref{eq:3_5_0} then provides the
(on-shell)
baryon-baryon interaction due to correlated $\pi\pi$ and $K\anti{K}$
exchange.

\subsubsection{The effective coupling constants}
\label{sec:3_5_1}

The results will be presented in terms of the effective coupling
strengths $G^{\sigma,\rho}_{AB\to CD}(t)$, which have been introduced
in Eqs.~\ref{eq:3_effccsig}, \ref{eq:3_effccrho} to parametrize
the correlated processes by (sharp mass) $\sigma$  and $\rho$ exchange.
We stress once more that this parametrization does not involve any
approximations as long as the full $t$-dependence of the effective
coupling strengths are taken into account. The parameters of $\sigma$
resp.\  $\rho$ exchange (mass of the exchanged particle: $m_\sigma$,
$m_\rho$; cutoff mass:
$\Lambda_\sigma$, $\Lambda_\rho$) are chosen to have the same values
in all particle channels.
$m_\sigma$ and  $m_\rho$ have been set to the values used in the
Bonn-J\"ulich models
of the $NN$~\cite{MHE} and $YN$~\cite{Holz} interactions, i.e.\
$m_\sigma=550\ MeV$, $m_\rho=770\ MeV$.
The cutoff masses      have been chosen such that the
coupling strengths in the $S=0, -1$ baryon-baryon channels vary only
weakly
with $t$. The resulting values ($\Lambda_\sigma=2.8\ GeV$,
$\Lambda_\rho=2.5\ GeV$) are quite large
compared to the values of the phenomenological parametrizations used in
Refs.~\cite{MHE,Holz}  and thus represent very hard form factors. Note
that, unless
stated otherwise, the upper limit $t'_{max}$ in the dispersion integrals
is put to $120 m_\pi^2$. Fig.~\ref{fig:5_6_1}  shows the effective
coupling strengths $G^{\sigma }_{AB}(t)$
in the baryon-baryon channels considered here, as function of $-t$. With
the exception of $G^{\sigma }_{\Sigma\Sigma}(t)$ (dash-dotted curve)
the effective
$\sigma$ coupling strengths vary only weakly with $-t$, which proves
that $550\ MeV$
is a realistic choice for the $\sigma$ mass. In the $\Sigma\Sigma$
channel the suitable
mass lies somewhat higher due to the strong coupling to the $K\anti{K}$
channel
generated by $\Delta$ exchange, cf.\ Fig.~\ref{fig:5_ssborn}.
If we compare the relative strengths of effective $\sigma$ exchange in
the
various baryon-baryon channels we observe the same features observed
already for the spectral functions: The scalar-isoscalar part of
correlated $\pi\pi$ and $K\anti{K}$  exchange is in the $ NN$  channel
about twice as
large as in both $YN$  channels and by a factor 3-4 larger than in the
$S=-2$  channels.

For sharp mass $\sigma$ exchange the following equation holds
\[
G^{\sigma }_{NN\to NN}(t) G^{\sigma }_{\Sigma\Sigma\to\Sigma\Sigma}(t)
=
[G^{\sigma }_{N\Sigma\to N\Sigma}(t)]^2\quad.
\]
In other words, the three processes are determined by two  coupling
constants
$g_{\sigma NN} \equiv \sqrt{G^{\sigma }_{NN\to NN}}$
and $g_{\sigma\Sigma\Sigma}\equiv\sqrt{G^{\sigma }_{\Sigma\Sigma\to
\Sigma\Sigma}}$,
such that $G^{\sigma }_{N\Sigma\to N\Sigma}(t)=g_{\sigma NN}
g_{\sigma\Sigma\Sigma}$.
For correlated
exchanges this is not necessarily true anymore; here we have in general
\[
G^{\sigma }_{NN\to NN}(t) G^{\sigma }_{\Sigma\Sigma\to\Sigma\Sigma}(t)
\geq
[G^{\sigma }_{N\Sigma\to N\Sigma}(t)]^2 \quad,
\]
which is just a consequence of the Schwarz inequality relation
\[
\; |\int f(t) g(t) dt |^2 \leq \int f(t)^2 dt \times \int g(t)^2 dt
\quad.
\]
The equality holds if both functions have the same $t$-dependence. This
is roughly fulfilled for $NN$ and $\Lambda\Lambda$, but not for
$\Sigma\Sigma$;
therefore the
equality approximately holds in the first but not in the second case.
Consequently, in general, effective vertex couplings for correlated
exchanges
are not well defined since they might take different values in
different baryon-baryon channels.

Tab.~\ref{tab:5_6_10} contains the coupling strengths at $t=0$.
Besides the results
for our full model for correlated $\pi\pi$ and $K\anti{K}$ exchange
the table
includes results obtained when neglecting the
$B\anti{B}\to K\anti{K}$  Born amplitudes.
Obviously the inclusion of these amplitudes provides only $15\%$ of the
coupling strengths in the $NN$ channel; it is much more important in the
channels with strangeness, in fact providing the dominant part in the
$S=-2$ channels. Furthermore the table contains results obtained when
uncorrelated contributions involving spin-1/2 baryons only are
subtracted from the discontinuities of the invariant baryon-baryon
amplitudes in order to avoid double counting in case a simple OBE-model
is used in the $s$-channel. For the full Bonn $NN$ model contributions
involving spin-3/2 baryons have to be subtracted too (as done in
general in this work) since corresponding contributions are already
treated explicitly in the s-channel. Obviously processes involving
spin-3/2 baryons increase the `true' correlated contribution by about
$30\%$
in all channels.

In the $\rho$-channel the spectral functions are dominated by the
resonant
contributions in the region of the $\rho$ resonance. Therefore the
effective coupling strengths $^{ij}G^\rho_{AB\to CD}(t)$ $(ij=VV,VT,TV,
TT)$
vary even more
slowly with $t$ than those in the $\sigma$-channel. Because of this weak
$t$-dependence it is for the moment sufficient to consider only the
values of coupling strengths at $t=0$. They are shown for the $NN$
channel in
Tab.~\ref{tab:5_6_11}, for $YN$ in Tab.~\ref{tab:5_6_12} and in
Tab.~\ref{tab:5_6_13}
 for the $S=-2$ baryon-baryon channels.


Note that for the equal (unequal) mass case the results are given in
terms of 3 (4) coupling strengths. For equal masses the present
description in terms of correlated $\pi\pi$ exchange is more involved
compared to sharp mass $\rho$-exchange since the latter can be
characterized by two  parameters only, the vector and tensor coupling
constant. Also, as in the scalar channel, there is no definite relation
between coupling strengths in the various channels so that vertex
coupling constants cannot be uniquely extracted but depend on the
channel chosen. (Thus it is not surprising that our $\rho NN$ coupling
strengths, which are determined in the $NN$ system, are not fully
consistent with the vector and tensor coupling constants derived by
H\"ohler and Pietarinen~\cite{Hoehler2,Dumbrajs} from $\pi N$
scattering, though both calculations
agree of course qualitatively.)

Tables~\ref{tab:5_6_11}-\ref{tab:5_6_13} include results obtained when
only the $\rho$-pole terms are
considered in the $B\anti{B'}\to\mu\anti{\mu}$
Born amplitudes. In this case the effective
coupling strengths are still $SU(3)$ symmetric, i.e.\  they roughly
fulfill
the relations
\begin{equation}
\begin{array}{l@{\qquad}l}
g_{\Sigma\Lambda\rho}=0\quad, &
f_{\Sigma\Lambda\rho}={2\sqrt{3}\over 5} \left(g_{NN\rho}+f_{NN\rho}
\right)\quad, \\
g_{\Sigma\Sigma\rho}=2g_{NN\rho}\quad,&
f_{\Sigma\Sigma\rho}=-{2\over5}\left(3g_{NN\rho}-2f_{NN\rho}\right)
\quad,\\
g_{\Xi\Xi\rho}= g_{NN\rho}\quad,&
f_{\Xi\Xi\rho}= -{1\over5}\left(6g_{NN\rho}+f_{NN\rho}\right)\quad.\\
\end{array}
\end{equation}
with $\alpha_v^e=1$ and $\alpha_v^m=0.4$.
(We remind the reader that we have chosen the
bare couplings to exactly obey $SU(3)$ symmetry.) Obviously the
influence
of ($SU(3)$ broken) baryon masses on the $\rho$-pole contributions is
small,
probably because our calculations are performed in the pseudophysical
region, far below the baryon-baryon thresholds. As expected however the
effective coupling strengths of the complete calculation do not respect
$SU(3)$ symmetry anymore, due to the sizable influence of the (non-pole)
baryon-exchange processes.

Doing again a restricted subtraction of spin-1/2 baryon contributions
only (suitable for an OBE model) we now do not have a unique trend for
the change of coupling strengths in all channels, as found before in
the scalar case. In case of the vector coupling strength ($VV$) in the
$NN$ system the restricted subtraction leads even to a smaller value.
This
is not too surprising if one realizes that the $\rho$-vector coupling
strengths are obtained from differences of approximately equal
spectral functions so that well-controlled changes in the
spectral functions can lead sometimes to large modifications of coupling
strengths, in arbitrary direction.

At this point we would like to make a remark about the sensitivity of
our results to the upper limit in the dispersion integral, $t'_{max}$.
It
is in fact quite small: Lowering $t'_{max}$ from the generally used
value of
$120m_\pi^2$ to $80m_\pi^2$ the effective $\sigma$-coupling strengths
are
increased by less than $10\%$; in the $\rho$-channel the variations
are even
smaller. Moreover, ratios of coupling strengths in the various channels
are practically unchanged.


Two points remain to be addressed:
\begin{description}
\item[(i)]
 For the couplings of baryons to
pseudoscalar mesons we have assumed in our calculations $SU(3)$ symmetry
to be realized for pseudoscalar-type coupling, cf.\
Sect.~\ref{sec:bornsp}. Results are
different when the same symmetry is assumed for couplings of
pseudovector type, since (apart from the $NN\pi$ coupling) $BB'\mu$
couplings are then increased by a factor $(M_B+M_B')/2M_N$ leading to
much stronger $\sigma$-coupling strengths in the strange baryon-baryon
channels, see Table~\ref{tab:5_6_4}.
\item[(ii)]
 We have assumed for the $F/(F+D)$ ratios
$\alpha_p=\alpha_v^m=0.4$
predicted by the static quark model. If we change $\alpha_p$
by about $10\%$ (to 0.45) the resulting changes in the effective
$\sigma$-coupling
strengths are much less than $10\%$. The reason is that part of the Born
amplitudes are increased while others are decreased, so that the total
effect is quite small. The situation is completely different for
$\alpha_v^m$,
which determines the bare tensor couplings $f_{BB'\rho}^{(0)}$ in the
$\rho$ pole graph.
The same increase of $\alpha_v^m$ (to 0.45) leads to strong
modifications in
the $\rho$ coupling strengths, especially for the tensor-tensor part.
\end{description}


\subsubsection{Comparison with other models}

In the $NN$ channel we can also determine the effective
$\sigma$- and $\rho$-coupling strengths from the quasiempirical
$N\anti{N}\to \pi\pi$ amplitudes~\cite{Hoehler2,Dumbrajs}.
Corresponding spectral functions have already been discussed before.
We now show in Fig.~\ref{fig:5_6_2}  the results for the
product of
effective coupling strengths and form factors obtained from our
microscopic model and the quasiempirical
amplitudes~\cite{Hoehler2,Dumbrajs}, in comparison
to values used in the (full)  Bonn potential~\cite{MHE}. Since the
quasiempirical
amplitudes are  available only up to $t'_{max}=50m_\pi^2$, a
corresponding
cutoff is used in the dispersion integral. If we use the same lower
cutoff also for our model corresponding results essentially agree with
those obtained for the quasiempirical amplitudes. Obviously the slightly
stronger increase of the spectral function $\rho^\sigma_S$ of the
microscopic
model at small $t'$ roughly compensates for the larger maximum of the
quasiempirical spectral function at $t'\approx 20m_\pi^2$. Furthermore
we obtain
a considerable reduction of strength in the $\sigma$-channel from
inclusion of the repulsive contributions above $t'=50 m_\pi^2$ whereas
in the $\rho$-channel such pieces have only a very small influence.
Especially in the $\rho$-channel our results are considerably larger
than the values used in the Bonn potential. The reason is the form
factor
with $\Lambda_{NN\rho}=1.4\,GeV$ used in the Bonn potential, which
reduces the
strength at $t=0$ by as much as $50\%$.
One final point remains to be addressed: Our present results differ
considerably (by up to $30\%$) from our former calculations based on
a different microscopic model for the $N\anti{N}\to\pi\pi$ amplitudes.
The reason
for these discrepancies (in the spectral functions and effective
coupling
strengths) is that the subtraction of the uncorrelated terms from
the discontinuities is model-dependent. Both models differ in the
parametrization of $\Delta$ exchange (\ref{eq:Del_exch}); furthermore
the model of Ref.~\cite{Kim}
does not include the $\rho$-pole term in the transition amplitude. Still
both models provide a similar description of the quasiempirical data.

The average size of the effective coupling strengths is only a rough
measure of the strength of correlated $\pi\pi$ and $K\anti{K}$
exchange in the
various particle channels. The precise energy dependence of the
correlated exchange as well as its relative strength in the different
partial waves of the $s$-channel reaction is determined by the spectrum
of the exchanged invariant masses, i.e. the spectral functions, leading
to a different $t$-dependence of the effective coupling strengths.

Fig.~\ref{fig:5_6_3} shows the on-shell $NN$ potentials in spin-singlet
states with
angular momentum $L=0,2$, and 4, which are generated by the
scalar-isoscalar
part of correlated $\pi\pi$ and $K\anti{K}$ exchange. As expected it is
attractive
throughout. Slight differences occur between the potentials derived from
the microscopic model for the $B\anti{B'}\to\mu\anti{\mu}$
 amplitudes and those determined
from the quasiempirical $N\anti{N}\to\pi\pi$ amplitudes, which can be
traced to
differences in the spectral function $\rho^\sigma_S(t)$
(cf.\ Fig.~\ref{fig:5_5_4}):
For small $t'$ the
microscopic input is larger; therefore the corresponding potential in
high
partial waves (which is dominated by the small-$t'$ behavior) is by
about
$20\%$ larger than the quasiempirical result. In the $^1S_0$ partial
wave, on the
other hand, medium and short ranged exchange processes characterized
by larger
$t'$-values contribute. In this region the microscopic amplitudes are
considerably weaker; furthermore they contain the repulsive
contributions
above the $K\anti{K}$ threshold (cf.\ Fig.~\ref{fig:5_5_1}).
Consequently the resulting potential
is somewhat less attractive in the $^1S_0$  partial wave.
In agreement with Ref.~\cite{Kim} our present results (evaluated either
from the
microscopic model or the quasiempirical amplitudes) are stronger
than $\sigma'$-exchange of the full Bonn potential. The difference is
especially large in high partial waves since $\sigma'$-exchange, which
corresponds to a spectral function proportional to
$\delta(t'-m^2_\sigma)$, does not
contain the long-range part of the correlated processes. Indeed if we
parametrize our results derived from the microscopic model by
$\sigma$-exchange
as before (Sect.~\ref{sec:3_5_1}) but use for the effective coupling
strength
$G^\sigma_{NN\to NN}(t)$
the constant value at $t=0$ we obtain rough agreement with our
unapproximated
result in the $^1S_0$ partial wave but underestimate it considerably
in high
partial waves. Obviously the replacement of correlated $\pi\pi$ and
$K\anti{K}$
 exchanges
by an exchange of a sharp mass $\sigma$ meson with $t$-independent
coupling
strength cannot provide a simultaneous description of low and high
partial
waves.


It is interesting to compare our results for the effective $\sigma$-
and $\rho$-
coupling strengths in baryon-baryon channels with non-vanishing
strangeness
with those used in the hyperon-nucleon interaction models of the
Nijmegen~\cite{NijII,NijIII,NijIV}
and J\"ulich~\cite{Holz} groups. These are (with the exception of the
J\"ulich
model B which will not be considered in the following) OBE models, i.e.\
$\sigma$ (and $\rho$) exchange effectively include uncorrelated
processes
involving the $\Delta$-isobar. Therefore we have to use for comparison
dispersion-theoretic results in which only uncorrelated processes
involving spin--1/2 particle intermediate
states have been subtracted.
Table~\ref{tab:5_6_6} shows the relative coupling strengths in the
different
baryon-baryon channels for various models, in the $\sigma$-channel.
Apart from the
Nijmegen model D~\cite{NijII}, in which the scalar $\epsilon$-meson is
treated as an $SU(3)$
singlet and therefore couples with the same strength to all channels,
the
interaction is by far strongest in the $NN$ channel for all remaining
models,
and it becomes weaker with increasing strangeness. Obviously, the
Nijmegen
soft core model~\cite{NijIV} is nearest to the dispersion-theoretic
predictions.
Table~\ref{tab:5_6_6b} shows the analogous results in the
$\rho$-channel for the
vector-vector ($VV$) and tensor-tensor ($TT$) components. There are
sizable differences
between the effective coupling constants  from correlated exchange and
the
coupling constants of the  OBE-models as well
as among the OBE-models themselves. The latter models assume the vector
coupling to the isospin current to be universal, which fixes its
relative
strength in the different particle channels (apart from form factors
included in the J\"ulich model): In the $N\Sigma$ channel it is twice
as large as in
the $NN$ channel and vanishes for the transition $N\Lambda\to N\Sigma$.
The correlated
exchange result deviates strongly, which is another manifestation that
$SU(3)$ symmetry does not hold for correlated exchanges, even if it is
assumed for the bare $\rho BB'$ couplings present in our microscopic
model.

\section{Summary and Outlook}

An essential part of baryon-baryon interactions is the strong attraction
of medium range, which in one-boson-exchange models is parametrized by
an exchange of a fictitious scalar-isoscalar meson with a mass of about
$500 MeV$. In extended meson exchange models this part is naturally
generated by two-pion-exchange processes. Besides uncorrelated processes
correlated terms have to be considered in which both pions interact
during their exchange; in fact these terms provide the main contribution
to the intermediate-range interaction.

In the scalar-isoscalar channel of the $\pi\pi$ interaction the coupling
to
the $K\anti{K}$ channel plays a strong role, which has to be explicitly
included
in any model meant to be realistic for energies near and above the
$K\anti{K}$
threshold. As kaon exchange is an essential part of hyperon-nucleon
interactions a simultaneous investigation of correlated $\pi\pi$ and
$K\anti{K}$
exchanges is clearly suggested. In this work we have therefore derived
the correlated $\pi\pi$ as well as $K\anti{K}$ exchange contributions
in various
baryon-baryon channels. Starting point of our calculations was a
microscopic model for the transition amplitudes of the baryon-antibaryon
system ($B\anti{B'}$) into two pseudoscalar mesons ($\pi\pi$,
$K\anti{K}$) for energies below
the $B\anti{B'}$ threshold. The correlations between the two mesons
have been
taken into account by means of $\pi\pi-K\anti{K}$ amplitudes (determined
likewise fieldtheoretically~\cite{Lohse,Pearce_piN}), which provide an
excellent reproduction
of empirical $\pi\pi$ data up to $1.3 GeV$. With the help of unitarity
and
dispersion-theoretic methods we have then determined the baryon-baryon
amplitudes for correlated $\pi\pi$ and $K\anti{K}$ exchange in the
$J^P=0^+$ ($\sigma$) and
$J^P=1^-$ ($\rho$) $t$-channel.

In the $\sigma$-channel the strength of correlated $\pi\pi$ and
$K\anti{K}$ exchange
decreases with the strangeness of the baryon-baryon channels becoming
more negative. In the $NN$ channel the scalar-isoscalar part of
correlated
exchanges is by about a factor of 2 stronger than in both
hyperon-nucleon
channels ($\Lambda N$, $\Sigma N$) and by a factor 3 to 4 stronger than
in the $S=-2$
channels ($\Lambda \Lambda$, $\Sigma\Sigma$, $N\Xi$).
The influence of $K\anti{K}$ exchange is strong in
baryon-baryon channels with non-vanishing strangeness while it is
small in the $NN$ channel. This feature can be traced to different
coupling constants and isospin factors and especially to the different
masses involved in the various baryon-antibaryon channels.

The role of correlated $K\anti{K}$ exchange is small in the
$\rho$-channel. Here
the correlations are dominated by the (genuine) $\rho$-resonance in the
$\pi\pi$ interaction. Among the various $B\anti{B'} - \pi\pi,
K\anti{K}$ Born amplitudes
the direct coupling of the $\rho$-resonance to the baryons in the form
of a $\rho$-pole graph provides the dominant contribution to correlated
exchange.

It turns out that our results depend only slightly on the upper limit
(cutoff) introduced in the dispersion integral. Some uncertainty results
from applying $SU(3)$ resp.\ $SU(6)$ relations to either pseudoscalar or
pseudovector $\pi$ and $K$ coupling constants. Note that the same
problem
occurs already in OBE-models of the hyperon-nucleon interaction.
Ultimately it has to be decided by comparison with experiment which
procedure is to be preferred. Moreover, if instead of $SU(6)$ symmetry
$SU(3)$ symmetry is assumed only, the results for correlated exchanges
depends on the $F/(F+D)$ ratios $\alpha_p$ and $\alpha_v^m$. While the
dependence on $\alpha_p$
is only weak variation of$\alpha_v^m$ leads to noticeable changes in
the
model predictions for the correlated exchange in the $\rho$-channel.
Also
here a final decision about the correct choice of $\alpha_p$ and
$\alpha_v^m$ can be
made only by comparison with experiment. Again these parameters
occur already in OBE hyperon-nucleon models. Therefore no new
parameters are introduced when including correlated $\pi\pi$ and
$K\anti{K}$
exchange in baryon-baryon interaction models. On the contrary, the
elimination of single $\sigma$ and $\rho$ exchange reduces the number of
free parameters and thus enhances the predictive power of corresponding
interaction models.

Our results can be represented in terms of suitably defined effective
coupling strengths. It turns out that the resulting values in the
various baryon-baryon channels are not connected by $SU(3)$ relations.
For
example, although we have even assumed $SU(6)$ symmetry for the coupling
strength of the bare $\rho$ to the baryon current sizable baryon
exchange
processes destroy this symmetry in the final effective couplings.
Consequently the assumption of $SU(3)$ symmetry for single $\sigma$- and
$\rho$-exchange is not supported by our findings.

With this model constructed in the present work it is now possible to
take correlated $\pi\pi$ and $K\anti{K}$ exchange reliably into account
in the
various baryon-baryon channels. Especially in channels in which only
little
empirical information exists the elimination of phenomenological
$\sigma$-
and $\rho$-exchange considerably enhances the predictive power of
baryon-baryon interaction models. Clearly the inclusion of correlated
exchange
in existing interaction models (e.g.\ the Bonn $NN$
potential~\cite{MHE} and the
J\"ulich $YN$ models~\cite{Holz}) requires readjustment of free model
parameters to
the empirical data. Having fixed these parameters in the $NN$ and $YN$
channel the interaction model can then be extended parameter-free to
other baryon-baryon channels with strangeness $S=-2$ using $SU(3)$
arguments
for the genuine couplings. In this way in the frame of the Bonn-J\"ulich
models the possibility arises for the first time to make sensible
statements about the existence of bound baryon-baryon states with
strangeness $S=-2$, which should be of some importance regarding the
analysis of H-dibaryon experiments.

\begin{appendix}
\vskip1cm
\noindent {\Large \bf {Appendix}}

\section{Conventions}
\label{sec:AnhA}

As far as possible  the conventions in this work are chosen in
accordance with Ref.~\cite{BjDr}.
The helicity spinors $u(\vec p,\lambda)$ ($v(\vec p,\lambda)$)
of a Dirac particle of spin 1/2 and mass $M$
are solutions of the free Dirac equation in momentum space for
positive (negative) energy and helicity $\lambda$
\begin{eqnarray}
(\not{\!p} - M) u(\vec p,\lambda) &=& 0\quad,
\nonumber \\
(\not{\!p} + M) v(\vec p,\lambda) &=& 0\quad,
\end{eqnarray}
with the 4-momentum $p^\mu$
($p^0 \equiv  E_p = + \sqrt{M^2 + \vec p\,^2}$) and
$\not\!{ p} \equiv { p}^\mu \gamma_\mu$.
With the phase convention of Ref.~\cite{Hipp} the helicity spinors
read ($\lambda=\pm 1/2$)
\begin{eqnarray}
u(\vec p,\lambda) &=& \sqrt{\epsilon_p \over 2 M}
\left( \begin{array}{c}
1 \\
2\lambda |\vec p\,|  /  \epsilon_p
\end{array} \right)
\ket{\lambda}  \quad  ,
 \nonumber \\
v(\vec p,\lambda) &=& \sqrt{\epsilon_p \over 2 M}
\left( \begin{array}{c}
- |\vec p\,| / \epsilon_p   \\
2\lambda
\end{array} \right)
\ket{-\lambda} \quad,
\label{eq:app2a}
\end{eqnarray}
where $\epsilon_p \equiv  M + E_p$
They are normalized according to
\begin{eqnarray}
\anti{u}(\vec p,\lambda) u(\vec p,\lambda') &=& \delta_{\lambda\lambda'}
\quad,
\nonumber \\
\anti{v}(\vec p,\lambda) v(\vec p,\lambda') &=&
-\delta_{\lambda\lambda'} \quad,
\end{eqnarray}
with $\anti{u}(\vec p,\lambda) =
u^\dagger (\vec p,\lambda) \gamma_0$.

If  $\vec p$ lies in the $xz$-plane and encloses the  polar angle
$\theta$ with the $\check{e}_z$-axis
the eigenstates  $\ket{\lambda} $
 of the helicity operator,
\begin{equation}
{\vec\sigma \over 2} \cdot {\vec p \over |\vec p\,| } \ket{\lambda}
=
\lambda \ket{\lambda}\quad,
\end{equation}
are related to the  Pauli spinors $\chi_m$ by
\begin{equation}
\ket{\lambda} =
\exp \left( -{i\over2} \sigma_2 \theta \right) \chi_\lambda
\quad.
\end{equation}
where
\begin{equation}
\vec \sigma\cdot \check{e}_z \; \chi_{\pm{1\over2}}
 = \pm  \chi_{\pm{1\over2}}
\quad.
\end{equation}

The same phase convention as in Ref.~\cite{Hipp} is adopted for the
helicity spinors~\ref{eq:app2a}; namely, the particle and antiparticle
spinors are related by the charge conjugation $\cal C$~\cite{BjDr}:
\begin{equation}
v(\vec p,\lambda)= {\cal C} \anti{u}^T (\vec p,\lambda)
=i \gamma_2 {u}^* (\vec p,\lambda)\quad.
\end{equation}

Helicity eigenstates of two-particle systems in the center-of-mass
frame are built as a product of two helicity spinors according to the
phase convention of Jacob and Wick~\cite{JW}.
In case of two spin-1/2 particles 1 and 2, the helicity
spinors~\ref{eq:app2a} are used for particle  1 (momentum $\vec p\,$)
and for particle 2 (momentum  $-\vec p\,$) the following spinors are
used:
\begin{eqnarray}
u(-\vec p,\lambda_2) &=& \sqrt{\epsilon_p \over 2 M}
\left( \begin{array}{c}
1 \\
2\lambda_2 |\vec p\,|  /  \epsilon_p
\end{array} \right)
\ket{\lambda_2}   \quad ,
 \\
v(-\vec p,\lambda_2) &=& \sqrt{\epsilon_p \over 2 M}
\left( \begin{array}{c}
 |\vec p\,| / \epsilon_p \\
-2\lambda_2
\end{array} \right)
\ket{-\lambda_2}   \quad,
\label{eq:app3a}
\end{eqnarray}
where
\begin{equation}
\ket{\lambda_2} = \exp \left( -{i\over2} \sigma_2 \theta \right)
\chi_{-\lambda_2}\quad.
\end{equation}

The hadronic field operators  can be expanded in momentum space
solutions
of the corresponding equation of motion. For Dirac particles
($J^P=\half^+$) and (pseudo)scalar particles ($J^P=0^\pm$), see
Ref.~\cite{BjDr}.
For spin-1 and spin-3/2 particles the field operators $\phi^\mu$ and
$\psi^\mu$, respectively, then read:
\begin{equation}
\phi^\mu(x)={1\over (2\pi)^{3/2}}
\sum_\lambda \int d^3k
{1 \over \sqrt {2 \omega_k}}
\epsilon^\mu(\vec k,\lambda)
\left[
a(\vec k,\lambda)  e^{-i k \cdot x}
+
a^\dagger (\vec k,\lambda)  e^{i k \cdot x}
\right]
\end{equation}
\begin{equation}
\psi^\mu(x)={1\over (2\pi)^{3/2}}
\sum_\Lambda \int d^3p
\sqrt{M \over E_p}
\left[
b(\vec p,\Lambda) u^\mu(\vec p,\Lambda) e^{-i p \cdot x}
+
d^\dagger (\vec p,\Lambda) v^\mu(\vec p,\Lambda) e^{i p \cdot x}
\right]
\end{equation}
with the polarization vector $\epsilon^\mu(\vec k,\lambda)$,
cf.~\cite{Lurie}.
Note that the hermitian conjugated component of the spherical operator
$a$ is
related to the component of the hermitian operator $a^\dagger$
by~\cite{Edmonds}:
\begin{equation}
a(\vec k,\lambda)^\dagger =  (-1)^\lambda a^\dagger (\vec k,-
\lambda)\quad.
\label{eq:app4a}
\end{equation}
The Rarita-Schwinger spinors~\cite{Lurie} $ u^\mu(\vec p,\Lambda)$ and
 $v^\mu(\vec p,\Lambda)$
are solutions of the Rarita-Schwinger equation
\begin{equation}
\begin{array}{l@{\qquad}rcl@{\qquad\qquad}rcl}
&(\not{\!p} - M) u^\mu(\vec p,\Lambda) &=& 0,&
(\not{\!p} + M) v^\mu(\vec p,\Lambda) &=& 0 \\
\mbox{and} &&&&&& \\
&\gamma_\mu u^\mu(\vec p,\Lambda) &=& 0,&
\gamma_\mu v^\mu(\vec p,\Lambda) &=& 0.
\end{array}
\label{eq:AnhARS1}
\end{equation}
The creation and destruction operators of bosons  (fermions) follow
the usual (anti-)commutator relations.

For spin-1 and spin-3/2 particles~\cite{Read}
the Feynman propagators  $S_v(k)$ and $S_D^{\mu\nu}(p;A)$,
which are derived from the time-ordered product
of the corresponding field operators~\cite{BjDr}, read in momentum
space:
\begin{equation}
 S_v(k) = {-g^{\mu\nu} + k^\mu k^\nu/m^2  \over k^2 - m^2 +
i\epsilon}
\label{eq:rhoprop}
\end{equation}
\begin{eqnarray}
 S_D^{\mu\nu}(p;A)& =& {\not\! p + M   \over p^2 - M^2 + i\epsilon}
\left[
-g^{\mu\nu}
+ {\gamma^\mu \gamma^\nu \over 3}
+{2\over 3 M^2} p^\mu p^\nu
-{p^\mu \gamma^\nu - p^\nu \gamma^\mu \over 3M}
\right]
\nonumber \\ [\baselineskip]
&&+ S^{\mu\nu}_{\rm non-pole} (p;A)
\label{eq:AnhARS}
\end{eqnarray}
with the non-pole part
\begin{equation}
S^{\mu\nu}_{\rm non-pole} (p;A) =
{1 \over 3 M^2} {A+1 \over 2A+1}
\left[
\gamma^\mu {{(A+1)\over2} \not\!p   + A M \over 2A+1} \gamma^\nu
-(p^\mu \gamma^\nu + p^\nu \gamma^\mu )
\right]\quad.
\end{equation}
Without restricting the general validity of the results
the parameter $A$ is set in our work to $A=-1$ so that the non-pole
part vanishes (see Sect.~\ref{sec:bornsp}).

\section{ Matrix elements of $B \overline{B'} \to \mu\overline{\mu}$
Born amplitudes}
\label{sec:AnhC}

In order to evaluate the helicity amplitudes~\ref{eq:born_heli} of the
$B \overline{B'} \to \mu\anti{\mu}$ $
(\mu\anti{\mu}=\pi\pi,\;K\anti{K})$ Born
terms~\ref{eq:4born32}-\ref{eq:4born33} the coordinate system is
conveniently chosen such that the relative momentum $\vec q$ of the
$B \overline{B'}$ state points along the $z$-axis and the relative
momentum $\vec k$ of the two pseudoscalar mesons $\mu\overline{\mu}$
lies in the $xz$-plane; i.e., the components of the two momenta read
\begin{equation}
\vec q=\left(
\begin{array}{c}
0 \\ 0 \\ q
\end{array}
\right)\quad,
\qquad
\vec k=\left(
\begin{array}{c}
k \sin\vartheta \\ 0 \\ k \cos\vartheta
\end{array}
\right)
\quad,
\end{equation}
with the scattering angle $\vartheta=\winkel(\vec p,\vec k\,)$.

As explained in Sect.~\ref{sec:kap4} the $B \overline{B'} \to
\mu\overline{\mu}$ amplitudes need to be evaluated only for
$B\overline{B'}$ states being on their mass-shell:
\begin{equation}
\sqrt{t} = E_B +E_{B'} = \sqrt{M_B+q^2} + \sqrt{M_{B'}+q^2}
\end{equation}
Therefore, in the following,  $t$ is suppressed  as an argument of the
helicity amplitudes.

In case of the $B \overline{B'} \to \pi\pi$ amplitudes
one has in principle to take into account also the exchange graph (with
the external pion lines exchanged)  arising
from the symmetrized $\pi\pi$ states.
However, this can be simply done by considering the selection rule for
the $\pi\pi$ states:
$ (-1)^{J+I}=+1$ ($J$: total angular momentum, $I$: total isospin);
i.e., multiplying the direct, partial wave decomposed
 $B \overline{B'} \to \pi\pi$ amplitude by
a factor $(1+(-1)^{J+I})$.

In the final results for the $B \overline{B'} \to
\mu\overline{\mu}$ Born amplitudes given below we introduced the
following
abbreviations:
\begin{equation}
\begin{array}{rcl@{\qquad}rcl}
\alpha_+ & \equiv& \epsilon_B + \epsilon_{B'}\;,&
\alpha_- & \equiv  & \epsilon_B - \epsilon_{B'}\;, \\
\beta_+ & \equiv & \epsilon_B \epsilon_{B'} + q^2\;, &
\beta_- & \equiv & \epsilon_B \epsilon_{B'} - q^2\;, \\
\end{array}
\end{equation}
with $\epsilon_B\equiv M_B + E_B$.

\subsection{ Exchange of a $J^P= {1\over2}^+$ baryon}

\begin{eqnarray}
\bra{\mu\overline{\mu},\vec k} V_X \ket{B \overline{{B'}},\vec q ++} &=&
 C  \{
-q [ M_X (t + 4 k^2) \alpha_+
           - q_0 (t - 4 k^2) \alpha_-
           + (t + 4 k^2) \beta_-  ]
\nonumber \\
&& +k\cos\theta    [  4 M_X \sqrt{t} \beta_+
                 + (t - 4 k^2) \beta_- ]
\nonumber \\
&& +8 k^2 q\cos^2\theta   \beta_-  \}
\quad,
\end{eqnarray}

\begin{equation}
 \bra{\mu\overline{\mu},\vec k} V_X \ket{B \overline{{B'}},\vec q +-}=
-C k \sin\theta
     \{       - 4 q^2 \sqrt{t} \alpha_+
        + (t - 4 k^2) \beta_+
        + 4 M_X \sqrt{t} \beta_-
  + 8 k q \cos\theta  \beta_+     \}
\; ,
\end{equation}
where
\begin{equation}
C:=
 - {f_{BX\mu} f_{{B'}X\mu}\over 8 m_\mu^2
    \sqrt{\epsilon_B \epsilon_{B'} M_B M_{B'} }}
{F^2_{X}(p^2)
\over q_0^2 - E_X^2} \quad,
\end{equation}
with
$E_X=\sqrt{\vec p\,^2 + M_X^2}$ with $\vec p=\vec q -\vec k$.

\subsection{
Exchange of a  $J^P={3\over2}^+$ baryon}
\renewcommand{\arraystretch}{1.3}
\begin{equation}
\begin{array}{rrl}
\multispan{3}{
$ \bra{\mu\overline{\mu},\vec k} V_X \ket{B \overline{{B'}},\vec q ++}=
$\hfill
}
 \\
 C  \Biggl\{ {1 \over q_0^2 - E_X^2 }
\{ & q & [
       M_X A(t,k) \alpha_+
   + q_0 A(t,k) \alpha_-
   + 2 M_X  k^2 \sqrt{t} \beta_+
+ A(t,k) \beta_- ]
 \\
&+ k\cos\theta & [
      8 M_X q^2 k^2 \alpha_+
  + 8 q_0 q^2 k^2 \alpha_-
\\ &&
   + 2 M_X \sqrt{t} (q_0^2 - M_X^2 - q^2 - k^2)
                 \beta_+
+ ( 8 k^2 q^2 - A(t,k) ) \beta_- ]
 \\
&+ 2 k^2 q \cos^2\theta & [
     - 2 M_X q^2 \alpha_+
 - 2 q_0 q^2 \alpha_-
+  M_X \sqrt{t} \beta_+
 - 2 ( q^2 + 2 k^2) \beta_- ]
 \\
 & + 4 k^3 q^2\cos^3\theta & \beta_- \}
 \\
+x_\Delta   \{ & -q & [
       M_X (t + 4 k^2)   \alpha_+
   - 2  q_0 t \alpha_- ]
 \\
&+ 4 k \cos\theta & [
      M_X \sqrt{t} \beta_+
  - 2 k^2  \beta_- ]
 \\
&+ 8 k^2 q \cos^2\theta &
     \beta_-  \}
 \\
+x_\Delta^2   \{ & -q & [
       2 M_X (t + 4 k^2)  \alpha_+
   - q_0  (t - 4 k^2) \alpha_-
 + (t + 4 k^2)  \beta_- ]
 \\
&+ k\cos\theta & [
   + 8 M_X \sqrt{t} \beta_+
 + (t - 4 k^2)  \beta_- ]
 \\
&+ 8 k^2 q \cos^2\theta &
     \beta_-  \}
\Biggr\}
\end{array}
\end{equation}
\begin{equation}
\begin{array}{rrl}
\multispan{3}{
$\bra{\mu\overline{\mu},\vec k} V_X \ket{B \overline{{B'}},\vec q +-}=$
\hfill
}
 \\
 C k \sin\theta
\times
\Biggl\{  {1 \over q_0^2 - E_X^2 }  \{ & -&  [
      2 M_X^2 \sqrt{t} q^2  \alpha_+
    + 2 M_X q_0  q^2 \sqrt{t} \alpha_-
 \\&&
   - A(t,k) \beta_+
 + 2 M_X \sqrt{t} (q_0^2 - M_X^2 - k^2)
    \beta_- ]
 \\
&+ 2 k q \cos\theta & [
   4 k^2  \beta_+
 - M_X \sqrt{t} \beta_- ]
 \\
&- 4 k^2 q^2 \cos^2\theta &
\beta_+
 \}
 \\
+4 x_\Delta    \{ &&
        2 k^2  \beta_+
    - M_X \sqrt{t} \beta_-
 \\
& - 2  k q \cos\theta &
     \beta_+
        \}
 \\
+x_\Delta^2   \{   & - & [
     - 4 q^2 \sqrt{t} \alpha_+
    + (t - 4 k^2)  \beta_+
 + 8 M_X \sqrt{t} \beta_- ]
 \\
& - 8 k q \cos\theta &
    \beta_+ \}
\Biggr\} \quad,
\end{array}
\end{equation}
\renewcommand{\arraystretch}{1.}
with
\begin{equation}
C:=
 - {f_{BX\mu} f_{{B'}X\mu}\over 12 m_\mu^2 M_X^2
\sqrt{\epsilon_B \epsilon_{B'} M_B M_{B'} }}
F^2_{X}(p^2) \quad,
\end{equation}
\begin{equation}
A(t,k)  := q_0^2 t - 4 k^2 M_X^2 - M_X^2 t  - 4 k^4 \quad.
\end{equation}

\subsection{ $\rho$-pole graph}
\label{sec:AnhC_rho}

For the $\rho$-pole graph the $\rho$-propagator and the form factor
$F_{\mu\mu\rho}(k^2)$
(cf.\ Eq.~\ref{eq:rhoFF}) do not depend on the scattering angle
$\theta$.
Therefore, the partial wave decomposition can easily be performed
analytically.
The results are:
\begin{eqnarray}
\bra{\mu\overline{\mu}, k} V^J_\rho \ket{B \overline{{B'}}, q ++}
&=&  \delta_{J1}
C_\rho  k
[g^{(0)}_{B{B'}\rho} \beta_-
+ {f^{(0)}_{B{B'}\rho} \over 2 M_N }  \sqrt{t}
\beta_+ ],
 \\
\bra{\mu\overline{\mu}, k} V^J_\rho \ket{B \overline{{B'}}, q +-}&=&
\delta_{J1}
\sqrt{2} C_\rho  k
[g^{(0)}_{B{B'}\rho} \beta_+
+ {f^{(0)}_{B{B'}\rho} \over 2 M_N }  \sqrt{t}
\beta_-],
\end{eqnarray}
where
\begin{equation}
C_\rho:=
{ 4\pi \over 3}
{ g^{(0)}_{\mu\mu\rho}  \over \sqrt{\epsilon_B \epsilon_{B'} M_B M_{B'}
}}
{F_{\mu\mu\rho}(k^2) \over t - (m^{(0)}_\rho)^2}
\quad.
\end{equation}


\section{$SU(3)$ relations for coupling constants}
\label{sec:borniso}

The microscopic model for the $B\anti{B'}\to \pi\pi, K\anti{K}$
amplitudes presented in Sect.~\ref{sec:kap4} imposes $SU(3)$ symmetry to
the coupling constants at the hadronic vertices in order to keep the
number of free parameters as low as possible.
The $SU(3)$ relations for the coupling constants are derived by
constructing an $SU(3)$-invariant interaction Lagrangian from
particle field operators, which possess a well-defined behavior under
$SU(3)$
transformations. This method is well-established~\cite{deSw}
 and makes use of the
$SU(3)$ Clebsch-Gordan coefficients and the so-called isoscalar factors
which are tabulated in Ref.~\cite{deSw}. Therefore, we give in the
following only the final results relevant for this work.

The coupling of the pseudoscalar meson octet ($\pi$, $\eta_8$, $K$,
$\anti{K}$) to the current of the $J^P=1/2^+$ baryon octet ($N$,
$\Lambda$, $\Sigma$, $\Xi$)  is described
by the Lagrangian~\cite{deSw}
\renewcommand{\arraystretch}{1.3}
\begin{equation}
\begin{array}{rcl@{\!\!\quad}cl}
{\cal L}_{B'Bp}&=\;\;\;& g_{NN\pi} (N^\dagger \vec\tau N) \cdot
 \vec\pi & +& g_{NN\eta_8} (N^\dagger N) \eta_8 \\ &+&g_{\Lambda NK}
 \left[ (N^\dagger K) \Lambda + \Lambda^\dagger(K^\dagger N) \right] &
 +&g_{\Sigma NK} \left[ (N^\dagger\vec\tau K)\cdot\vec\Sigma +
 \vec\Sigma^\dagger\cdot(K^\dagger\vec\tau N) \right] \\
 &+&g_{\Sigma\Lambda\pi} \left[\vec\Sigma^\dagger\cdot\vec\pi \Lambda
 +\Lambda^\dagger \vec\Sigma\cdot\vec\pi\right] &
 -&ig_{\Sigma\Sigma\pi}\left(\vec\Sigma^\dagger\times\vec\Sigma\right)
\cdot\vec\pi
 \\ &+&g_{\Lambda\Lambda\eta_8}\Lambda^\dagger\Lambda\eta_8 &
 +&g_{\Sigma\Sigma\eta_8} \vec\Sigma^\dagger\cdot\vec\Sigma \eta_8 \\
 &+&g_{\Xi\Lambda K}\left[ \Lambda^\dagger(\anti{K}^\dagger \Xi) +
 (\Xi^\dagger \anti{K}) \Lambda\right] & +&g_{\Xi\Sigma K} \left[
 \vec\Sigma^\dagger\cdot(\anti{K}^\dagger\vec\tau\, \Xi) +
 (\Xi^\dagger\vec\tau \anti{K})\cdot\vec\Sigma
\right]
 \\ &+&g_{\Xi\Xi\pi} (\Xi^\dagger \vec\tau \, \Xi) \cdot \vec\pi &
+&g_{\Xi\Xi\eta_8} (\Xi^\dagger \Xi) \eta_8\quad,
\end{array}
\label{eq:su3_11a}
\end{equation}
\renewcommand{\arraystretch}{1.}
where the coupling constants being of the pion and the kaon
 are given by
\renewcommand{\arraystretch}{1.5}
\begin{equation}
\begin{array}{l@{\qquad}l}
g_{NN\pi}=g\quad, & g_{\Lambda NK}=-{1\over\sqrt 3}g (1+2\alpha)\quad,\\
g_{\Sigma NK}= g (1-2\alpha)\quad,& g_{\Sigma\Lambda\pi}={2\over\sqrt 3}
g(1-\alpha)\quad,\\
g_{\Sigma\Sigma\pi}=2g \alpha\quad,g_{\Xi\Lambda K}={1\over\sqrt 3} g
(4\alpha-1)\quad,\\
g_{\Xi\Sigma K}=-g \quad,g_{\Xi\Xi\pi}= -g (1-2\alpha)\quad.\\
\end{array}
\label{eq:su3_11b}
\end{equation}
\renewcommand{\arraystretch}{1.}
Hence, the coupling of $\pi$ and $K$ to the baryon octet is determined
by two
parameters: the coupling strength $g_8$ and the so-called
$F/(F+D)$-ratio $\alpha_p$.

The transition from pseudoscalar to vector mesons can be simply made by
the replacement
\begin{equation}
\pi\to\rho ,\quad K\to K^* ,\quad  \eta\to\phi ,\quad \eta'\to\omega.
\nonumber
\end{equation}
However, as can be seen from the spin-momentum part
of the interaction Lagrangian Eq.~\ref{eq:BBvcoup2}
the vector mesons couple to the baryon octet in two different
ways: via the vector and the tensor coupling with coupling constants
$g$ and
$f$, respectively.
Now, it is not clear which coupling constants underly $SU(3)$ symmetry
relations. Besides the canonic assumption~\cite{NijII,NijIV}
that the $SU(3)$ relations apply to
$g$ and $f$, sometimes~\cite{NijIII,Holz} the electric ($g$) and the
magnetic ($G=g+f$) coupling constants are subject to $SU(3)$.
In fact, both assumptions are in a certain sense equivalent~\cite{Reu}.
For given $SU(3)$ parameters $(g_1,g_8,\alpha_v^g)$ and
 $(f_1,f_8,\alpha_v^f)$ the parameters $(G_1,G_8,\alpha_v^m)$ of the
magnetic coupling can be chosen such that both coupling schemes lead to
the same tensor couplings $f=G-g$.

Extending the $SU(3)$ symmetry to $SU(6)$,   predictions for
the $F/(F+D)$-ratios of the different multiplet couplings can be
derived~\cite{Pais}:
\begin{equation}
\alpha_p=0.4 ,\qquad
\alpha_v^e=1 ,\qquad
\alpha_v^m=0.4\quad.
\label{eq:alpha_su6}
\end{equation}
 $\alpha_v^e=1$ corresponds to the usual assumption of
universal electric coupling of the  $\rho$ meson to the isospin
current~\cite{Sakurai}, demanding for instance the equality $g_{NN\rho}$
and
$g_{KK\rho}$.

The coupling of the pseudoscalar meson octet to the current for the
transition between the $J^P=3/2^+$ baryon decuplet ($\Delta$, $Y^*$,
$\Xi^*$,
$\Omega$) and the $J^P=1/2^+$ baryon octet is given by the
Lagrangian~\cite{ReuDiss}:
\renewcommand{\arraystretch}{1.2}
\begin{equation}
\begin{array}{rcl@{\quad}cl@{\quad}cl}
{\cal L}_{DBp}& =\;& f_{N\Delta\pi} (N^\dagger \vec T
 \Delta)\cdot\vec\pi &&&& \\ &+& f_{NY^*K}(N^\dagger\vec\tau K)\cdot
 \vec Y^* & +& f_{\Sigma\Delta K}
 \vec\Sigma^\dagger\cdot(K^\dagger\vec T\Delta)&& \\ &+& f_{\Lambda
 Y^*\pi} \Lambda^\dagger\vec Y^*\cdot\vec\pi & -& i f_{\Sigma Y^*\pi}
 \left(\vec\Sigma^\dagger\times\vec Y^*\right)\cdot\vec\pi& +&
 f_{\Sigma Y^*\eta_8} \vec\Sigma^\dagger\cdot\vec Y^* \eta_8 \\ &+&
 f_{\Lambda \Xi^*K} \Lambda^\dagger(\anti{K}^\dagger \Xi^*) & +&
 f_{\Sigma \Xi^*K} \vec\Sigma^\dagger\cdot(\anti{K}^\dagger\vec\tau
 \Xi^*) & +& f_{\Xi Y^*K} (\Xi^\dagger\vec\tau \anti{K})\cdot\vec Y^*
 \\ &+& f_{\Xi\Xi^*\pi} (\Xi^\dagger \vec\tau \Xi^*) \cdot \vec\pi &
 +& f_{\Xi\Xi^*\eta_8} (\Xi^\dagger \Xi^*) \eta_8&& \\ &+&
 f_{\Xi\Omega K} (\Xi^\dagger K) \Omega &&&& \\ &+&\quad h.c.&&&&
\end{array}
\label{eq:su3_22}
\end{equation}
\renewcommand{\arraystretch}{1.}
with the $\pi$ and $K$ coupling constants
\renewcommand{\arraystretch}{1.5}
\begin{equation}
\begin{array}{l@{\qquad}l@{\qquad}l}
 f_{N\Delta\pi}=f\quad, && \\ f_{NY^*K}=-f/\sqrt{6}\quad, &
 f_{\Sigma\Delta K}=-f\quad, & \\
f_{\Lambda Y^*\pi}=f/\sqrt{2} \quad,
 & f_{\Sigma Y^*\pi}=-f/\sqrt{6} \quad,
& f_{\Lambda \Xi^*K}=f/\sqrt{2}\quad, \\
 f_{\Sigma \Xi^*K}=f/\sqrt{6} \quad, & f_{\Xi Y^*K}=-f/\sqrt{6} \quad,
& f_{\Xi\Xi^*\pi}=-f/\sqrt{6}\quad, \\
f_{\Xi\Omega K}=f\quad. &&
\end{array}
\end{equation}
Obviously, the coupling constants now depend only on one parameter,
the coupling strength $f$.


The $SU(3)$ part of the  transition amplitudes is obtained by evaluating
the matrix elements of the appropriate products of the
$SU(3)$-Lagrangians~\ref{eq:su3_11a} and \ref{eq:su3_22}.
If the transition amplitudes are evaluated between states of definite
total
isospin  the results for the $SU(3)$ part can be separated
into the product of the corresponding coupling constants and the
so-called isospin factor.
For instance, the isospin factor $T_{B_1\anti{B}_2\to \mu\mu'}^X (I)$
for the
transition of a baryon-antibaryon state $B_1\anti{B}_2$
(with $B_1,B_2 \in \{ 8_B\}$)
to a pseudoscalar meson $\mu$ and its antiparticle $\anti{\mu}$ follows
from
\begin{eqnarray}
\lefteqn{
T_{B_1\anti{B}_2\to \mu\mu'}^X (I) g_{B_1X\mu} g_{B_2X\mu'} \equiv
\bra{\mu\anti{\mu},Im} \left( {\cal LL'} \right)_X \ket{B_1\anti{B}_2,
Im}
} \nonumber \\ && =\sum_{m_{B_1},m_{B_2} \atop m_\mu,m_{\anti{\mu}}}
\bra{I_{B_1} I_{\bar B_2} m_{B_1} m_{\bar B_2}}\kket{Im}
\bra{I_\mu I_{ \anti{\mu}} m_\mu m_{\anti{\mu}}}\kket{Im}
\nonumber \\
&&\qquad
\bra{\mu I_\mu m_\mu , \anti{\mu} I_{ \anti{\mu}}  m_{\anti{\mu}}}
\left( {\cal LL'} \right)_X
\ket{B_1 I_{B_1} m_{B_1}, \anti{B}_2 I_{\bar B_2}  m_{\bar B_2}}\quad,
\label{eq:su3_22a}
\end{eqnarray}
with $m$ denoting the $z$-component of the total isospin $I$
and a corresponding notation for the particle isospins
$I_{B_1},I_{\bar B_2}, I_\mu,I_{\mu'}$.
The index $X$ in $\left( {\cal LL'}\right)_X $ indicates that the two
Lagrangians ${\cal L}$ and ${\cal L'}$ are coupled by contracting the
field operators of the exchanged baryon isomultiplet $X$ (or of the
$\rho$-meson in case of the $\rho$-pole diagram).

The isospin factors obtained in this way for the processes included in
the Born terms of the $B\anti{B'}\to\mu\anti{\mu}$ model (cf.\
Fig.~\ref{fig:4_0b})  are listed in
 Tab.~\ref{tab:4_1}. Note that due to the phase
convention~\ref{eq:SU3phase}
introduced in connection with the isospin-crossing matrix in
Section~\ref{sec:kap3_6} some isospin
factors (e.g. those for $N\anti{N}\to\pi\pi$) differ in sign  from the
ones usually used in the literature~\cite{Kim,Durso}.

\end{appendix}
\def\Nucl{Nucl.\ }
\def\Phys{Phys.\ }
\def\Rev{Rev.\ }
\def\Lett{Lett.\ }
\def\PL{\Phys\Lett}
\def\PLB{\Phys\Lett B}
\def\NP{\Nucl\Phys}
\def\NPA{\Nucl\Phys A}
\def\NPB{\Nucl\Phys B}
\def\NPBS{\Nucl\Phys (Proc.\ Suppl.\ )B}
\def\PR{\Phys\Rev}
\def\PRL{\Phys\Rev\Lett}
\def\PRC{\Phys\Rev C}
\def\PRD{\Phys\Rev D}
\def\RMP{\Rev  Mod.\ \Phys}
\def\ZP{Z.\ \Phys}
\def\ZPA{Z.\ \Phys A}
\def\ZPC{Z.\ \Phys C}
\def\AOP{Ann.\ \Phys}
\def\PRep{\Phys Rep.\ }
\def\ANP{Adv.\ in \Nucl\Phys Vol.\ }
\def\PTP{Prog.\ Theor.\ \Phys}
\def\PTPS{Prog.\ Theor.\ \Phys Suppl.\ }
\def\PL{\Phys \Lett}
\def\JPF{J.\ Physique}
\def\FBSS{Few--Body Systems, Suppl.\ }
\def\IJMP{Int.\ J.\ Mod.\ \Phys A}
\def\NuCi{Nuovo Cimento~}


%
%
%
\renewcommand{\arraystretch}{1.8}
\begin{table}[p]
\begin{displaymath}
\begin{array}{|cc|cc||cc|c||c|c|}
\hline
A&B&C&D&I_s&I_t&X(I_s,I_t)& F_\sigma & F_\rho \\
\hline\hline
N	&N	&N	&N	&0,1	&0,1	&
\left(
\begin{array}{cc}
1/2 & -3/2	\\
1/2 & 1/2
\end{array} \right)&
1/2&1/2\\
\hline
N	&\Lambda&N	&\Lambda&1/2		&0	& -1/\sqrt{2}
&-1/\sqrt{2}&\mbox{---}\\
\hline
N	&\Sigma	&N	&\Sigma	&1/2,3/2	&0,1 &
\left(
\begin{array}{cc}
1/\sqrt{6}	& -1	\\
1/\sqrt{6}	& 1/2
\end{array} \right)
&1/\sqrt{6}&1/2 \\
\hline
N	&\Sigma&N	&\Lambda&1/2		&1	& -\sqrt{3/2}
&\mbox{---}&-1/\sqrt{2}\\
\hline
\Lambda	&\Lambda&\Lambda&\Lambda&0		&0	& 1
&1&\mbox{---}\\
\hline
\Sigma	&\Sigma	&\Sigma	&\Sigma	&0,1,2		&0,1,2 &
\left(
\begin{array}{ccc}
1/3	&-1	&5/3	\\
1/3	&-1/2	&-5/6	\\
1/3	&+1/2	&1/6	\\
\end{array} \right)
&1/3&1/2\\
\hline
N	&\Xi	&N	&\Xi	&0,1	&0,1	&
\left(
\begin{array}{cc}
-1/2 & 3/2	\\
-1/2 & -1/2
\end{array} \right)
&-1/2&-1/2\\
\hline
\end{array}
\end{displaymath}
\caption{Isospin-crossing matrices $X(I_s,I_t)$ for the various
baryon-baryon channels $A+B\to C+D$ considered in this paper.
The  $t$-channel reaction related by crossing reads  $A+\anti{C}\to
D+\anti{B}$.
In the matrix $X(I_s,I_t)$  $I_s$ increases from top to bottom and $I_t$
from left to right.
The last two columns contain the factors $F_\sigma$ and $F_\rho$
(see text for further explanation).
}
\label{tab:3_1}
\end{table}
\renewcommand{\arraystretch}{1.}

\renewcommand{\arraystretch}{1.3}
\begin{table}[p]
\begin{center}
\begin{tabular}{|c|r@{\qquad}c@{\qquad}c@{\qquad}c@{\qquad}c@{\qquad}|}
\hline
particle & mass ($MeV$) & $B$ & $S$ & $J^P$ & $I$  \\
\hline
$\pi$		& 139.57 	& 0	&$0$	&$0^-$	&$1$	\\
$K$		& 495.82 	& 0	&$+1$	&$0^-$	&${1\over2}$
	\\
$\rho$		& 770.	 	& 0	&$0$	&$1^-$	&$1$	\\
	&	\multispan{1}{($m_\rho^{(0)}: 1151.26 $)\hfill} &&&& \\
\hline
$N$		& 938.919 	& 1	&$0$	&${1\over2}^+$	&
${1\over2}$\\
$\Lambda$	& 1115.68 	& 1	&$-1$	&${1\over2}^+$	&$0$
     \\
$\Sigma$	& 1193.1  	& 1	&$-1$	&${1\over2}^+$	&$1$
    \\
$\Xi$		& 1318.1  	& 1	&$-2$	&${1\over2}^+$	&
${1\over2}$\\
$\Delta$	& 1232.   	& 1	&$0$	&${3\over2}^+$	&
${3\over2}$\\
$Y^*\equiv\Sigma(1385)$
		& 1385.    	& 1	&$-1$	&${3\over2}^+$	&
$ 1$       \\
$\Xi^*$		& 1533.4  	& 1	&$-2$	&${3\over2}^+$	&
${1\over2}$\\
$\Omega^-$	& 1672.5  	& 1	&$-3$	&${3\over2}^+$	&
$0$        \\
\hline
\end{tabular}
\end{center}
\caption{
Particles considered in the model for the $B\anti{B'}\to\mu\anti{\mu}$
Born amplitudes.
Given are their masses and the relevant quantum numbers
($B$: baryon number, $S$: strangeness, $J$: spin, $P$: parity, $I$:
isospin).}
\label{tab:4_0}
\end{table}

\renewcommand{\arraystretch}{1.3}
\begin{table}[p]
\begin{center}
\begin{tabular}{c@{\qquad}c@{\qquad}|@{\qquad}c@{\qquad}c@{\qquad}c}
\hline
$B\anti{B'}\to\mu\anti{\mu}$ & process &  $T(0)$ & $T(1)$ & $T(2)$ \\
\hline\hline
$N\anti{N}\to\pi\pi$			& $N$		& $+\sqrt{6}$
	& $+2$		&	\\
		 			& $\Delta$	& $+\sqrt{8/3}$
	& $-2/3$	&	\\
		 			& $\rho$	&
	& $-2$		&	\\
\hline
$N\anti{N}\to K\anti{K}$		& $\Lambda$	& $+1$
	& $+1$		&	\\
					& $\Sigma, Y^*$	& $+3$
	& $-1$		&	\\
		 			& $\rho$	&
	& $-2$		&	\\
\hline\hline
$\Lambda\anti{\Lambda}\to\pi\pi$	& $\Sigma, Y^*$	& $-\sqrt{3}$
	& 		&	\\
\hline
$\Lambda\anti{\Lambda}\to K\anti{K}$	& $N$		& $-\sqrt{2}$
	& 		&	\\
					& $\Xi, \Xi^*$	& $-\sqrt{2}$
	& 		&	\\
\hline\hline
$\Sigma\anti{\Sigma}\to\pi\pi$		& $\Lambda$	& $+1$
	& $+1$		& $+1$	\\
					& $\Sigma, Y^*$	& $+2$
	& $+1$		& $-1$	\\
		 			& $\rho$	&
	& $-2$		&	\\
\hline
$\Sigma\anti{\Sigma}\to K\anti{K}$	& $N$		& $+\sqrt{6}$
	& $-2$		&	\\
					& $\Delta$	& $+\sqrt{8/3}$
	& $+2/3$	&	\\
					& $\Xi, \Xi^*$	& $+\sqrt{6}$
	& $+2$		&	\\
		 			& $\rho$	&
	& $-2$		&	\\
\hline\hline
$\Lambda\anti{\Sigma}\to\pi\pi$		& $\Sigma, Y^*$	&
	& $-\sqrt{2}$	& 	\\
		 			& $\rho$	&
	& $+\sqrt{2}$	&	\\
\hline
$\Lambda\anti{\Sigma}\to K\anti{K}$	& $N$		&
	& $-\sqrt{2}$	&	\\
					& $\Xi, \Xi^*$	&
	& $+\sqrt{2}$	&	\\
		 			& $\rho$	&
	& $+\sqrt{2}$	&	\\
\hline\hline
$\Xi\anti{\Xi}\to\pi\pi$		& $\Xi, \Xi^*$	& $-\sqrt{6}$
	& $-2$		& 	\\
		 			& $\rho$	&
	& $+2$		&	\\
\hline
$\Xi\anti{\Xi}\to K\anti{K}$		& $\Lambda$	& $-1$
	& $+1$		& 	\\
					& $\Sigma, Y^*$	& $-3$
	& $-1$		& 	\\
					& $\Omega^-$	& $-1$
	& $-1$		& 	\\
		 			& $\rho$	&
	& $+2$		&	\\
\hline
\end{tabular}
\end{center}
\caption{Isospin factors  $T(I)$ for the various
$B\anti{B'}\to\mu\anti{\mu}$ Born amplitudes.
In the column labelled `process'
the exchanged baryon
($Y^*\equiv \Sigma(1385)$) or a
`$\rho$' in case of a $\rho$-pole diagram is given.
}
\label{tab:4_1}
\end{table}
\renewcommand{\arraystretch}{1.}

\renewcommand{\arraystretch}{1.3}
\begin{table}[p]
\[
\begin{array}{|c|cl|}
\hline
\kappa_{\rho}^{(0)}	&4.21	&	\\
 x_\Delta 		&-0.823	&	\\
\Lambda_8 		&1779.1	&MeV	\\
\Lambda_{10}		&1704.7	&MeV	\\
\hline
\alpha_{p}		&2/5 &(SU(6)\protect\cite{Pais}) \\
\alpha_v^m		&2/5 &(SU(6)\protect\cite{Pais})  \\
\hline
\end{array}
\]
\caption{Parameter of the  microscopic model
for the  $B\anti{B'}\to\mu\anti{\mu}$  Born amplitudes.
The four parameter  $\kappa_{\rho}^{(0)}$, $x_\Delta$, $\Lambda_8$ and
$\Lambda_{10}$  are adjusted to the quasiempirical  $N\anti{N}\to\pi\pi$
amplitudes of Refs.~\protect\cite{Hoehler2,Dumbrajs}.}
\label{tab:5_param}
\end{table}
\renewcommand{\arraystretch}{1.}

\renewcommand{\arraystretch}{1.3}
\begin{table}[p]
\[
\begin{array}{|r|r|rr|rrr|}
\hline
\multicolumn{7}{|c|}{G^\sigma_{AB\to AB}/4\pi } \\
\hline
&NN&N\Lambda&N\Sigma&\Lambda\Lambda&\Sigma\Sigma&N\Xi	\\
\hline
\mbox{full model}			&5.87	&2.82	&2.58	&1.52
	&1.72	&1.19	\\
\mbox{without $K\anti{K}$ Born terms}	&5.07	&1.80	&1.06	&0.64
	&0.22	&0.35	\\
\mbox{subtractions for OBE model}	&7.77	&3.81	&3.15	&2.00
	&2.31	&1.52	\\
\hline
\end{array}
\]
\caption{Effective $\sigma$ coupling strengths $G^\sigma_{AB\to AB}
(t=0)$
for correlated  $\pi\pi$ and $K\anti{K}$ exchange in the various
baryon-baryon channels.
(The meaning of the several rows is given in the text.)
}
\label{tab:5_6_10}
\end{table}
\renewcommand{\arraystretch}{1.}


\renewcommand{\arraystretch}{1.3}
\begin{table}[p]
\[
\begin{array}{|r|rcr|}
\hline
\multicolumn{4}{|c|}{^{ij}G^\rho_{NN\to NN}/4\pi\quad (ij=VV,VT,TV,TT) }
 \\
\hline
 &\multicolumn{3}{c|}{NN}
\\
&VV&VT,TV&TT \\
\hline
\mbox{full model}
&1.00	&5.35	&28.91	\\
\mbox{without baryon exchange}
&0.52	&2.17	&9.13	\\
\mbox{subtraction for OBE model}
&0.33	&5.58	&35.23	\\
\hline
\end{array}
\]
\caption{Effective $\rho$ coupling strengths $^{ij}G^\rho_{NN\to NN}$
$(ij=VV,VT,TV,TT)$ at $t=0$
for correlated $\pi\pi$ and $K\anti{K}$ exchange in the $NN$ channel.
(The meaning of the several rows is given in the text.)
}
\label{tab:5_6_11}
\end{table}
\renewcommand{\arraystretch}{1.}


\renewcommand{\arraystretch}{1.3}
\begin{table}[p]
\[
\begin{array}{|r|rrrr|rrrr|}
\hline
\multicolumn{9}{|c|}{^{ij}G^\rho_{NY\to NY'}/4\pi\quad (ij=VV,VT,TV,TT)
 } \\
\hline
&\multicolumn{4}{c|}{N\Sigma}
&\multicolumn{4}{c|}{N\Lambda\to N\Sigma}	\\
 &VV&VT&TV&TT	&VV&VT&TV&TT\\
\hline
\mbox{full model}
&1.64	&1.92	&8.95	&10.15	&-0.15	&3.97	&-0.81	&21.42	\\
\mbox{without baryon exchange}
&1.03	&1.12	&4.34	&4.70	&0.00	&1.86	&0.00	&7.83	\\
\mbox{subtraction for OBE model}
&1.53	&1.52	&9.87	&9.85	&-0.52	&4.54	&-1.61	&25.60	\\
\hline
\end{array}
\]
\caption{The same as  Tab.~\protect\ref{tab:5_6_11},
for the hyperon-nucleon channels  $N\Sigma$ and $N\Lambda$-$N\Sigma$.
}
\label{tab:5_6_12}
\end{table}
\renewcommand{\arraystretch}{1.}


\renewcommand{\arraystretch}{1.3}
\begin{table}[p]
\[
\begin{array}{|r|rcr|rrrr|}
\hline
\multicolumn{8}{|c|}{^{ij}G^\rho_{AB\to AB}/4\pi\quad (ij=VV,VT,TV,TT) }
 \\
\hline
 &\multicolumn{3}{c|}{\Sigma\Sigma}
&\multicolumn{4}{c|}{N\Xi}	\\
&VV&VT,TV&TT &VV&VT&TV&TT	\\
\hline
\mbox{full model}		&2.78	&3.14	&4.02 	&0.87	&-1.92
	&4.71	&-10.38	\\
\mbox{without baryon exchange}	&2.06	&2.23	&2.42	&0.52	&-1.05
	&2.17	&-4.43	\\
\mbox{subtraction for OBE model}&3.09	&2.64	&4.38	&0.99	&-2.30
	&5.40	&-12.09	\\
\hline
\end{array}
\]
\caption{The same as  Tab.~\protect\ref{tab:5_6_11},
for the baryon-baryon channels with strangeness $S=-2$.
}
\label{tab:5_6_13}
\end{table}
\renewcommand{\arraystretch}{1.}

\renewcommand{\arraystretch}{1.3}
\begin{table}[p]
\[
\begin{array}{|cr|r|rr|rrr|}
\hline
\multicolumn{8}{|c|}{G^\sigma_{AB\to AB}/4\pi ( \alpha_p)  } \\
\hline
&\alpha_p &NN&N\Lambda&N\Sigma&\Lambda\Lambda&\Sigma\Sigma&N\Xi	\\
\hline
		&0.35	&5.86	&2.83	&2.52	&1.49	&1.69	&1.21
	\\
\mbox{\sl ps}	&0.40	&5.87	&2.82	&2.58	&1.52	&1.72	&1.19
	\\
		&0.45	&5.90	&2.84	&2.70	&1.59	&1.80	&1.19
	\\
\hline
		&0.35	&6.00	&3.36	&3.16	&2.04	&2.37	&1.64
	\\
\mbox{\sl pv}	&0.40	&6.00	&3.31	&3.28	&2.04	&2.46	&1.57
	\\
		&0.45	&6.00	&3.31	&3.45	&2.10	&2.62	&1.54
	\\
\hline
\end{array}
\]
\caption{Effective $\sigma$ coupling strengths
$G^\sigma_{AB\to AB} (t=0)$
depending on the $F/(F+D)$-ratio  $\alpha_p$.
{\sl ps} ({\sl pv}) indicates that $SU(3)$-symmetry
is assumed for the   pseudoscalar (pseudovector) coupling constants
 $g_{BB'\mu}$ ($f_{BB'\mu}$).
The column `{\sl ps}, $\alpha_p=0.40$' agrees with the results of the
full model in  Tab.~\protect\ref{tab:5_6_10}.
 }
\label{tab:5_6_4}
\end{table}
\renewcommand{\arraystretch}{1.}

\renewcommand{\arraystretch}{1.3}
\begin{table}[p]
\[
\begin{array}{|r|rrrrrr|}
\hline
&NN&N\Lambda&N\Sigma&\Lambda\Lambda&\Sigma\Sigma&N\Xi	\\
\hline
\pi\pi+ K\anti{K}			&1	&0.49	&0.41	&0.26
	&0.30	&0.19	\\
\hline
\mbox{OBEPT~\protect\cite{MHE} \& J\"ulich A~\protect\cite{Holz}}
					&1	&0.45	&0.63	&0.34
	&0.66	&	\\
\mbox{Nijmegen D~\protect\cite{NijII}}	&1	&1   	&1   	&1
	&1   	&1   	\\
\mbox{Nijmegen F~\protect\cite{NijIII}}	&1	&0.74	&0.61	&0.55
	&0.37	&0.41	\\
\mbox{Nijmegen SC~\protect\cite{NijIV}}	&1	&0.58	&0.45	&0.34
	&0.20	&0.10	\\
\hline
\end{array}
\]
\caption{
Strength of the  $\sigma$-like contributions (at $t=0$)
to the various baryon-baryon interactions relative to the  $NN$ channel.
In case of the dispersiontheoretic result for $\pi\pi$ and $K\anti{K}$
exchange only those uncorrelated contributions are subtracted which
are generated already in the $s$-channel by the iteration of an $OBE$
potential
(cf.\  Tab.~\protect\ref{tab:5_6_10}).
In case of the  OBE models the numbers are extracted from the
coupling constants  of the  $\sigma$- (OBEPT \& J\"ulich A with
inclusion of form factors and averaging over the several isospin
channels)
or the  $\epsilon$-meson
(Nijmegen models and the $\Xi\Xi\epsilon$ coupling from $SU(3)$).
}
\label{tab:5_6_6}
\end{table}
\renewcommand{\arraystretch}{1.}


\renewcommand{\arraystretch}{1.3}
\begin{table}[p]
\[
\begin{array}{|r|rr|rr|rr|}
\hline
&\multicolumn{2}{c|}{NN}
&\multicolumn{2}{c|}{N\Sigma}
&\multicolumn{2}{c|}{N\Lambda\to N\Sigma} \\
&VV&TT	&VV&TT	&VV&TT	\\
\hline
\pi\pi+ K\anti{K}
&0.28&28.87&1.25&8.07&-0.43&20.97\\
\hline
\mbox{OBEPT~\protect\cite{MHE} \& J\"ulich A~\protect\cite{Holz}}
&0.50&18.57&0.79&8.87&0&9.84 \\
\mbox{Nijmegen D~\protect\cite{NijII}}	&
0.35&23.20&0.71&15.51&0&17.82 \\
\mbox{Nijmegen F~\protect\cite{NijIII}}	&
0.63&27.34&1.25&28.76&0&14.96\\
\mbox{Nijmegen SC~\protect\cite{NijIV}}	&
0.79&14.16&1.59&7.79&0&11.85 \\
\hline
\end{array}
\]
\caption{
Strength of the  $\rho$-like contributions (at $t=0$)
to the  interaction in the various baryon-baryon channels with $S=0,-1$.
The values for  $\pi\pi$ and $K\anti{K}$ exchange correspond to the
effective
$\rho$ coupling strengths in the third row of
Tabs.~\protect\ref{tab:5_6_11}--\protect\ref{tab:5_6_13}
times the form factors  for effective $\rho$ exchange.
For the  OBE models the values are determined from the coupling
constants of the $\rho$ meson (in case of OBEPT and J\"ulich A under
consideration of form factors).
}
\label{tab:5_6_6b}
\end{table}
\renewcommand{\arraystretch}{1.}
\clearpage
%
%
%
%
%
\begin{figure}
\caption{Two-pion exchange in the nucleon-nucleon interaction:
a) iterative boxes,
b) crossed boxes,
c) correlated two-pion exchange.
The iterative box with an $NN$ intermediate state is generated  in the
scattering equation  by iterating the one-pion exchange. In OBE models
all other contributions are parametrized by $\sigma_{OBE}$ and
$\rho$ exchange. In the Bonn $NN$ potential~\protect\cite{MHE} the
uncorrelated contributions a) and b) are evaluated explicitly whereas
the correlated $\pi\pi$ exchange is parametrized by  $\sigma'$ and
$\rho$ exchange.
 }
\label{fig:1_2}
\end{figure}

\begin{figure}
\caption{
Two-pion and two-kaon exchange in the baryon-baryon process
$A+B\to C+D$.
The unshaded ellipse denotes the direct coupling of the two
pseudoscalar mesons $\mu\anti{\mu}=\pi\pi,K\anti{K},\anti{K}K$ to the
baryons without any correlation effects (cf.\
Fig.~\protect\ref{fig:1_3}).
The shaded circle in the lower diagram for the correlated exchange
stands for the full off-shell amplitude of the process
$\mu\anti{\mu}\to\mu'\anti{\mu'}$.
}
\label{fig:1_1}
\end{figure}

\begin{figure}
\caption{
Microscopic model for the $B\anti{B'}\to\pi\pi,K\anti{K}$ Born
amplitudes.
The solid lines denote (anti-)baryons, the dashed lines the
pseudoscalar mesons  $\pi\pi$ or $K\anti{K}$.
The sum over exchanged baryons $X$ contains all members of the
$J^P={1\over2}^+$ octet and the $J^P={3\over2}^+$ decuplet which can
be exchanged in accordance with the conservation of strangeness and
isospin.
}
\label{fig:1_3}
\end{figure}

\begin{figure}
\caption{Two-particle scattering process.}
\label{fig:3_1_1}
\end{figure}

\begin{figure}
\caption{Model  for the $B\anti{B'} \to \mu\anti{\mu}$ amplitudes
($\mu\anti{\mu}=\pi\pi,K\anti{K}$).}
\label{fig:4_0a}
\end{figure}

\begin{figure}
\caption{Born amplitudes included in the model of
Ref.~\protect\cite{Pearce_piN}
for the $\pi\pi-K\anti{K}$ interaction.}
\label{fig:4_5}
\end{figure}

\begin{figure}
\caption{Contributions to the Born amplitudes
$\Sigma\anti{\Sigma}\to\pi\pi,K\anti{K}$.}
\label{fig:4_0b}
\end{figure}

\begin{figure}
\caption{$\pi\pi$ phase shifts in the  $\sigma$ and $\rho$ channel and
the corresponding inelasticity in the  $\sigma$ channel.
(From Ref.~\protect\cite{Pearce_piN}.)
}
\label{fig:4_6}
\end{figure}

\begin{figure}
\caption{$N\anti{N}\to\pi\pi,K\anti{K}$ helicity amplitudes in
Frazer-Fulco normalization.
The solid lines are the results of our microscopic model.
The squares denote the quasiempirical
data~\protect\cite{Hoehler2,Dumbrajs} obtained by analytic
continuation of the  $\pi N$ and $\pi\pi$ scattering amplitudes.
}
\label{fig:5_2}
\end{figure}

\begin{figure}
\caption{ $N\anti{N}\to\pi\pi,K\anti{K}$ Born amplitudes.}
\label{fig:5_nnborn}
\end{figure}

\clearpage

\begin{figure}
\caption{ $\Sigma\anti{\Sigma}\to\pi\pi,K\anti{K}$ Born amplitudes.}
\label{fig:5_ssborn}
\end{figure}

\begin{figure}
\caption{ $\pi\pi$ and $K\anti{K}$ amplitudes in the $\sigma$ channel
($J=I=0$)
as a function of the squared c.m.\ energy $t$. The dashed (solid) line
represents the real (imaginary) part of the amplitude calculated with
the fieldtheoretical model of Ref.~\protect\cite{Pearce_piN}.
 }
\label{fig:5_tmumu}
\end{figure}

\begin{figure}
\caption{ Helicity amplitudes  $T^J_\pm$ for
 $N\anti{N}\to\pi\pi$ (solid) and
$N\anti{N}\to K\anti{K}$ (dotted).
For $J=1$ the small $N\anti{N}\to K\anti{K}$ amplitudes are suppressed.
The dashed lines in the $\sigma$ channel denote the
$N\anti{N}\to\pi\pi$ amplitudes,
that are calculated without the  $N\anti{N}\to K\anti{K}$ Born
amplitudes. In the $\rho$ channel the dash-dotted lines are obtained if
only the $\rho$-pole graphs is included in the
$N\anti{N}\to \mu\anti{\mu}$ Born amplitudes.}
\label{fig:5_tnn}
\end{figure}

\begin{figure}
\caption{ $\Sigma\anti{\Sigma}\to\pi\pi,K\anti{K}$
helicity amplitudes $T^J_\pm$. For the meaning of the curves, see
Fig.~\protect\ref{fig:5_tnn}; in addition, the $\Sigma\anti{\Sigma}\to
K\anti{K}$ amplitudes are included for $J=1$.
}
\label{fig:5_tss}
\end{figure}

\begin{figure}
\caption{
 $N\anti{N}\to\pi\pi$ amplitude $T^0_+$ calculated with the full
model in the energy range around the $K\anti{K}$ threshold (dotted).
The solid (dashed) line denotes the real (imaginary) part of $T^0_+$.
}
\label{fig:5_thr}
\end{figure}

\begin{figure}
\caption{ Spectral function $\rho^\sigma_S(t)$ for the scalar
component of correlated
$\pi\pi$ and $K\anti{K}$ exchange in the scalar-isoscalar channel of
various baryon-baryon processes:}
\vspace{.3cm}
\protect\parbox{15cm}{
\hspace{.5cm} a) $NN$  (solid), $N\Lambda$ (short dashed),
$N\Sigma$ (dotted);\hfill \\
\phantom{}\hspace{.5cm} b) $\Lambda\Lambda$ (long dashed),
$\Sigma\Sigma$ (dash-dotted), $N\Xi$ (dash-double-dotted).\hfill
}
\label{fig:5_5_1}
\end{figure}

\begin{figure}
\caption{ Spectral function $\rho^\sigma_S(t)$
in the $NN$ (solid) and $N \Sigma$ (dotted) channel derived with the
full model (cf.\ Fig.~\protect\ref{fig:5_5_1}).
If the contributions of the
 $B\anti{B'}\to K\anti{K}$ Born amplitudes are neglected the dashed
($NN$) and the dash-dotted lines ($N\Sigma$) are obtained.
}
\label{fig:5_5_1b}
\end{figure}

\begin{figure}
\caption{ Spectral functions  $\rho^\rho_i(t)$ ($i=S,V,T,6$)
for the contribution of correlated  $\pi\pi$ and $K\anti{K}$ exchange
in the  $\rho$ channel to the $NN$ (solid), $N\Sigma$ (dotted),
 $\Sigma\Sigma$ (dash-dotted)
 and $N\Xi$ interaction (dash-double-dotted)
 as well as to the
 $N\Lambda-N\Sigma$ transition amplitude (short dashed).
}
\label{fig:5_5_2}
\end{figure}

\begin{figure}
\caption{$NN$ spectral functions  $\rho^\sigma_S(t)$ and
$\rho^\rho_i(t)$ ($i=S,V,T$).
The solid lines are derived with the microscopic model of
Sect.~\protect\ref{sec:kap4} for the
$B\anti{B'}\to \mu\anti{\mu}$ amplitudes. The results obtained with
the quasiempirical $N\anti{N}\to\pi\pi$ amplitudes of
Refs.~\protect\cite{Hoehler2,Dumbrajs} are denoted by the dashed
lines.
}
\label{fig:5_5_4}
\end{figure}

\begin{figure}
\caption{
Effective $\sigma$ coupling strengths
$G^\sigma_{AB\to AB}/(4\pi)$
for correlated  $\pi\pi$ and $K\anti{K}$ exchange as a function
of the squared 4-momentum transfer $t<0$ in the baryon-baryon channels:
$NN$ (solid), $N\Lambda$ (short dashed),
$N\Sigma$ (dotted), $\Lambda\Lambda$ (long dashed),
$\Sigma\Sigma$ (dash-dotted)
and $N\Xi$ (dash-double-dotted).
 }
\label{fig:5_6_1}
\end{figure}

\clearpage

\begin{figure}
\caption{ Effective  strength of the  $NN$ interaction due to
correlated $\pi\pi$ and $K\anti{K}$ exchange  in the  $\sigma$ and
$\rho$ channel as a function of the 4-momentum transfer $t<0$.
Shown are
$g^2_{\sigma NN}\!\equiv\! G^\sigma_{NN\to NN}$,
$g^2_{\rho NN}\!\equiv\, ^{VV}G^\rho_{NN\to NN}$,
$f^2_{\rho NN}\!\equiv \, ^{TT}G^\rho_{NN\to NN}$ and
$f_{\rho NN}/g_{\rho NN} \!\equiv \!
[^{TT}G^\rho_{NN\to NN}/\,  ^{VV}G^\rho_{NN\to NN}]^{1/2}$
including the form factors.
The solid (dotted) line is derived from  the microscopic model for
correlated
 $\pi\pi$ and $K\anti{K}$ exchange using $t'_{max}=120m_\pi^2$
($t'_{max}=50m_\pi^2$, only for $g_{\sigma NN}^2$).
The dashed line follows from the  quasiempirical $N\anti{N}\to\pi\pi$
amplitudes~\protect\cite{Hoehler2,Dumbrajs}.
The effective strength of $\sigma'$ and $\rho$ exchange in the Bonn
potential~\protect\cite{MHE}
is denoted by the dash-dotted line.
 }
\label{fig:5_6_2}
\end{figure}

\begin{figure}
\caption{ The $\sigma$-like part of the $NN$ on-shell potential in
various partial waves as a function of the kinetic energy in the
laboratory system.
The solid line is derived from our microscopic model for correlated
$\pi\pi$ and $K\anti{K}$ exchange (with $t'_{max}=120m_\pi^2$).
The dotted lines are obtained if this dispersiontheoretic result is
parametrized by $\sigma$ exchange
and the coupling strength $G^\sigma_{NN\to NN}(t)$ is subsequently set
to the constant value at $t=0$
The  dispersiontheoretic calculation using the  quasiempirical
$N\anti{N}\to\pi\pi$ amplitudes of
Refs.~\protect\cite{Hoehler2,Dumbrajs} gives the dashed lines.
Finally, the dash-dotted lines correspond to
$\sigma'$ exchange used in the Bonn potential~\protect\cite{MHE}.
 }
\label{fig:5_6_3}
\end{figure}

\vfill

\clearpage
\end{document}